\documentclass[reprint, amsmath,amssymb,aps,pre]{revtex4-2}

\usepackage[pdfborder={0 0 0.5 [3 2]}, plainpages=false]{hyperref}%
\usepackage[left=1in,right=1in,top=1in,bottom=1in]{geometry}%

\usepackage{enumerate}
\usepackage{amsfonts}
\usepackage{amsthm}
\usepackage{bbm}
\usepackage[table,xcdraw]{xcolor}
\usepackage{mathtools}
\usepackage{cool}
\usepackage{graphicx,epsfig}

\usepackage[capitalize,nameinlink]{cleveref}
\crefname{section}{section}{sections}
\crefname{subsection}{subsection}{subsections}
\Crefname{section}{Section}{Sections}
\Crefname{subsection}{Subsection}{Subsections}

\Crefname{figure}{Figure}{Figures}

\crefformat{equation}{\textup{#2(#1)#3}}
\crefrangeformat{equation}{\textup{#3(#1)#4--#5(#2)#6}}
\crefmultiformat{equation}{\textup{#2(#1)#3}}{ and \textup{#2(#1)#3}}
{, \textup{#2(#1)#3}}{, and \textup{#2(#1)#3}}
\crefrangemultiformat{equation}{\textup{#3(#1)#4--#5(#2)#6}}%
{ and \textup{#3(#1)#4--#5(#2)#6}}{, \textup{#3(#1)#4--#5(#2)#6}}{, and \textup{#3(#1)#4--#5(#2)#6}}

\Crefformat{equation}{#2Equation~\textup{(#1)}#3}
\Crefrangeformat{equation}{Equations~\textup{#3(#1)#4--#5(#2)#6}}
\Crefmultiformat{equation}{Equations~\textup{#2(#1)#3}}{ and \textup{#2(#1)#3}}
{, \textup{#2(#1)#3}}{, and \textup{#2(#1)#3}}
\Crefrangemultiformat{equation}{Equations~\textup{#3(#1)#4--#5(#2)#6}}%
{ and \textup{#3(#1)#4--#5(#2)#6}}{, \textup{#3(#1)#4--#5(#2)#6}}{, and \textup{#3(#1)#4--#5(#2)#6}}

\crefdefaultlabelformat{#2\textup{#1}#3}

\hypersetup{hidelinks}

\graphicspath{ {images/} }

\begin{document}

\title{Standing and Traveling Waves in a Model of Periodically Modulated One-dimensional Waveguide Arrays}

\author{Ross Parker}
\affiliation{Department of Mathematics, Southern Methodist University, Dallas, TX 75275, USA}
\email{rhparker@smu.edu}

\author{Jes\'us Cuevas-Maraver}
\affiliation{Grupo de F\'{\i}sica No Lineal, Departamento de F\'{\i}sica Aplicada I,
Universidad de Sevilla. Escuela Polit\'{e}cnica Superior, C/ Virgen de Africa, 7, 41011-Sevilla, Spain}
\affiliation{Instituto de Matem\'{a}ticas de la Universidad de Sevilla (IMUS). Edificio
Celestino Mutis. Avda. Reina Mercedes s/n, 41012-Sevilla, Spain}

\author{P.\,G. Kevrekidis} 
\affiliation{Department of Mathematics and Statistics, University of Massachusetts, Amherst MA 01003, USA}
\email{kevrekid@math.umass.edu}

\author{Alejandro Aceves}
\affiliation{Department of Mathematics, Southern Methodist University, Dallas, TX 75275, USA}
\email{aaceves@smu.edu}

\begin{abstract}
In the present work, we study coherent structures in a one-dimensional discrete nonlinear 
Schr{\"o}dinger lattice  in which the coupling between waveguides is periodically modulated. Numerical experiments with single-site initial conditions show that, depending on the power, the system exhibits two fundamentally different behaviors. At low power, initial conditions with intensity concentrated in a single site give rise to transport, with the energy moving unidirectionally along the lattice, whereas high power initial conditions yield stationary solutions. We explain these two behaviors, as well as the nature
of the transition between the two regimes, by analyzing a simpler model where the couplings between waveguides are given by step functions. For the original model, we numerically construct both stationary and moving coherent structures, which are solutions reproducing themselves exactly after an integer multiple of the coupling period. For the stationary solutions, which are true periodic orbits, we use Floquet analysis to determine the parameter regime for which they are spectrally stable. Typically, the traveling solutions are characterized by having small-amplitude, oscillatory tails, although we identify a set of parameters for which these tails disappear. These parameters turn out to be independent of the lattice size, and our simulations suggest that for these parameters, numerically exact
traveling solutions are stable.
\end{abstract}

\maketitle

\section{Introduction}\label{sec:intro}

The study of nonlinear lattice dynamics has been fundamental in advancing our understanding of light propagation in nonlinear 
optics~\cite{LEDERER20081} and the wavefunction properties
of atomic condensates~\cite{RevModPhys.78.179},
among others. In the former realm, the relevant models consider the propagation of light
in coupled arrays of optical waveguides, while
in the latter setting, they explore the
evolution of the mean-field wavefunction in the
context of deep optical lattices.
In both scenarios, the universal model
of interest (also considered as an envelope wave
model in other discrete settings, including
mechanical and electrical lattices~\cite{remoissenet,DP06}) has been the
prototypical discrete nonlinear Schr{\"o}dinger (DNLS)
lattice~\cite{kev09}.

Progressively, over the past few years, a
topic that has been gaining significant
traction has been the exploration of
topological features in both linear and
nonlinear systems exhibiting wave dynamics.
Indeed, recent studies in a diverse host of fields including,
but not limited to, photonics \cite{Ozawa2019},
phononics~\cite{Ma2019,Susstrunk2016}, 
metamaterials~\cite{Bertoldi}, and
atomic physics~ \cite{Cooper2019}, highlight unique dynamical properties resulting from the
interplay of nonlinearity
and topology. 
Relevant realizations of, e.g., SSH lattice systems and associated
anomalous edge states have also been 
recently proposed in the work of~\cite{ssh} with the potential
of application in the context of topoelectrical metamaterials.
Notably, their interplay has been leveraged to produce solitonic excitations and
domain walls~\cite{Lumer2013, Solnyshkov2017, Smirnova2019, Marzuola2019, Mukherjee2020,Chen2014, Hadad2017, Poddubny2018}, and to generate robust states propagating
on domain edges~\cite{Ablowitz2014, Leykam2016, Kartashov2016, Snee2019, Tao2020} that 
defy
discreteness-induced barriers such as the famous
Peierls-Nabarro barrier~\cite{Abl21a}. The resulting  ``topologically protected" states 
achieve unidirectional, uninhibited propagation around lattice defects in topological lattices~\cite{Abl19a}. These intense recent efforts
have been summarized, e.g., in~\cite{Smirnova2020,Ma2021}, and also in the very recent
and detailed review of~\cite{cole}. 

Among the many ongoing efforts in the field of topological photonics, we single out here a series of
highly influential recent experiments 
of Rechtsman and collaborators~\cite{Mukherjee2020,recht21,PhysRevLett.128.113901,Jurgensen2021}. Topological photonics has its roots in two seminal 2008 papers \cite{raghu1, raghu2}, where  the authors delineated in detail the one-on-one correspondence between condensed matter physics (CMP) and photonics. In particular, the propagation direction $z$ plays the role of time in the original CMP setting, and hence leads
    to a notion of ``pseudo-time''. For the same reason the conjugate wave-number is referred to as a ``pseudo-frequency''
($t \longleftrightarrow z; \omega \longleftrightarrow k$). 
The first of these works~\cite{Mukherjee2020} showcased the 
experimental realization of Floquet 
solitons in a topological bandgap, the numerical existence and stability of which we subsequently
explored in~\cite{PhysRevE.105.044211}. More recently,
such dispersive nonlinear systems
with a coupling dependent on the evolution variable
were proposed as a suitable realization of nonlinear
Thouless pumps~\cite{PhysRevLett.128.113901}, and the
topological  properties of the bands such as the Chern
number were argued to govern the resulting soliton motion.
In~\cite{Jurgensen2021}, the analogy with the quantum
Hall effect and the original proposal of the Thouless
pump~\cite{PhysRevB.27.6083} was taken further by studying how
nonlinearity acts to quantize transport via soliton formation and spontaneous symmetry-breaking bifurcations. In the present work, influenced by these studies, we consider the system analyzed
in~\cite{Jurgensen2021}, but we depart from the 
adiabatic regime of focus in that work. 
By doing so, we are able to capture topologically induced stationary and dynamic states beyond the adiabatic approximation. We do so by enabling the computation of numerically
exact {\it stationary} solutions, but importantly also {\it traveling}
solutions. Not only do we generate such waveforms
by ``generic'' dynamical evolution experiments, but we also study
a simple variant of the model which considers piecewise-constant
coupling strengths (in a way reminiscent of the celebrated
Kronig-Penney model~\cite{kronig}). There, it becomes evident
that at a qualitative level, the transition between standing and
traveling waves mirrors the self-trapping transition
of the DNLS dimer~\cite{Kenkre1986}. The latter may provide a quite relevant
insight towards understanding the symmetry breaking transitions
and dynamics within the intensely studied topic of nonlinear Thouless
pumps. 

Our findings are structured as follows. In \cref{sec:theory}, we present
the theoretical setup and our quantitative diagnostics used in the model of interest. 
In \cref{sec:rescaling}, we rescale the model so that we can use the propagation distance $L$ as a parameter.
We subsequently turn to numerical computations in \cref{sec:numerics}, starting
with the evolution of single-site initial conditions, and then gaining
insights from the simplified piecewise-constant coupling model.
In \cref{sec:singlesite}, we perform evolution experiments which demonstrate that there are two fundamental behaviors to the system. For low initial power, the initial intensity moves either to the left or to the right in the lattice. The direction of motion depends on the lattice site chosen for the initial condition. For high initial power, the intensity remains confined to a single lattice site. In addition, there does not appear to be a sharp transition between these two behaviors when the starting intensity of the single site is continuously varied.
In \cref{sec:simplified}, we consider a simplification of the model in which the coupling between waveguides is given by step functions. An analysis of this simplified model for an optical dimer explains these two observed behaviors, as well as the lack of a sharp transition between them.
In \cref{sec:coherent}, we numerically construct both stationary and traveling coherent structures. As opposed to what occurs with single-site initial conditions, these coherent structures reproduce themselves exactly after an integer multiple of the coupling period. We use Floquet theory to determine the spectral stability of the stationary coherent structures, which are periodic orbits of the system.
Finally, in section IV we
summarize our findings and present our conclusions, including a number
of directions for future study.

\section{Theoretical Analysis}\label{sec:theory}

\subsection{Mathematical Model}

As discussed above, and motivated by experiments such as those of~\cite{PhysRevLett.128.113901,Jurgensen2021}, we study light propagation in an array of coupled optical waveguides, where the coupling is periodically modulated along the axis of light propagation. Mathematically, this is described by the 
non-autonomous variant of the DNLS model of the form
\begin{equation}\label{eq:DNLSH}
i \frac{d u_n}{d z} + \sum_m H_{n,m}(z)u_m + g|u_n|^2 u_n = 0,
\end{equation}
where $u_n(z)$ is the complex amplitude of light propagating at the waveguide in the lattice site indexed $n$, $z$ is the propagation distance (in the direction along the waveguides), and $H$ is the linear, $z$-dependent (i.e., dependent on the
evolution variable) tight-binding Hamiltonian, or equivalently the lattice coupling profile. 
It is important to clarify here that the non-autonomous nature of the system
under study is in connection with the notion of pseudo-time (corresponding to the
propagation distance) as indicated above. It is with that sense of 
non-autonomy in mind that we will  proceed hereafter.
The parameter $g$ quantifies the strength of the cubic Kerr nonlinearity. For $H$, as in \cite{Jurgensen2021}, we use an off-diagonal implementation of the Aubry-Andr\'e-Harper model \cite{Aubry1980,Harper1955} with three sites per unit cell, resulting in the model
\begin{equation}\label{eq:model}
i \frac{d u_n}{d z} + J_n(z) u_{n+1} + J_{n-1}(z)u_{n-1} + g|u_n|^2 u_n = 0.
\end{equation}
The $z$-dependent coupling functions $J_n(z)$ are periodic in $z$ with spatial period $L$, which we will refer to as the coupling period. We note that \cite{Jurgensen2021} considers this model in the adiabatic regime, i.e. for very large $L$ (see, for example, \cite[Figure 2]{Jurgensen2021}, where $L=8000$), in which case the system is approximately at a ``frozen'' equilibrium for every $z$. 
This is a central point to the analysis presented therein, which is 
explicitly geared towards (and limited to) such an adiabatic regime.
By contrast, the parameter regime we consider herein is that of relatively small $L$, e.g. $L=2\pi$. In this case, stationary solutions are (genuine) periodic orbits of the system, which, in turn, enables us to use the tools of Floquet analysis to determine their spectral stability.

While we consider larger $L$ below (see, in particular, \cref{fig:statcontL}), we note that our methods do not allow us to compute exact periodic orbits for which $L$ is very large, i.e. ones which would approach the adiabatic regime (see the end of \cref{sec:statsol} for further details).

\begin{figure}
    \centering
    \includegraphics[width=8cm]{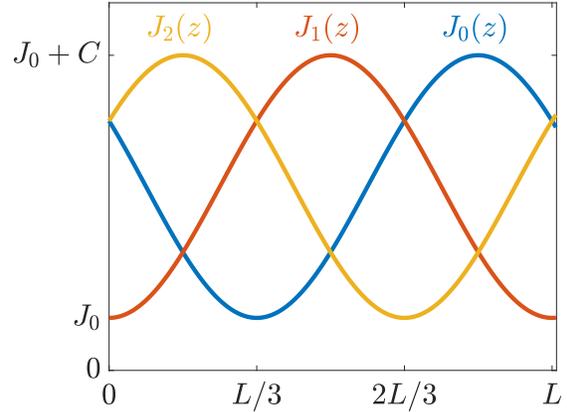}
    \caption{Coupling functions $J_0(z)$, $J_1(z)$, and $J_2(z)$ of $z$-dependent nearest-neighbor couplings over one spatial period $L$.}
    \label{fig:J}
\end{figure}

The choice of $J_n(z)$
\begin{equation}\label{eq:Jn}
J_n(z) = J_0 + C \cos^2\left( \frac{\pi}{L}z + \frac{4 \pi}{3} n + \frac{\pi}{6} \right)
\end{equation}
groups the lattice sites into unit cells comprising three waveguides each (\cref{fig:J}, see also \cite[Figure 1]{Jurgensen2021}), since $J_n(z) = J_m(z)$ for $m \equiv n\,(\text{mod}\,3)$. This choice of $J_n(z)$ is (slightly) modified from Equation (3) in the supplement of \cite{Jurgensen2021}; squaring the cosine function ensures that the coupling is always positive, which would be the case in a physical realization of the model. 
We note that if we do not square the cosine (as in the supplement of \cite{Jurgensen2021}), thus allowing for negative coupling values, we obtain the same qualitative behavior.
When $C = 0$, the nearest neighbor coupling is constant, and Eq.~\cref{eq:model} reduces to the ordinary DNLS with coupling via the discrete second difference operator \cref{eq:modelDNLS}, written in a co-rotating frame with frequency $2 J_0$, which is given by
\begin{equation}\label{eq:modelDNLS}
i \frac{d u_n}{d z} + J_0( u_{n+1} - 2 u_n + u_{n-1}) + g|u_n|^2 u_n 
+ 2 J_0 u_n 
= 0.
\end{equation}

\subsection{Model Rescaling and Density Evolution}\label{sec:rescaling}

In order to use the spatial period $L$ as a parameter, we rescale the propagation distance using the change of variables $z = L Z$ so that the coupling period is always 1.
\begin{equation}\label{eq:modelZ}
\begin{aligned}
i \frac{1}{L} \frac{d u_n}{d Z} &+ J_n(Z) u_{n+1} + J_{n-1}(Z)u_{n-1} + g|u_n|^2 u_n = 0 \\
J_n(Z) &= J_0 + C \cos^2\left( \pi Z + \frac{4 \pi}{3} n + \frac{\pi}{6} \right).
\end{aligned}
\end{equation}
At any propagation distance $Z$, the power of the solution is its squared $\ell^2$ norm 
\begin{equation}
P(u_n) = \sum_{n} | u_n(Z) |^2,
\end{equation} 
where the sum is taken over the entire lattice. The optical intensity at lattice site $n$ is the square amplitude $|u_n|^2$.
The power is conserved, i.e., $P(u_n)$ is independent of $Z$. Using equation \cref{eq:modelZ} and its complex conjugate, we derive the flux equations
for the density matrix elements $\rho_{mn} = \overline{u_n} u_m$
\begin{equation}\label{eq:rhoeq}
\begin{aligned}
\frac{d \rho_{mn}}{dZ} &= iL \Big[ J_m(Z) \rho_{m+1,n} + J_{m-1}(Z) \rho_{m-1,n} \\
&\qquad -J_n(Z) \rho_{m,n+1} - J_{n-1}(Z) \rho_{m,n-1} \big] \\
&\qquad + iLg\left( \rho_{mm} - \rho_{nn} \right) \rho_{mn}.
\end{aligned}
\end{equation}
The evolution of the optical intensity (or density) of the solution at lattice site $n$, which is given by $\rho_{nn} = | u_n |^2$, is
\begin{equation}\label{eq:powerevol}
\begin{aligned}
\frac{d\rho_{nn} }{dZ} &= iL \Big[ J_n(Z) \rho_{n+1,n} + J_{n-1}(Z) \rho_{n-1,n} \\
&\qquad -J_n(Z) \rho_{n,n+1} - J_{n-1}(Z) \rho_{n,n-1} \big] \\
&= -2L\,\text{Im}\Big[ J_n(Z) \rho_{n+1,n} + J_{n-1}(Z) \rho_{n-1,n} \Big],
\end{aligned}
\end{equation}
where we used the fact that $\rho_{n,m} = \overline{\rho_{m,n}}$.
We can split the last line of \cref{eq:powerevol} into two components, which we denote
\begin{equation}
\begin{aligned}
Q_n^L(Z) &= -2L\,J_{n-1}(Z)\,\text{Im}\,\rho_{n-1,n} \\
Q_n^R(Z) &= -2L\,J_n(Z)\,\text{Im}\,\rho_{n+1,n} 
\end{aligned}
\end{equation}
where $Q_n^L(Z)$ and $Q_n^R(Z)$ are the flow of intensity into site $n$ from the left and the right (respectively), and a positive sign indicates that intensity is flowing into site $n$ from the labeled neighboring site.

\section{Numerical Computations}\label{sec:numerics}

\subsection{Single site evolution}\label{sec:singlesite}

\begin{figure}
    \centering
    \includegraphics[width=9cm]{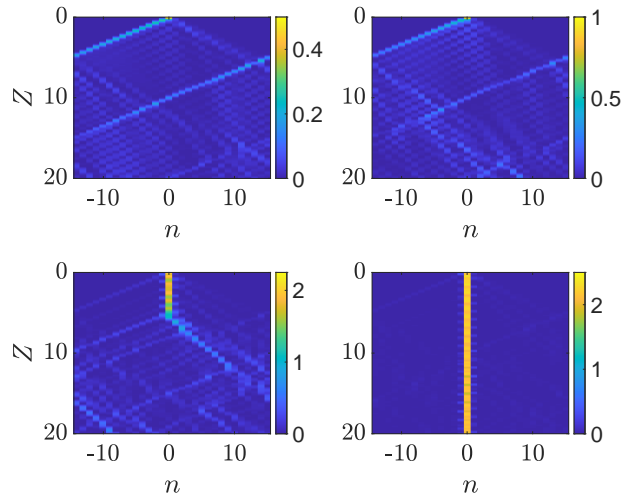}
    \caption{Colormap showing the intensity of the solution of equation \cref{eq:modelZ} evolving in $Z$, starting with a single excited site at $n=0$ with intensity $P=0.5$, $1$, $2.25$, and $2.5$ (left to right, top to bottom). }
    \label{fig:timestep0}
\end{figure}

\begin{figure}
    \centering
    \includegraphics[width=9cm]{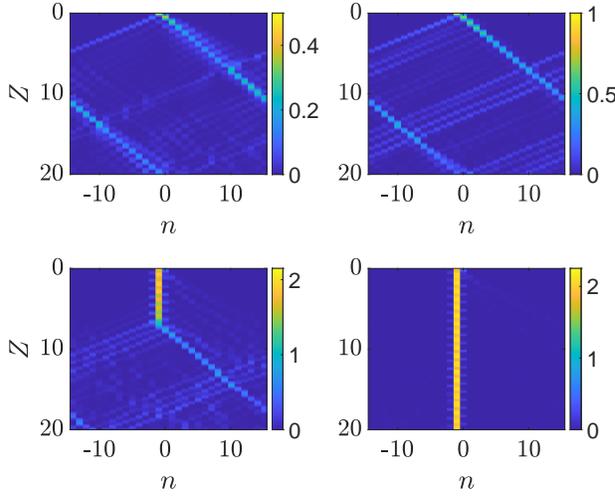}
     \caption{Colormap showing the intensity of the solution of equation \cref{eq:modelZ} evolving in $Z$, starting with a single excited site at $n=-1$ with intensity $P=0.5$, $1$, $2.15$, and $2.25$ (left to right, top to bottom).}
    \label{fig:timestep-1}
\end{figure}

As an initial experiment, we consider dynamical simulations starting with a single excited site at $Z=0$. Unless otherwise specified, the parameters in the section are $g=1$, $L=2\pi$, $J_0 = 0.05$, and $C=0.4$, and the simulations are run on a finite lattice with $m=30$ lattice points, with periodic boundary conditions on the ends of the lattice (i.e., a ring, which allows for waves to loop around when they reach the boundaries). Evolution in $Z$ is performed with \texttt{ode45} in Matlab using a tolerance of $10^{-9}$.
For a single-site initial condition of sufficiently high power (above a threshold between $P=2.25$ and $P=2.5$ for 
input intensity at $n=0$, and between $P=2.15$ and $P=2.25$ for input intensity at $n=-1$), the energy remains concentrated at a single site, and the resulting excitation appears to be stable (see bottom right panel of \cref{fig:timestep0} and \cref{fig:timestep-1}). 
We will address the associated slight difference in the power threshold between initial excitations at $n=0$ and $n=-1$ below.

As the initial power is lowered, this single-site solution becomes prone to mobility; in both cases, this leads the initially concentrated intensity to leak to the right within the lattice before dispersing throughout the lattice (bottom left panels of \cref{fig:timestep0} and \cref{fig:timestep-1}). For lower power initial conditions (between $P=0.5$ and $P=1$), the initial intensity moves either to the left in the lattice (for initial excited site at $n=0$, see top of \cref{fig:timestep0}) or to the right (for initial excited site at $n=-1$, see top of \cref{fig:timestep-1}) before dispersing
through the lattice. One explanation for this observation is as follows: for the first third of the period ($Z \in [0,1/3]$), the strongest coupling is between site $n=0$ and $n=-1$ via $J_{-1} = J_2$ (see Eq.~\cref{eq:Jn} and \cref{fig:J}), so intensity can flow to the left from $n=0$ to $n=-1$, which occurs when the initial power is sufficiently low. For $Z \in [1/3,2/3]$, the strongest coupling is between $n=-1$ and $n=-2$, and for $Z \in [1/3,2/3]$, the strongest coupling is between $n=-2$ and $n=-3$, thus we expect the intensity to travel three sites to the left over one period. Similarly, for the rightward moving solution starting at $n=-1$, the rightward coupling is strongest for a rightward sequence of sites on the $Z$ intervals $[0,1/3]$, $[2/3,1]$, and $[4/3,5/3]$, thus this solution moves to the right three sites in two periods. A similar rightward moving behavior occurs when the initial excited site is $n=1$ (not shown). The first coupling for $n=1$ is to the right on the interval $[1/3, 2/3]$; the behavior is then similar to that of the rightward moving solution for $n=-1$.
We thus conclude that
this fundamental difference between the leftward and rightward moving solutions and the
associated speeds is a direct consequence of the form of the periodic coupling function, together with the lattice site at which the initial intensity is placed; this is suggestive also towards the difference in power thresholds noted above.
In addition, we believe that this
intuitive explanation is quite straightforward and, thus,
appealing towards understanding solitary wave motion in such
non-autonomous discrete nonlinear systems.

\subsection{Simplified model}\label{sec:simplified}

\begin{figure}
    \centering
    \includegraphics[width=8cm]{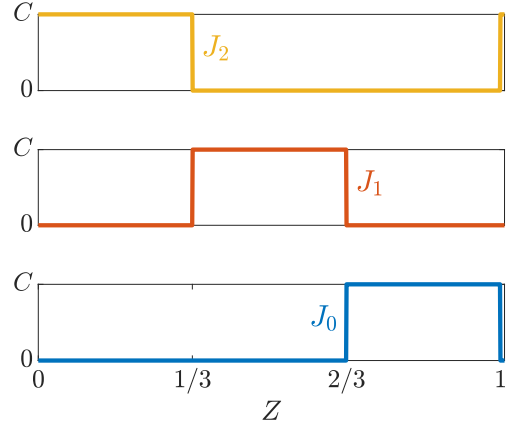}
    \caption{Simplified coupling functions \cref{eq:simpleJn} for $Z \in [0,1]$.}
    \label{fig:Jsimple}
\end{figure}

A further heuristic and qualitative, yet in our view informative and intuitive, explanation of these different behaviors can be obtained by considering the simplification of the system of \cref{eq:model} obtained by approximating the coupling functions $J_n(Z)$ with step functions, as is done, e.g.,
in the setting of the Kronig-Penney model~\cite{kronig}. 
Note that such a perspective has also been beneficial in
a quantitative
fashion in the case of nonlinearity (rather than dispersion)
management in works such as~\cite{PhysRevLett.97.033903,PhysRevLett.97.234101}.
Specifically, we define 
\begin{equation}\label{eq:simpleJn}
J_n(Z) = \begin{cases}
C\chi_{[0,1/3]}(Z) & n \equiv 2 \text{ (mod 3)} \\
C\chi_{[1/3,2/3]}(Z) & n \equiv 1 \text{ (mod 3)}\\
C\chi_{[2/3,1]}(Z) & n \equiv 0 \text{ (mod 3)}
\end{cases}
\end{equation}
for $Z \in [0,1]$, and extend periodically for all $Z$ (\cref{fig:Jsimple}). The function $\chi_{[a,b]}(Z)$ is the characteristic function of the interval $[a,b]$, defined by
\[
\chi_{[a,b]}(Z) = \begin{cases}
1 & Z \in [a,b] \\
0 & \text{otherwise}.
\end{cases}
\]

Using this approximation, on the interval $Z \in [0, 1/3]$, the only active coupling is between the sites $n \equiv0\,(\text{mod}\,3)$ and their left neighbors, which effectively creates a string of independent dimers. 
Our analysis follows Kenkre and Campbell's study of self-trapping in a DNLS dimer \cite{Kenkre1986}. Similar to what was done in that work, we will derive a second order differential equation for the difference in intensity between the two sites of the dimer. This ODE will have solutions in terms of Jacobi elliptic functions, and these will be used to explain the observed transition between mobile and trapped solutions as the coupling strength is increased. We note that this allows us to write the intensities $|u_n|^2$ (not the complex amplitudes $u_n$) in terms of Jacobi elliptic functions.

Looking only at the sites $n=0$ and $n=-1$, i.e., one of these dimers, the system of equations \cref{eq:powerevol} reduces to the four equations
\begin{equation}\label{eq:powerevolsimple}
\begin{aligned}
\frac{d\rho_{0,0}}{dZ} &= iLC \left( \rho_{-1,0} - \rho_{0,-1} \right) \\
\frac{d\rho_{-1,-1}}{dZ}  &= iLC \left( \rho_{0,-1} - \rho_{-1,0} \right) \\
\frac{d\rho_{-1,0} }{dZ} &= iL \big[ C(\rho_{0,0} - \rho_{-1,-1}) \\
&\qquad + g ( \rho_{-1,-1} - \rho_{0,0}) \rho_{-1,0} \big] \\
\frac{d\rho_{0,-1}}{dZ}  &= iL \big[ C(\rho_{-1,-1} - \rho_{0,0}) \\
&\qquad + g ( \rho_{0,0} - \rho_{-1,-1} ) \rho_{0,-1} \big].\\
\end{aligned}
\end{equation}
Let $p = \rho_{0,0} - \rho_{-1,-1}$ be the difference between the intensities of site $n=0$ and site $n=1$.
As in \cite{Kenkre1986}, we will derive a second-order ODE for $p$.
Following the analysis in \cite{Kenkre1989}, we define 
\begin{equation}
\begin{aligned}
q &= i(\rho_{-1,0} - \rho_{0,-1}) \\
r &= \rho_{-1,0} - \rho_{0,-1}.
\end{aligned}
\end{equation}
Using \cref{eq:powerevolsimple}, we obtain the system of first-order ODEs
\begin{align}
\frac{dp}{dZ} &= 2LCq \label{eq:peq}\\
\frac{dq}{dZ} &= -L(2Cp - gpr) \label{eq:qeq}\\
\frac{dr}{dZ} &= -Lgpq. \label{eq:req}
\end{align}
Since
\[
\frac{d}{dZ}p^2 = 2p \frac{dp}{dZ} = 4Lcpq,
\]
equation \cref{eq:req} becomes
\[
\frac{dr}{dZ} = -\frac{g}{4C}\frac{d}{dZ}p^2,
\]
which has solution
\begin{equation}\label{eq:rsol}
r = r_0 - \frac{g}{4C}\left( p^2 - p_0^2 \right),
\end{equation}
where $r_0$ and $p_0$ are the initial conditions for $r$ and $p$, respectively, at $Z=0$. Differentiating \cref{eq:peq} and substituting \cref{eq:qeq} and \cref{eq:rsol}, we obtain the second-order differential equation for $p$
\begin{equation}\label{eq:pode}
\begin{aligned}
\frac{d^2p}{dZ^2} &= L^2 \left( A p - B p^3 \right) \\
A &= -4C^2 + 2Cgr_0 + \frac{g^2}{2}p_0^2
\qquad\quad B = \frac{g^2}{2}.
\end{aligned}
\end{equation}
This second-order, autonomous differential equation with a linear term and a cubic nonlinearity has an exact solution in terms of Jacobi elliptic functions (See section 22.13(iii) of \cite{NIST:DLMF} as well as \cite{Carlson2006}).
We are interested in the case of a single-site initial condition with intensity $P$ at site $n=0$, for which $p_0 = P$ and $r_0 = 0$. Following \cite{Kenkre1986}, for fixed $C$ and $g$, equation \cref{eq:pode} has an exact solution
\begin{equation}\label{eq:psol}
p(Z) =
\begin{cases}
P\,\text{cn}\left(2CLZ ; k=\dfrac{g P}{4C} \right) & P < P^*\\
P\,\text{dn}\left(\dfrac{g P L}{2}Z ; k=\dfrac{4C}{g P} \right) & P > P^*,
\end{cases}
\end{equation}
where the critical intensity $P^*$ is given by
\begin{equation}\label{eq:pstar}
P^* = \dfrac{4C}{g}.
\end{equation}
The functions $\text{cn}(z ; k)$ and $\text{dn}(z ; k)$ are the Jacobi elliptic functions with elliptic modulus $k$ (we note that these are often written in terms of the elliptic parameter $m$, where $m = k^2$). Since the power of the solution is conserved, i.e. $|u_0(Z)|^2 + |u_{-1}(Z)|^2 = P$, 
the intensity at sites $n=0$ and $n=-1$ on the interval $Z \in [0,1/3]$ is given by
\begin{equation}\label{eq:u0u-1}
\begin{aligned}
|u_0(Z)|^2 &= \frac{1}{2}\left( P + p(Z) \right)\\
|u_{-1}(Z)|^2 &= \frac{1}{2}\left( P - p(Z) \right).
\end{aligned}
\end{equation}

There are two fundamental behaviors of the single-site initial condition in the simplified model, depending on whether $P < P^*$ or $P > P^*$. In the dimer, a sharp transition between the two behaviors occurs at $P = P^*$ (see \cite{Kenkre1986}). We note that this transition is somewhat blurred in this model, even with the simplified coupling function, since the initial dimer coupling is broken ($J_n(Z)=0$)
at $Z=1/3$. 
Below we will discuss $P < P^*$ and $P > P^*$ case-by-case.

\subsubsection{\texorpdfstring{Case 1: $P < P^*$}{Case 1: P < Pstar}}

\begin{figure}
    \centering
    \begin{tabular}{cc}
    \includegraphics[width=4cm]{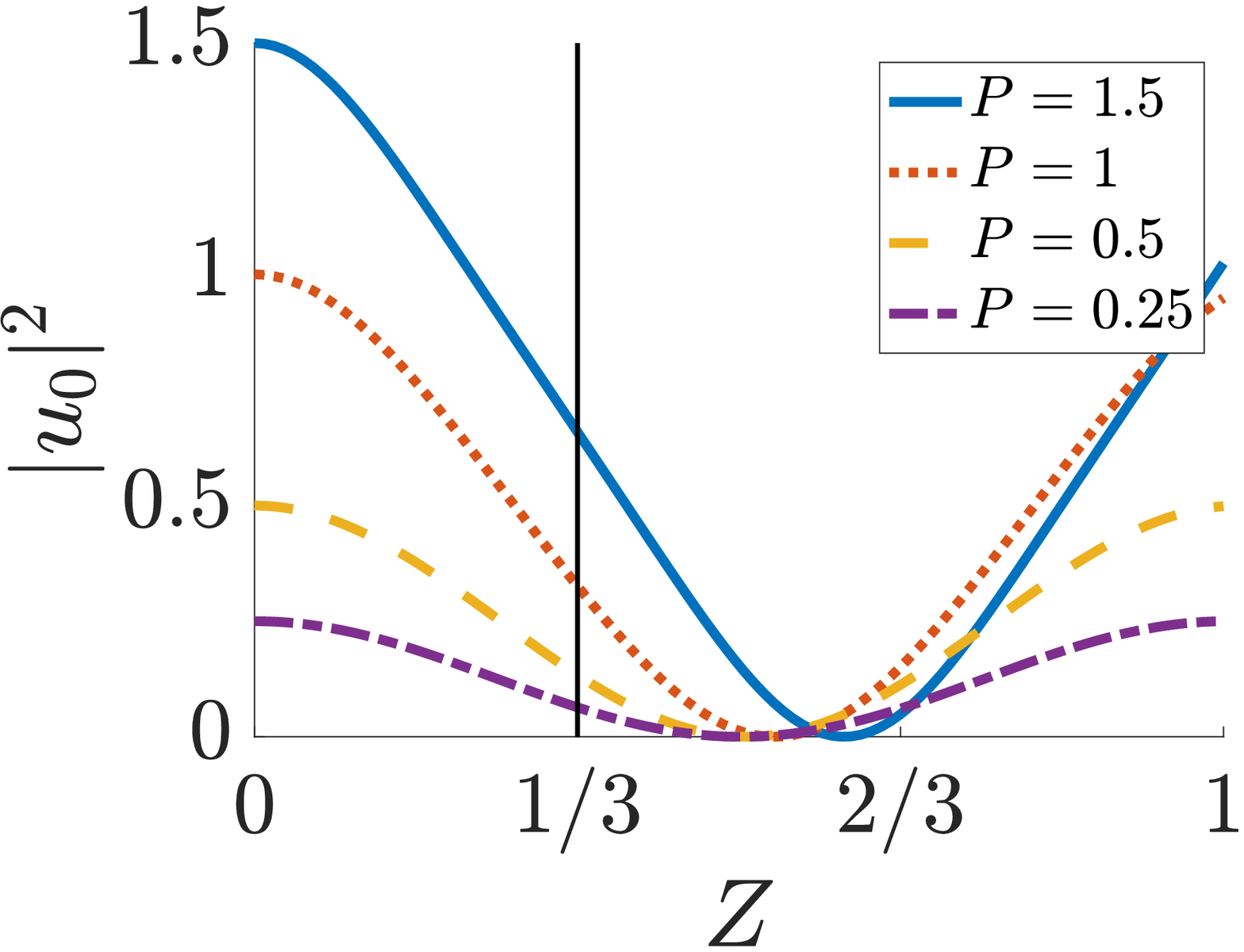} &
    \includegraphics[width=4cm]{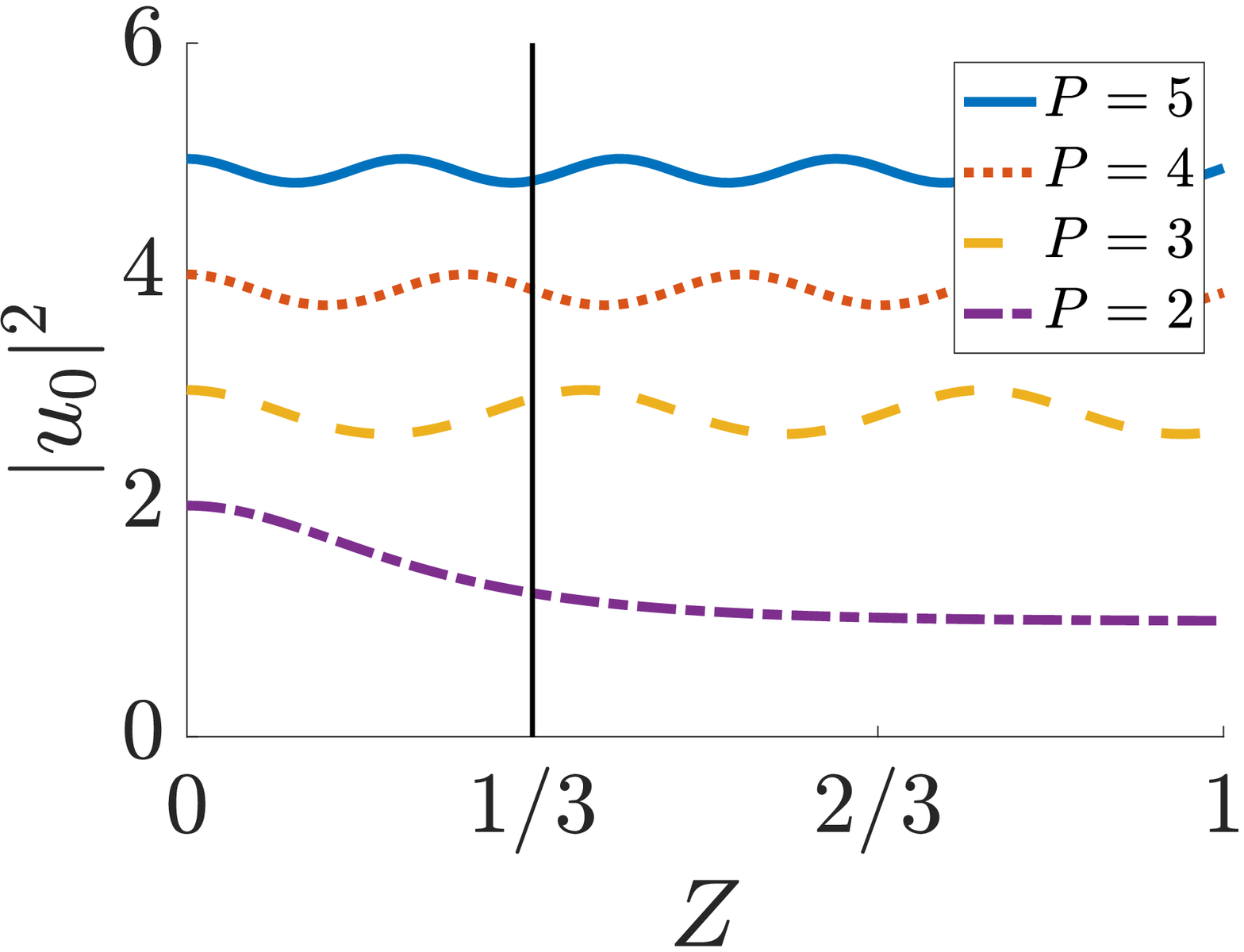}
    \end{tabular}
    \caption{Plot of $|u_0|^2$ in simplified model, given by \cref{eq:u0u-1}, vs. $Z$ for initial intensity $P<P^*$ (left) and $P>P^*$ (right). Although this solution only holds for $Z \in [0, 1/3]$ ($Z=1/3$ is marked with a solid vertical line), it is continued to $Z=1$ for illustrative purposes. $C=0.5$, $g=1$, $P^* = 2$.}
    \label{fig:simplemodel1}
\end{figure}

\begin{figure}
    \centering
    \begin{tabular}{cc}
    \includegraphics[width=4cm]{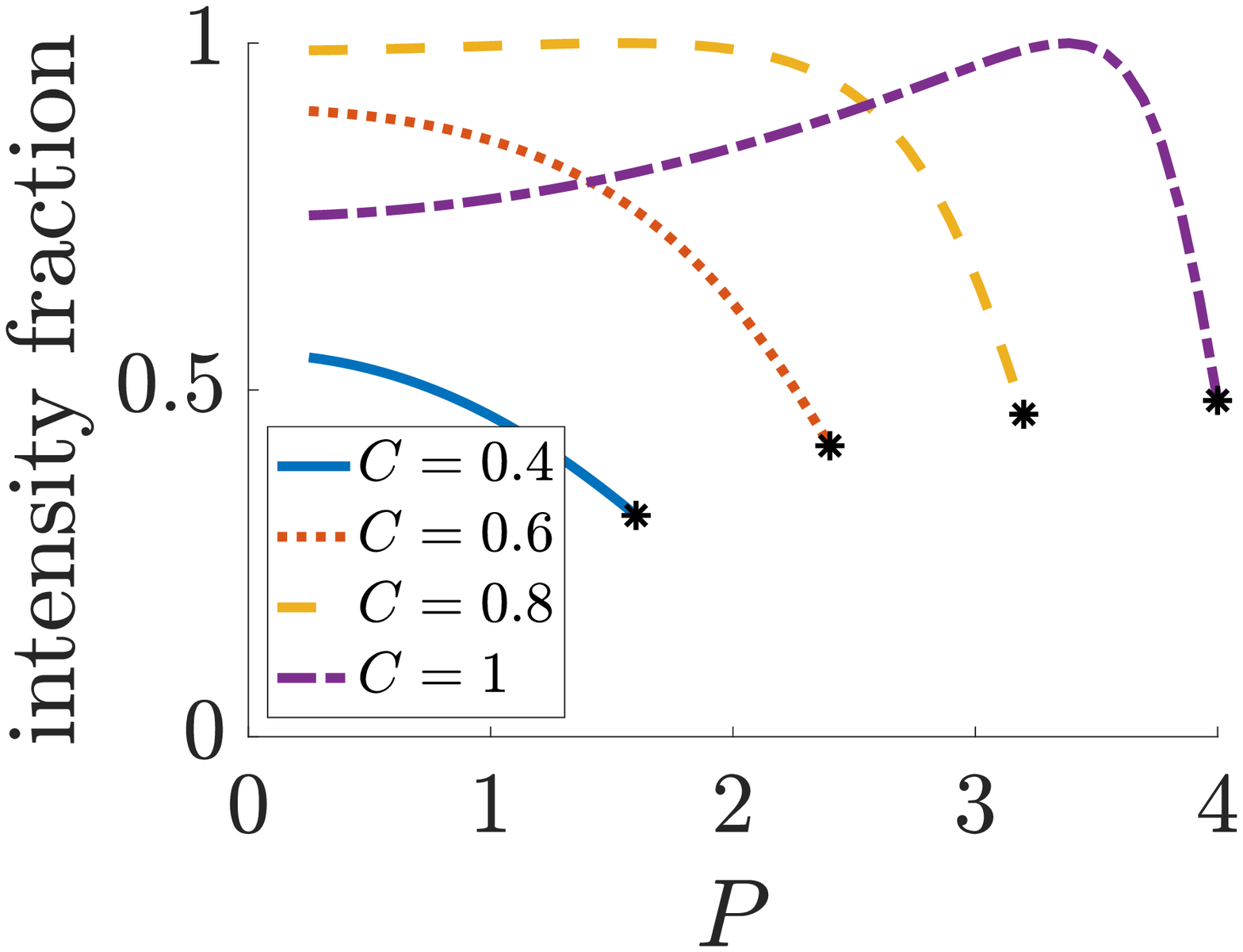} &
    \includegraphics[width=4cm]{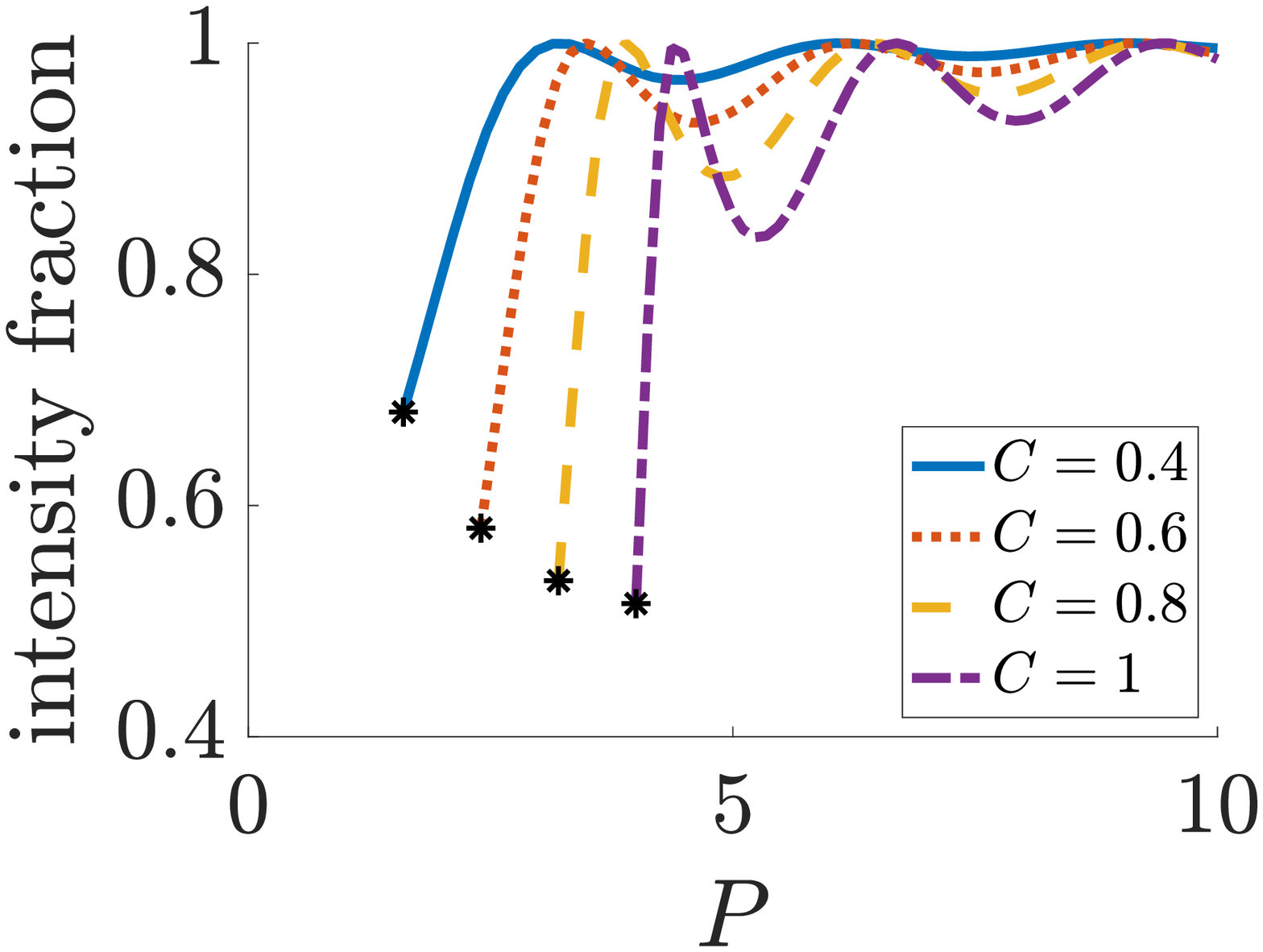}
    \end{tabular}
    \caption{Fraction of intensity transferred from site $n=0$ to site $n=-1$ at $Z=1/3$ for $P<P^*$ for varying $C$ (left). Fraction of intensity remaining at site $n=0$ at $Z=1/3$ for $P>P^*$ for varying $C$ (right). For each curve, $P^*$ is indicated with a star. $g=1$.}
    \label{fig:powertransfer}
\end{figure}

\begin{figure}
    \centering
    \includegraphics[width=9cm]{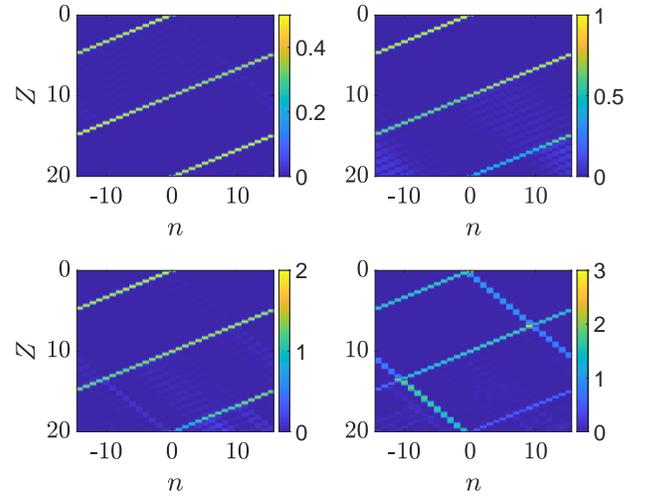}
    \caption{Colormap showing intensity of solution of equation \cref{eq:modelZ} with simplified coupling function \cref{eq:simpleJn} for $P<P^*$ evolving in $Z$, starting with a single excited site at $n=0$ with intensity $P=0.5$, $1$, $2$, and $3$ (left to right, top to bottom). Fraction of intensity transferred from site $n=0$ to site $n=-1$ at $Z=1/3$ is 0.9910, 0.9959, 0.9909, and 0.6636 (respectively). $P^*=3.2$, $C=0.8$, $g=1$.}
    \label{fig:timestepsimplebelowpstar}
\end{figure}

For $P < P^*$, the solution $p(Z)$ involves the Jacobi cn function, which oscillates about 0 with period $2K(k)/CL$, where
\begin{equation}\label{eq:Kellipticint}
K(k) = \int_0^{\pi/2} \frac{d\theta}{\sqrt{1-k^2 \sin^2 \theta}}
\end{equation}
is the complete elliptic integral of the first kind. (We note that the period of oscillation becomes infinite as $P$ approaches $P^*$ from below).
The intensity $|u_0|^2$, given by equation \cref{eq:u0u-1}, exhibits large amplitude oscillations with this period from 0 to $P$, centered at $P/2$ (\cref{fig:simplemodel1}, left).
Intensity initially flows to the left; if the coupling is not cut off at $Z=1/3$ (and no other couplings are activated), there is a critical value $Z_1^*$ of $Z$ at which the intensity from site $n=0$ has been completely transferred to site $n=-1$; after this point, intensity starts flowing back in the other direction. This critical value $Z_1^*$ is larger for larger starting intensity $P$ (see \cref{fig:simplemodel1}, left).
For most configurations, including all of the examples in \cref{fig:simplemodel1}, the critical value $Z_1^* > 1/3$, thus there is still intensity remaining in site 0 when the coupling switches off at $Z=1/3$. The left panel of \cref{fig:powertransfer} plots the fraction of intensity that has been transferred from site $n=0$ to site $n=-1$ at $Z=1/3$ as the starting intensity varies. If this fraction is close to 1, numerical evolution experiments find leftward-moving solutions starting from a single-site initial condition (see the first three panels of \cref{fig:timestepsimplebelowpstar}). As the initial intensity $P$ approaches $P^*$ from below, the fraction of intensity transferred at $Z=1/3$ decreases to approximately 0.5, and we do not expect to see pure leftward-moving solutions. Numerical evolution experiments in this case show that the initial intensity splits into a leftward and a rightward moving solution (bottom right panel of \cref{fig:timestepsimplebelowpstar}).

\begin{figure}
    \centering
    \includegraphics[width=9cm]{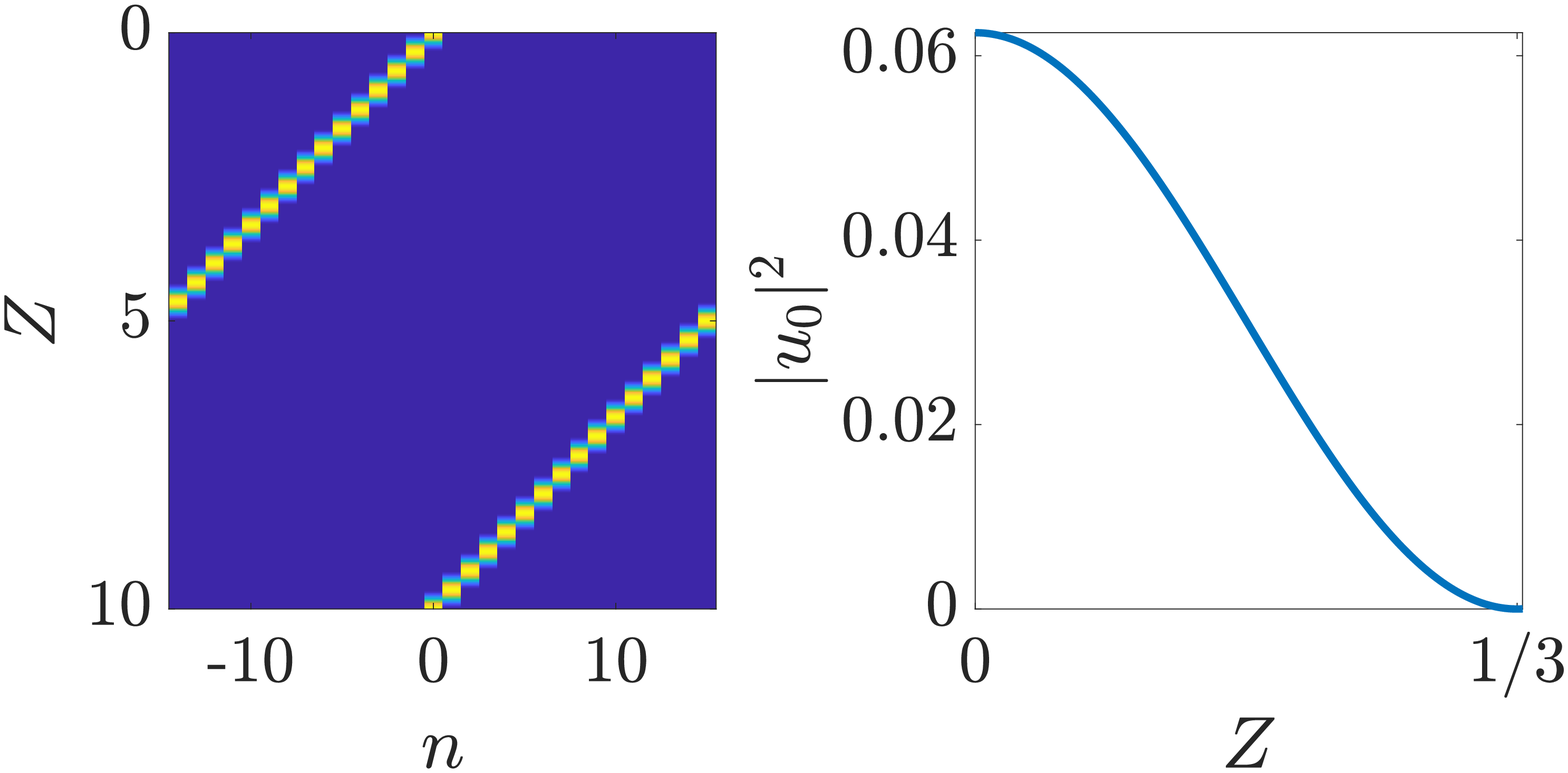}
    \includegraphics[width=9cm]{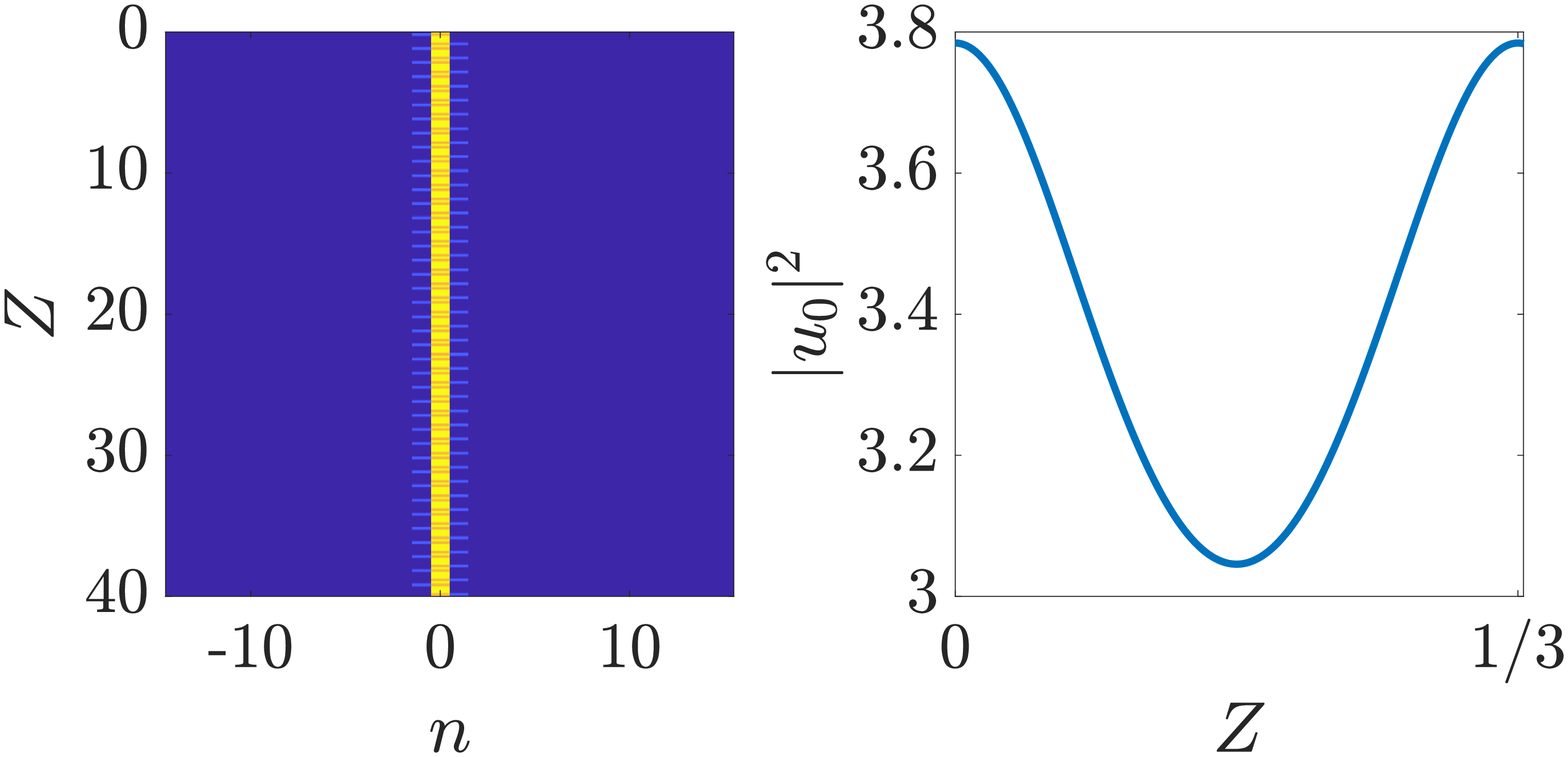}
    \caption{Colormap of evolution in $Z$ of single-site initial condition in simplified model (left) and intensity $|u_0|^2$ of site $n=0$ (right) on $Z\in[0,1/3]$ (right). Parameters chosen so that $Z_1^* = 1/3$ (top) and $Z_2^*=1/3$ (bottom).
    Starting intensity $P=0.0625$ (top) and $P=1.9453$ (bottom).
    $C=0.75$, $g=1$.}
    \label{fig:simplecomplete}
\end{figure}

\begin{figure}
    \centering
    \begin{tabular}{ccc}
    \includegraphics[width=2.5cm]{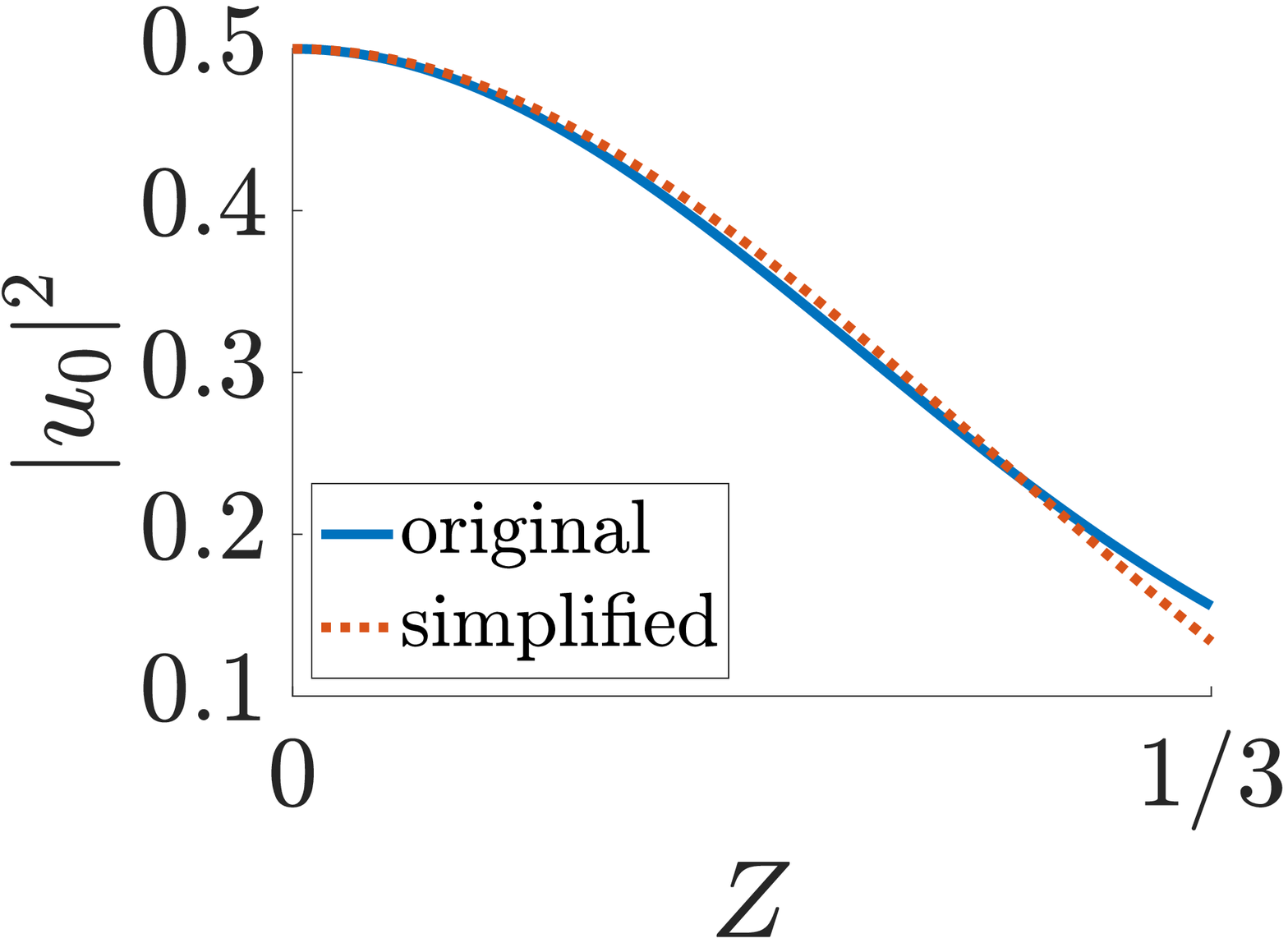} &
    \includegraphics[width=2.5cm]{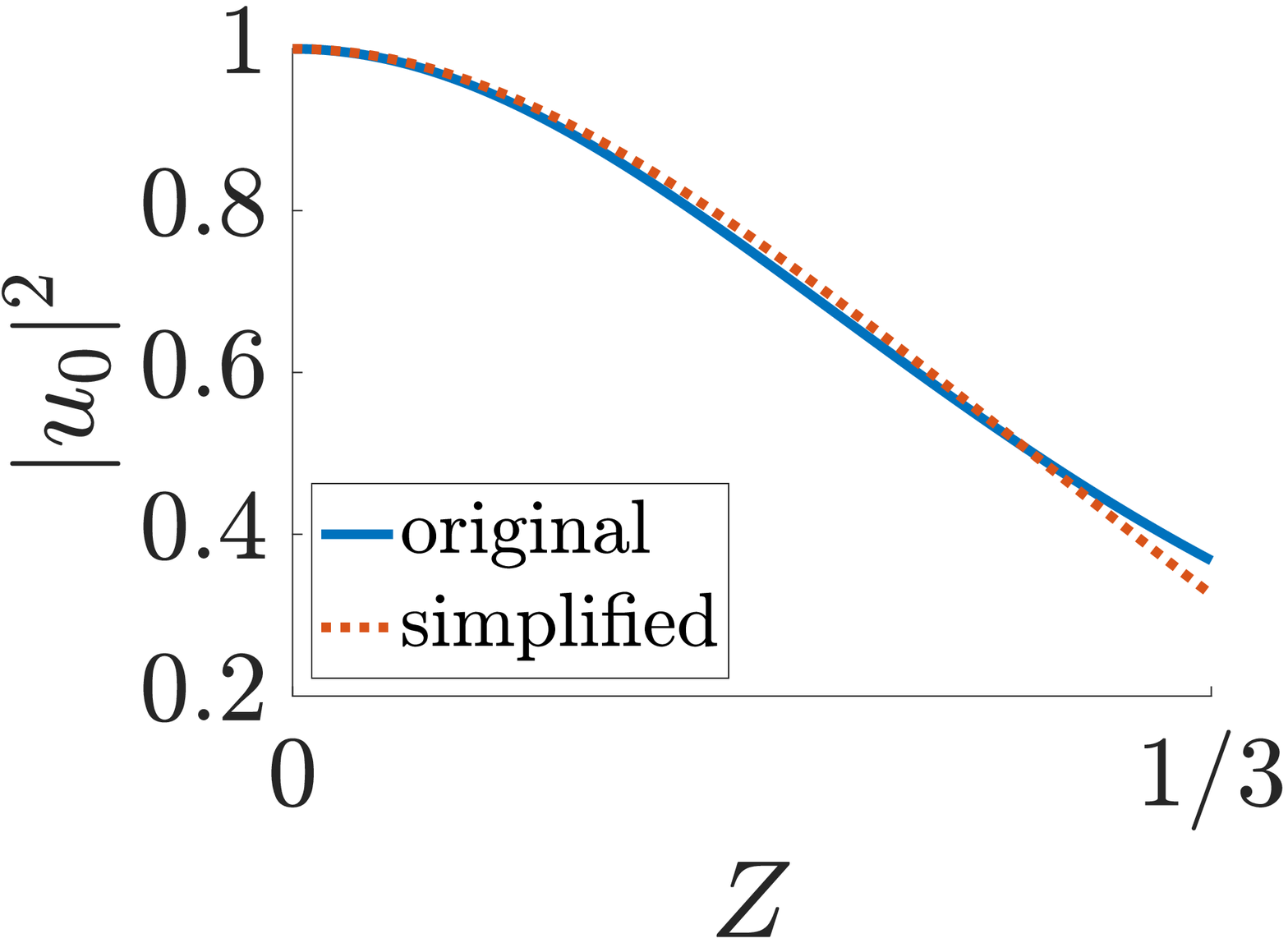}  &
    \includegraphics[width=2.5cm]{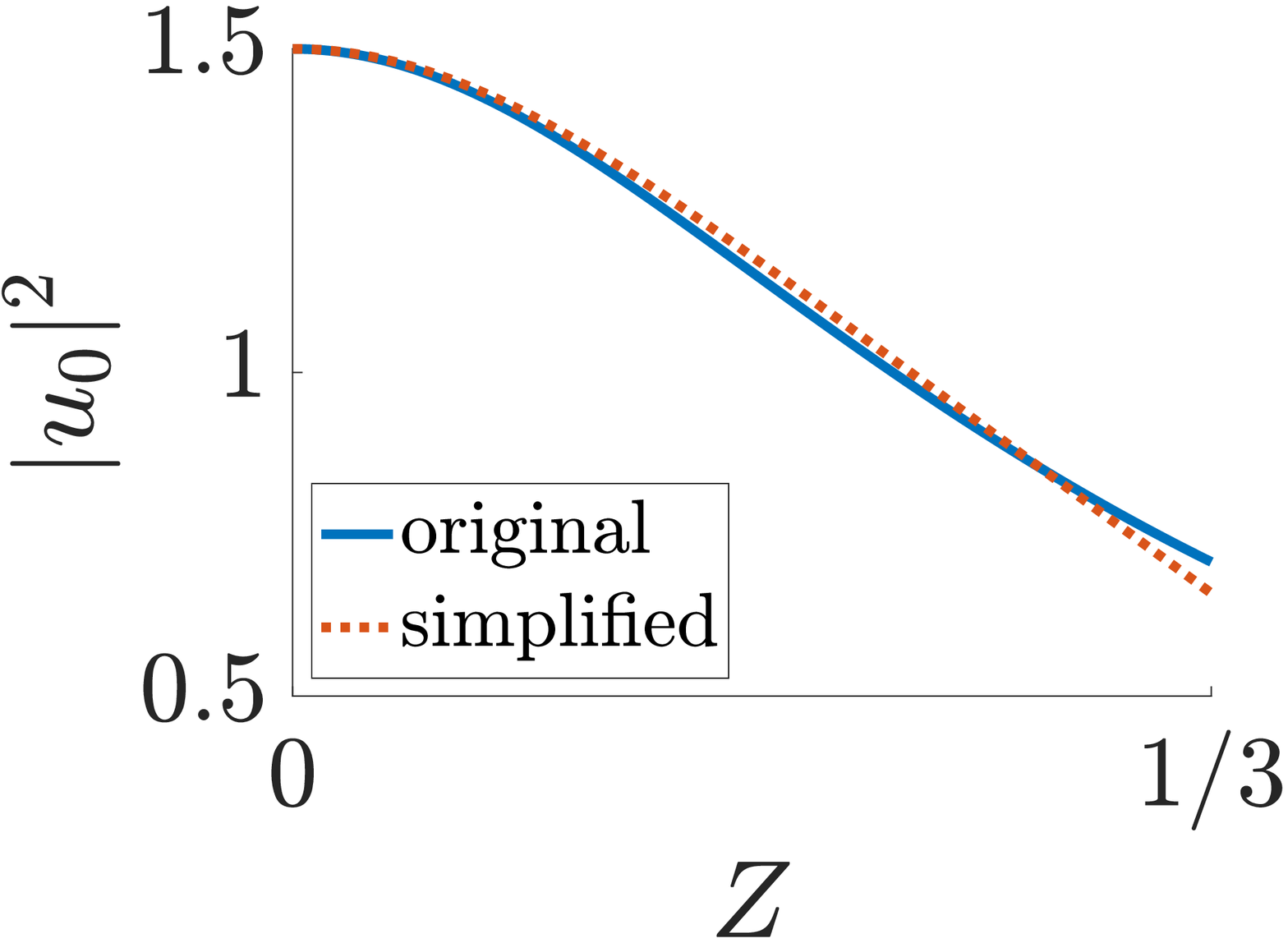} \\
    \includegraphics[width=2.5cm]{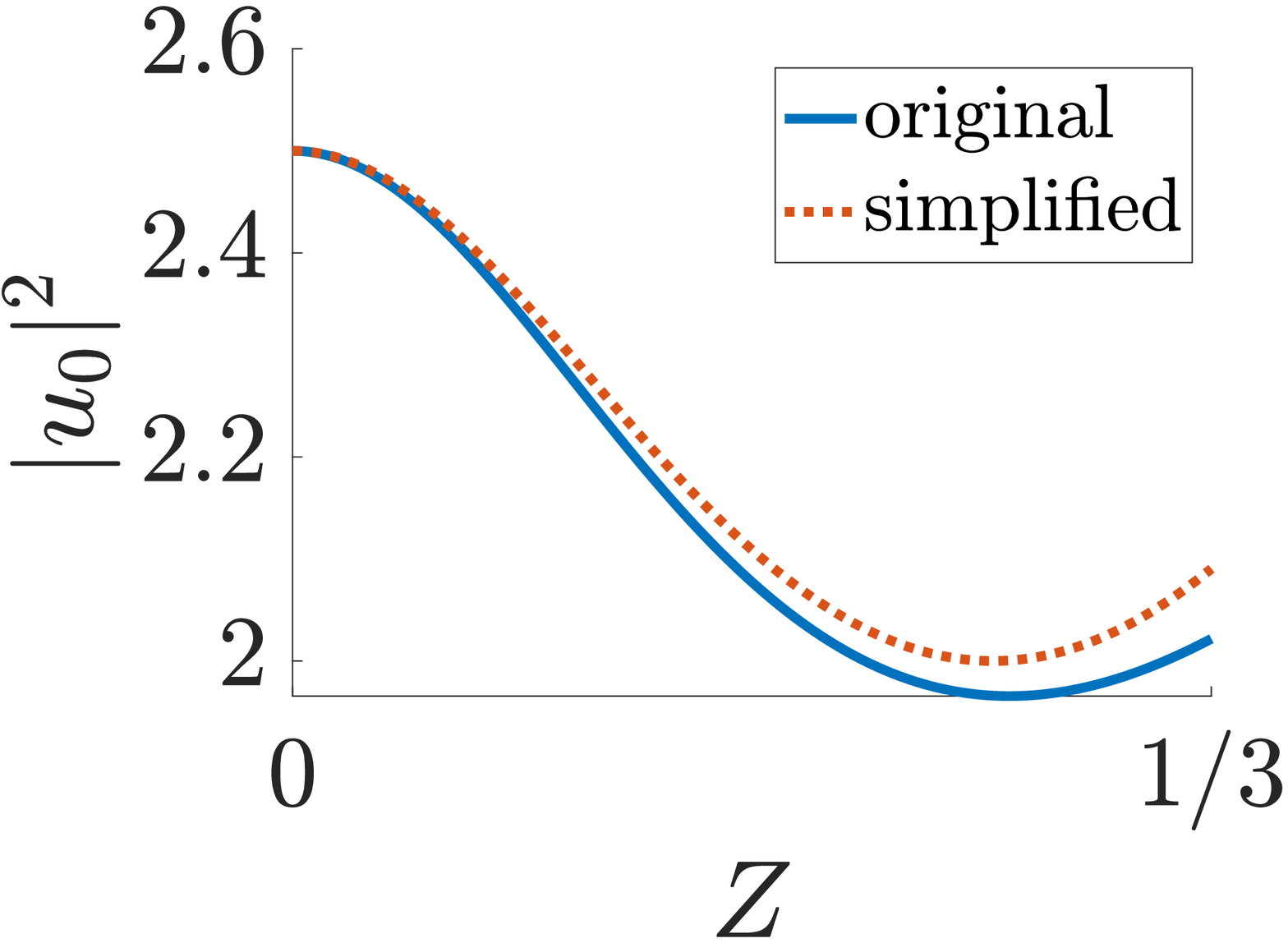} &
    \includegraphics[width=2.5cm]{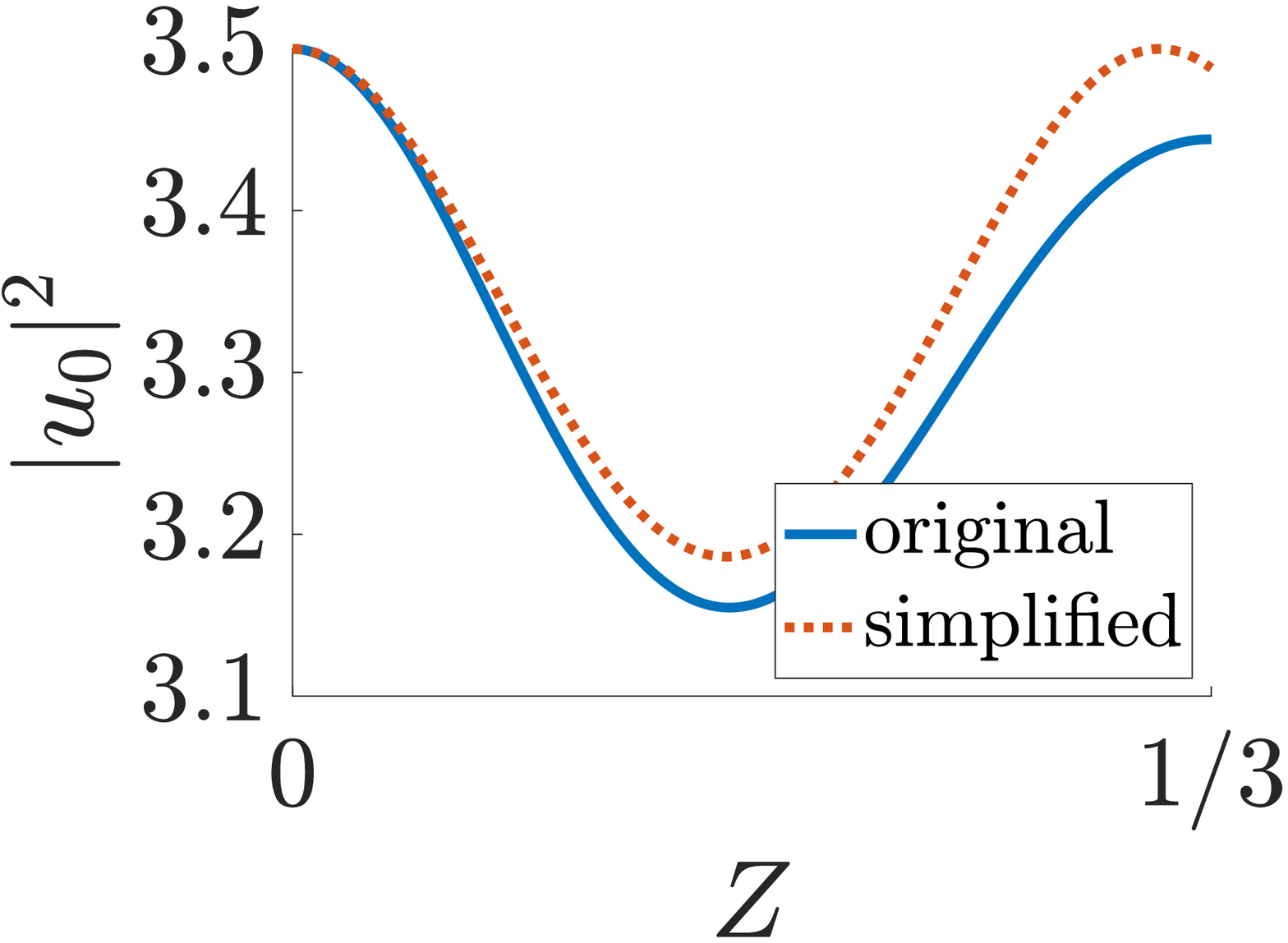}  &
    \includegraphics[width=2.5cm]{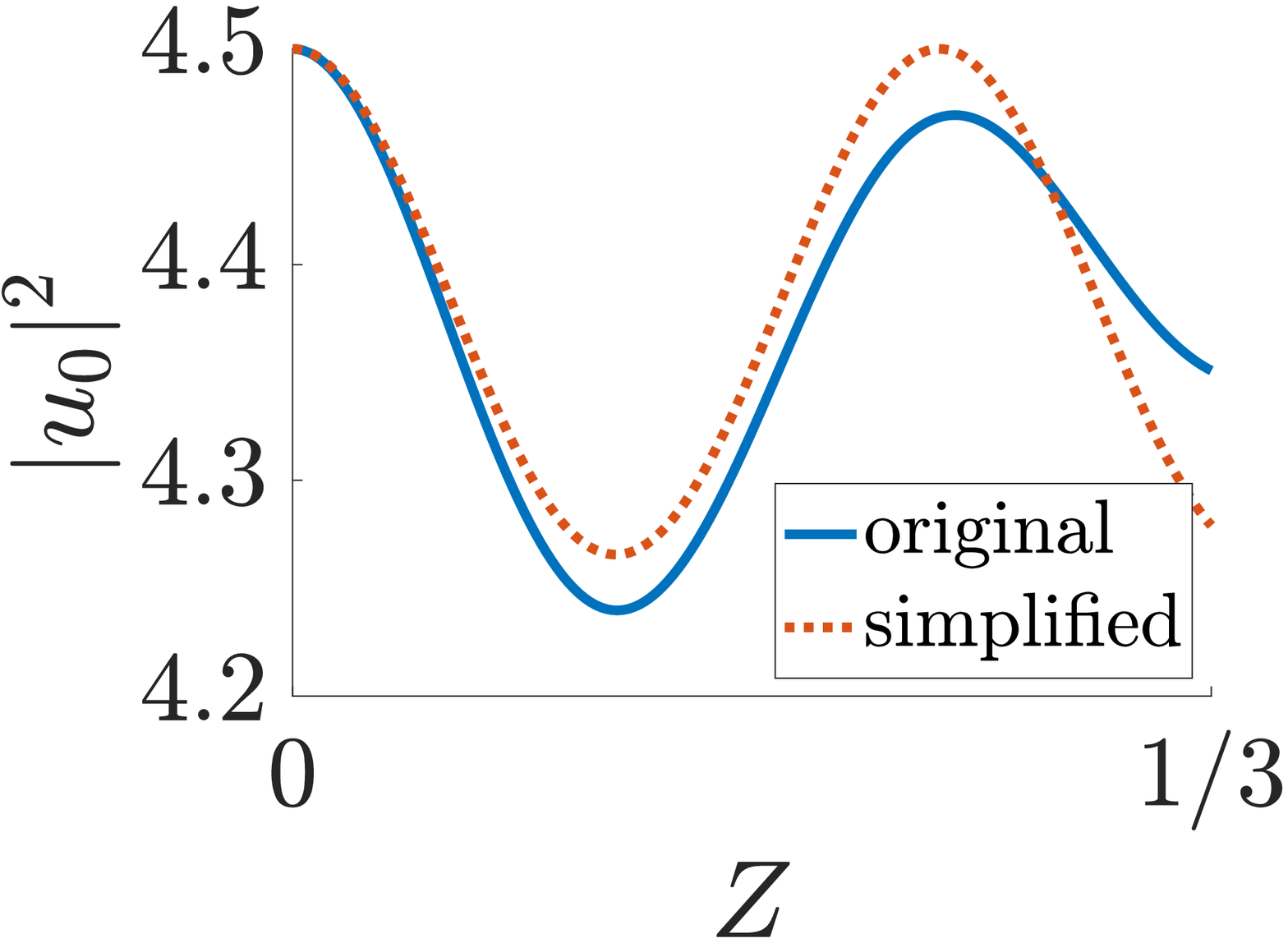}
    \end{tabular}
    \caption{Intensity $|u_0|^2$ of site $n=0$ on $Z\in[0,1/3]$ for $P<P^*$ (top, $P=0.5$, $1$, and $1.5$), and $P>P^*$ (bottom, $P=2.5$, $3.5$, and $4.5$) for original system (solid blue line) and simplified system (dotted orange line). $C=0.5$, $g=1$, $P^* = 2$.}
    \label{fig:simplecompare}
\end{figure}

We note that it is possible to choose parameters so that, for the simplified model, the period of $p(Z)$ is 2/3, i.e. $Z_1^* = 1/3$, so that the initial intensity has been completely transferred from $n=0$ to $n=-1$ when the coupling switches off. The next coupling is then between $n=-1$ and $n=-2$ for $Z \in [1/3,2/3]$. This pattern continues, and so for this choice of parameters, the simplified model supports 
an exact leftward-moving solution which persists for a large interval in $Z$
(\cref{fig:simplecomplete}, top). 
A comparison between the evolution of the original and simplified systems from single-site initial conditions for $P < P^*$, illustrating the similarity of the model
dynamics, is shown in the top panel of \cref{fig:simplecompare}. We will see below in \cref{sec:movingsol} that left-moving coherent structures exist in the full model, but these do not have pure single-site initial conditions. We also note that by symmetry of \cref{eq:powerevolsimple}, a similar analysis holds for rightward-moving solutions starting with single-site initial conditions at $n=-1$ on the interval $Z \in [0,1/3]$. 

\subsubsection{\texorpdfstring{Case 2: $P > P^*$}{Case 1: P > Pstar}}

\begin{figure}
    \centering
    \includegraphics[width=9cm]{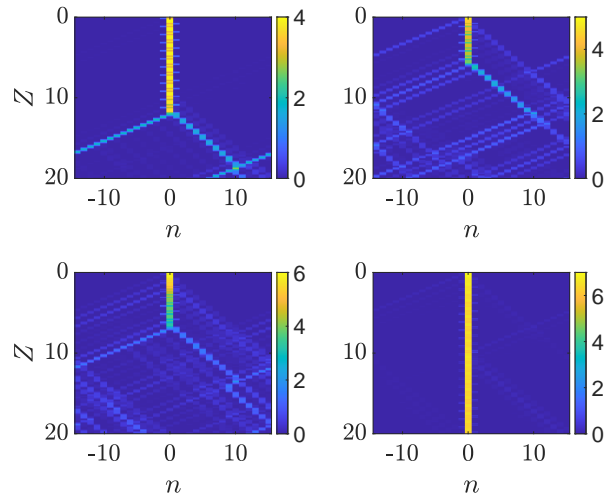}
    \caption{Colormap showing intensity of the solution of equation \cref{eq:modelZ} with simplified coupling function \cref{eq:simpleJn} for $P>P^*$ evolving in $Z$, starting with a single excited site at $n=0$ with intensity $P=4$, $5$, $6$, and $7$ (left to right, top to bottom). Fraction of intensity remaining at site $n=0$ at $Z=1/3$ is 0.9938, 0.8853, 0.9815, and 0.9797 (respectively).
    $P^*=3.2$, $C=0.8$, $g=1$.}
    \label{fig:timestepsimpleabovepstar}
\end{figure}

For $P > P^*$, the solution $p(Z)$ involves the Jacobi dn function, which oscillates about 1 with period $4 K(k)/gPL$, where $K(k)$ is defined by \cref{eq:Kellipticint}. (We again note that the period of oscillation becomes infinite as $P$ approaches $P^*$ from above).
The intensity $|u_0|^2$ exhibits small amplitude oscillations (which become progressively smaller as $P$ is increased) with this period about the initial intensity $P$ (\cref{fig:simplemodel1}, right).
As in Case 1, the intensity initially flows to the left.
If the coupling is not cut off at $Z=1/3$ (and no other couplings are activated), there is a critical value $Z_2^*$ of $Z$ at which point the system has returned to its initial state, i.e. the intensities at sites $n=0$ and $n=-1$ are once again $P$ and 0, respectively. 
For most configurations, including all of the examples in \cref{fig:simplemodel1}, the critical value $Z_2^* \neq 1/3$, thus when the coupling switches off, there has been some net transfer of intensity to the neighboring site $n=-1$. 
The right panel of \cref{fig:powertransfer} plots the fraction of intensity remaining at site $n=0$ at $Z=1/3$ for varying $C$. If this fraction is close to 1, numerical evolution simulations show stationary solutions starting from a single-site initial condition; the closer this fraction is to 1, the longer these stationary solutions will persist before breaking up; see the case examples in \cref{fig:timestepsimpleabovepstar}.

As in Case 1, we note that it is possible to choose parameters so that, for the simplified model, the system returns exactly to its starting condition at $Z=1/3$, i.e. $Z_2^* = 1/3$ (\cref{fig:simplemodel1}, bottom). In this case, for a single-site initial condition with a specific starting intensity, the simplified model supports a localized in space, time-periodic solution which persists for a large interval in $Z$. A comparison between the evolution of the original and simplified systems for $P > P^*$ is shown in the bottom panel of \cref{fig:simplecompare}. We discuss genuinely time-periodic (numerically exact) coherent structures in the full model in Subsection \cref{sec:statsol}. 
 
\subsection{Coherent Structures in the Full Model}
\label{sec:coherent}

These evolution experiments are strongly suggestive of the fact that the system \cref{eq:modelZ} supports two classes of coherent structures: localized in space, time-periodic solutions, which are centered at a particular lattice site, and moving solutions, which reproduce themselves exactly a specific number of sites to the left or to the right. Recall that for the rescaled system \cref{eq:modelZ}, the coupling period is 1. The stationary coherent structures will be periodic orbits whose period is a multiple of the coupling period, i.e., a positive integer $N$ (we refer to this $N$ below). We note that while it may be possible to find such solutions which have a non-integer period, Floquet analysis requires that the period of the solutions be commensurate with that of the coupling. Similarly, we will look for moving solutions that reproduce themselves, shifted left or right, after an integer period.

For appropriate choices of system parameters, we can compute both types of solutions numerically. For both localized (i.e., non-moving) and moving solutions, we use a shooting method with periodic boundary conditions imposed on $Z$, starting with a single-site initial guess. In addition, for the former 
case, we validate this method by using numerical parameter continuation with AUTO \cite{auto07p} to solve a periodic boundary value problem. Unless otherwise specified, the parameters in the section are the same as in the previous one. 

\subsubsection{Stationary (non-moving) solutions}\label{sec:statsol}

\begin{figure}
    \centering
    \includegraphics[width=9cm]{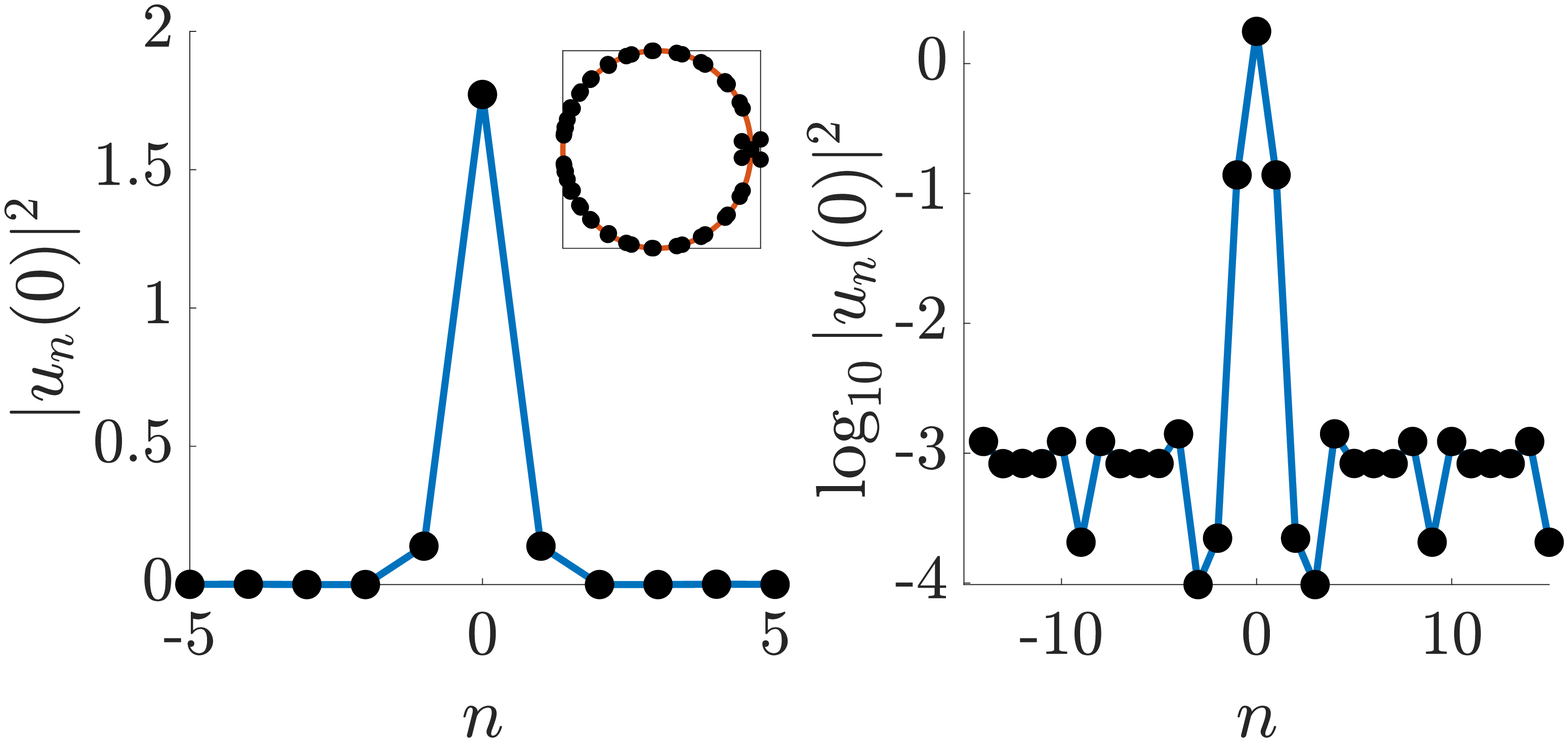}
    \includegraphics[width=9cm]{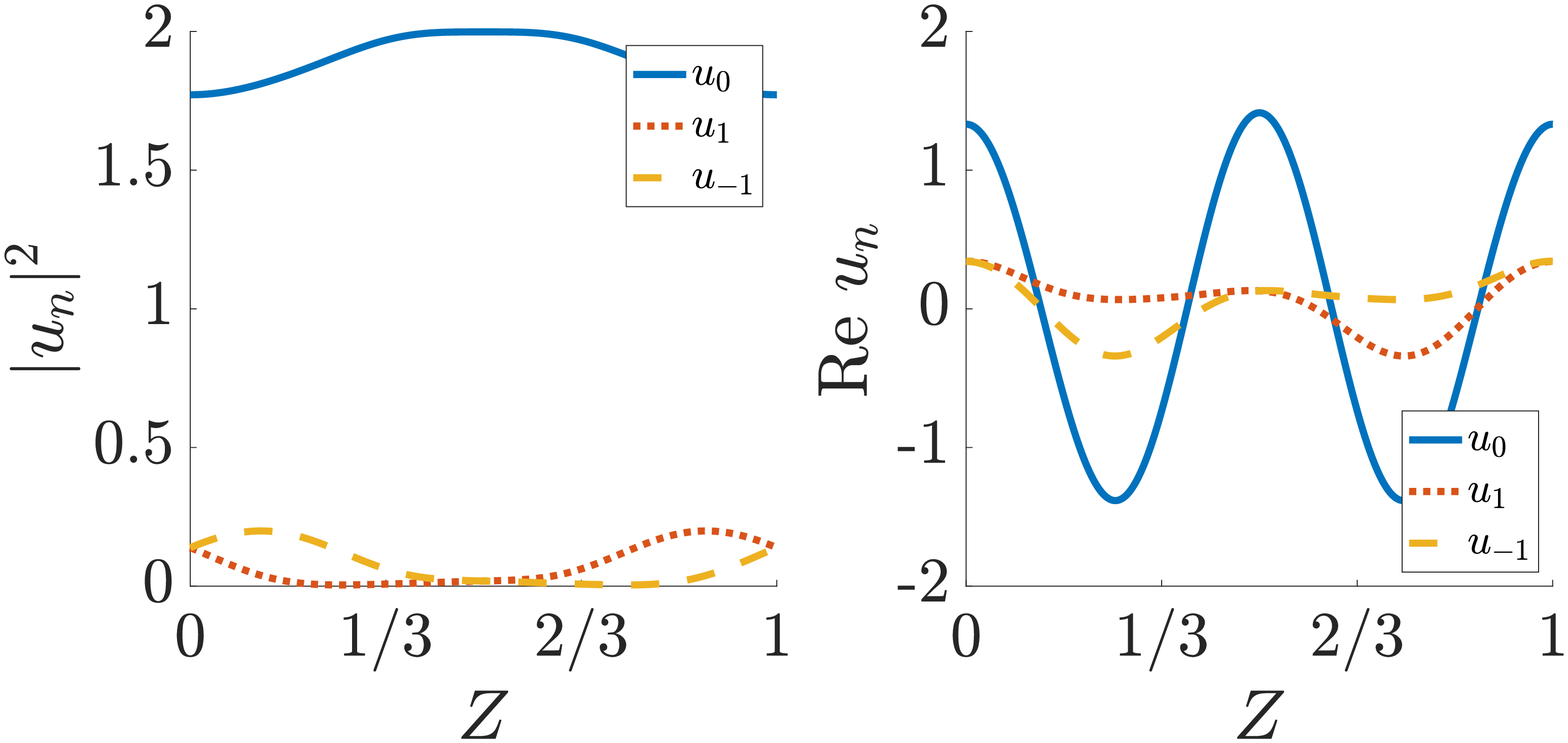}
    \includegraphics[width=9cm]{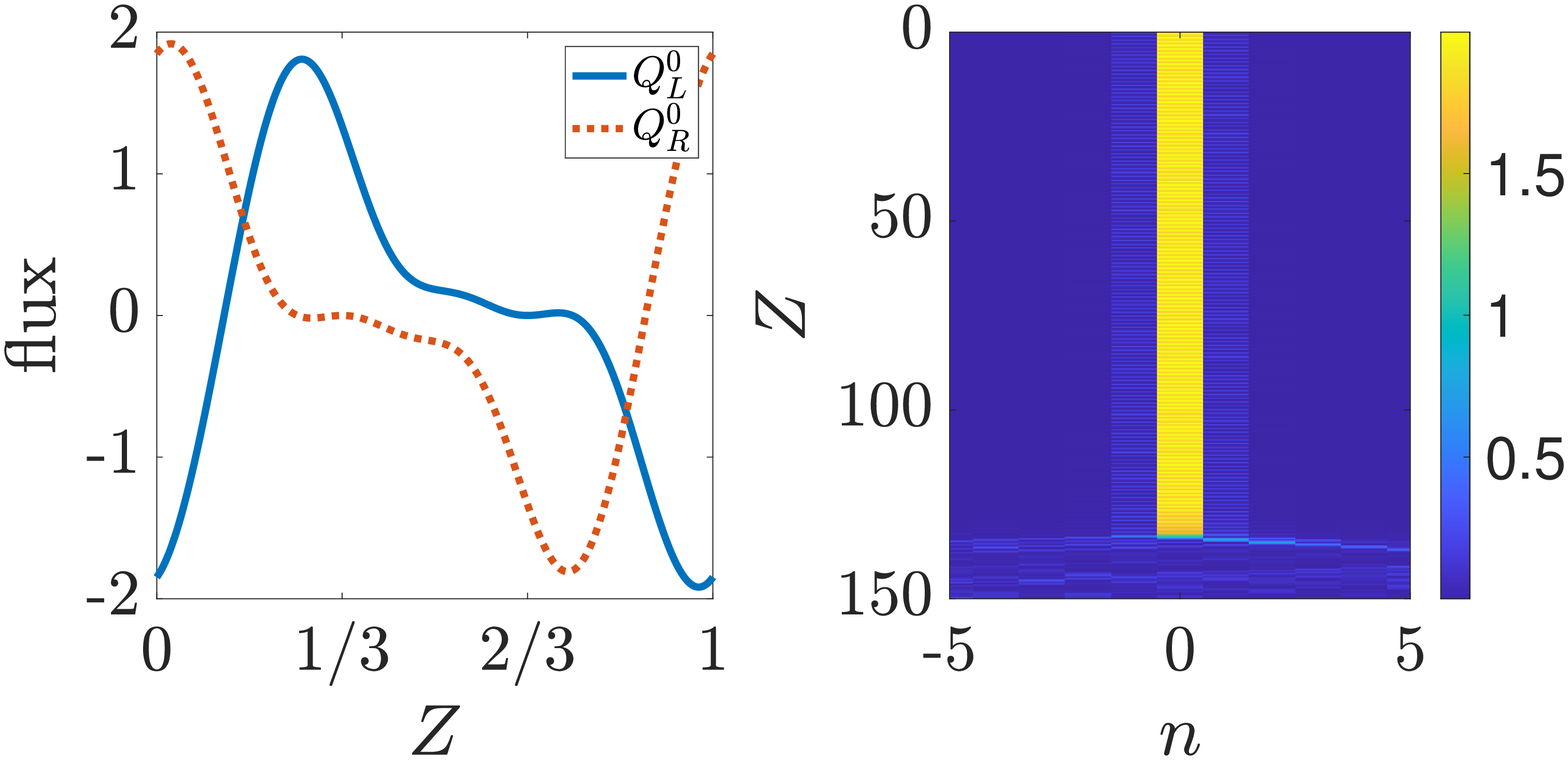}
    \caption{Top: initial intensity $|u_n(0)|^2$ (left) with inset showing Floquet multipliers, and log of initial intensity (right) for stationary solution with approximate power of 2. Middle: intensity (left) and real part (right) of three central sites over one period. Bottom: Leftward flow $Q_0^L$ and rightward flow $Q_0^R$ of intensity at central site $n=0$ (left), long term evolution in $Z$ (right). $C=0.4$, $J_0=0.05$.}
    \label{fig:stat2}
\end{figure}

\begin{figure}
    \centering
    \includegraphics[width=9cm]{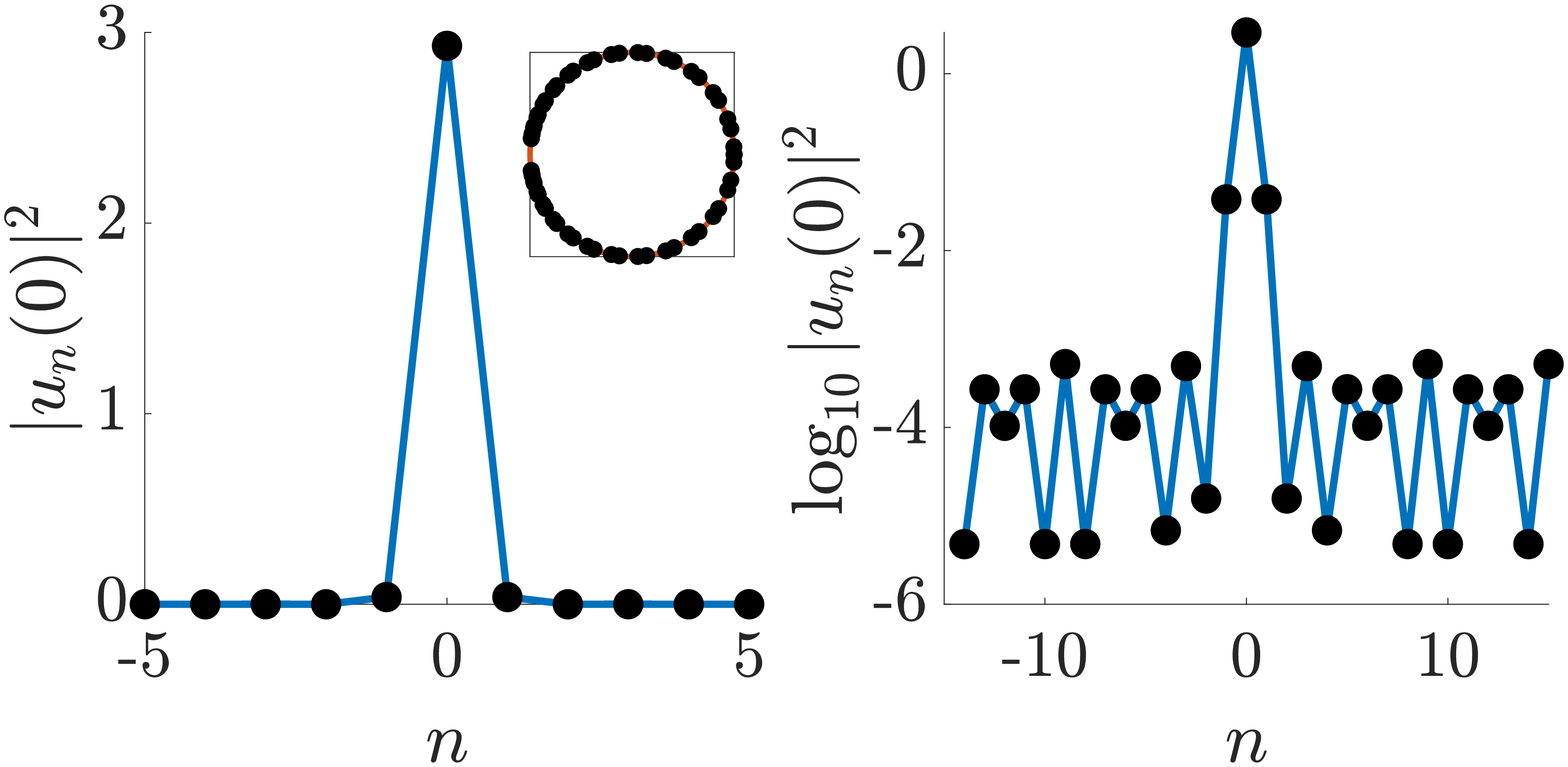}
    \includegraphics[width=9cm]{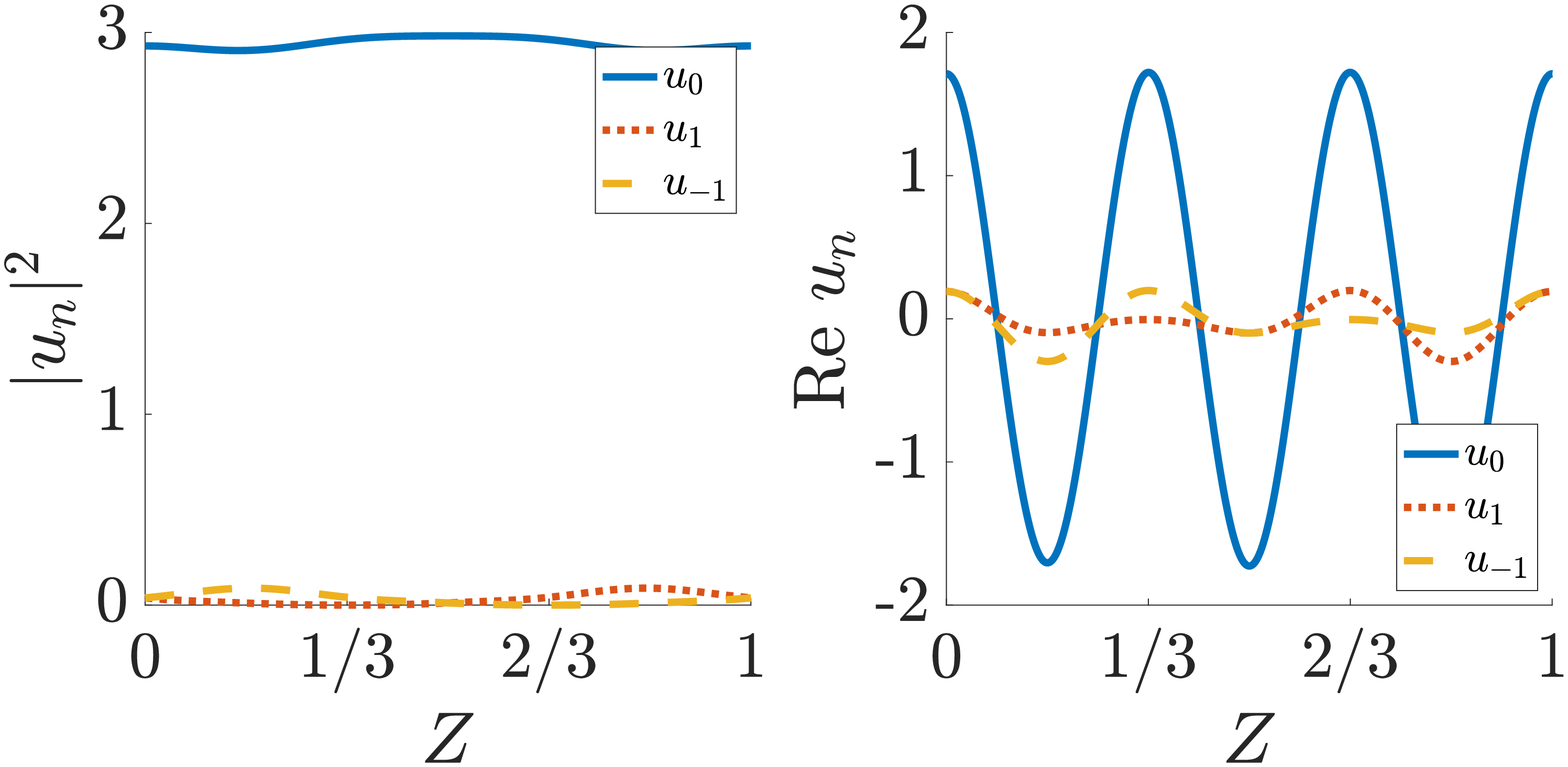}
    \includegraphics[width=9cm]{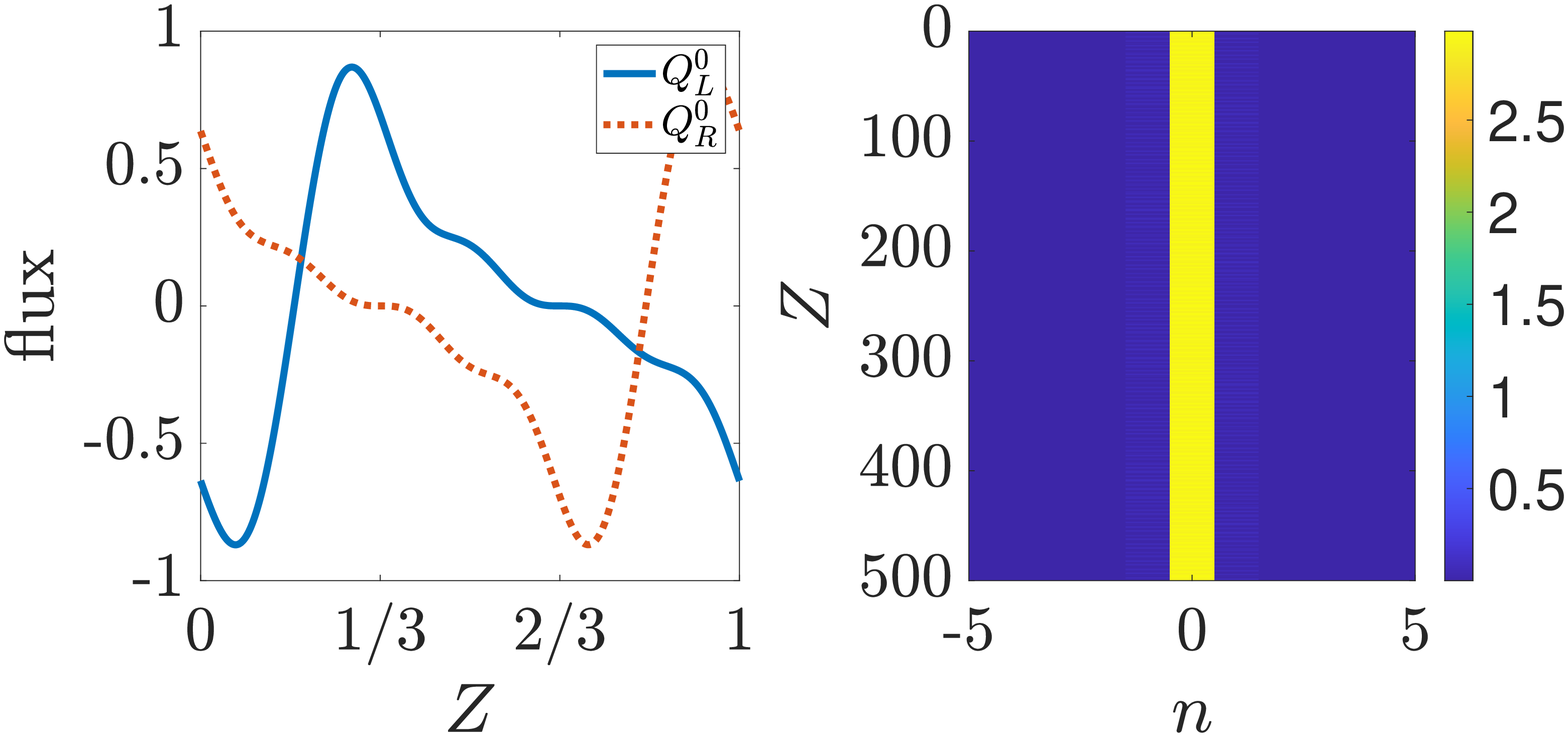}
    \caption{Same as \cref{fig:stat2}, but for the non-moving solution with approximate power of 3.}
    \label{fig:stat3}
\end{figure}

First, we look at the stationary solutions. At the anti-continuum (AC) limit ($J_0=0$ and $C=0$), the lattice sites are decoupled, and an initial intensity $P$ at lattice site $n$ will yield a standing wave solution of frequency $P$, i.e., of the form $u_n(Z) = \sqrt{P} e^{ 2 \pi i P Z}$. Since such a solution has period $1/P$, and stationary solutions must have an integer period, these solutions will exist in a discrete family for every integer period $N$, i.e., approximately $P = k/N$ for sufficiently large positive integer $k$. For period $N=1$ and the parameters in the previous section, for example, we expect to have
time-periodic, non-moving solutions for approximate integer intensities $P \geq 2$. See \cref{fig:stat2} and \cref{fig:stat3} for the first two of these solutions. By looking at the intensity and the real part of the central site $n=0$ (middle panel), we see that they are approximately standing waves with frequency 2 and 3 (respectively). We note that the stationary solutions do not decay to 0 with increasing $|n|$, but rather the tails exhibit small amplitude oscillatory patterns (see top right of \cref{fig:stat2} and \cref{fig:stat3}); the specific pattern of oscillations depends on the lattice size (not shown).
Looking at the sites adjacent to the central one, the left neighbor $u_{-1}$ peaks on the interval $[0,1/3]$ when the coupling $J_2(z)$ is most active, and the leftward flow $Q_0^L < 0$, indicating flow of intensity out of site 0 to the left. The right neighbor peaks on the interval $[2/3, 1]$ when the coupling $J_0(z)$ is most active, and the rightward flow $Q_0^R < 0$, indicating flow of intensity out of site 0 to the right. Both $Q_0^L$ and $Q_0^R$ are close to 0 on the interval $[1/3, 2/3]$, when neither nearest-neighbor coupling is strong.

Since these non-mobile solutions are true periodic orbits with integer period, so that their period is equal to or commensurate with that of the coupling, their spectral stability can be determined by Floquet theory. Numerical computation of the Floquet multipliers of the stationary solutions is shown in the insets of the top left plots in \cref{fig:stat2} and \cref{fig:stat3}. The lower power solution has two pairs of Floquet multipliers off of the unit circle, which is characteristic of an oscillatory instability. Long term evolution in $Z$ (bottom right plot of \cref{fig:stat2}) shows that this solution remains coherent until approximately $Z=130$. By contrast, the Floquet spectrum of the higher power solution lies on the unit circle, indicating spectral stability. Long term evolution in $Z$ (bottom right plot of \cref{fig:stat3}) shows that this solution is still coherent at $Z=500$.

\begin{figure}
    \centering
    \includegraphics[width=8cm]{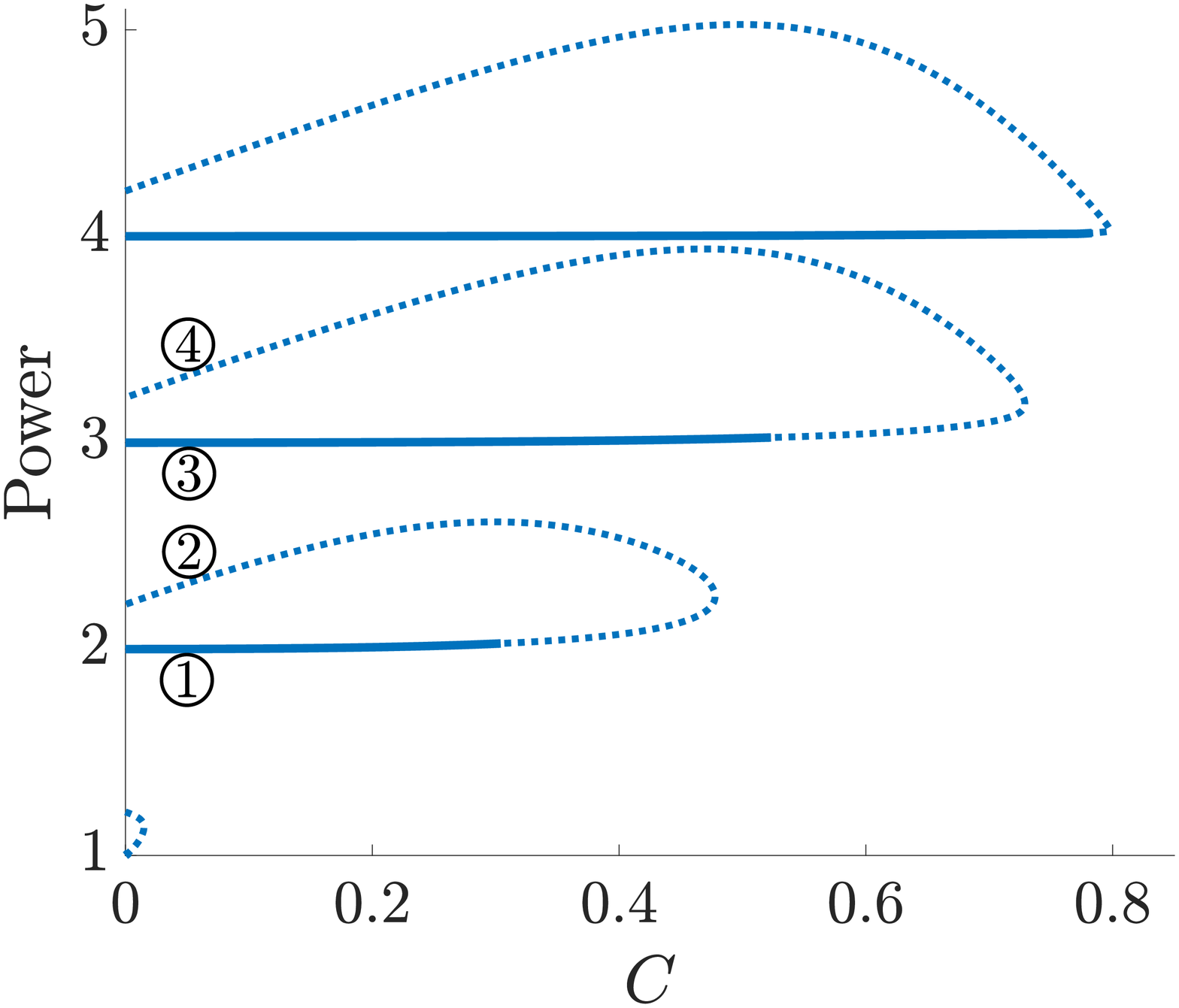}
    \includegraphics[width=8cm]{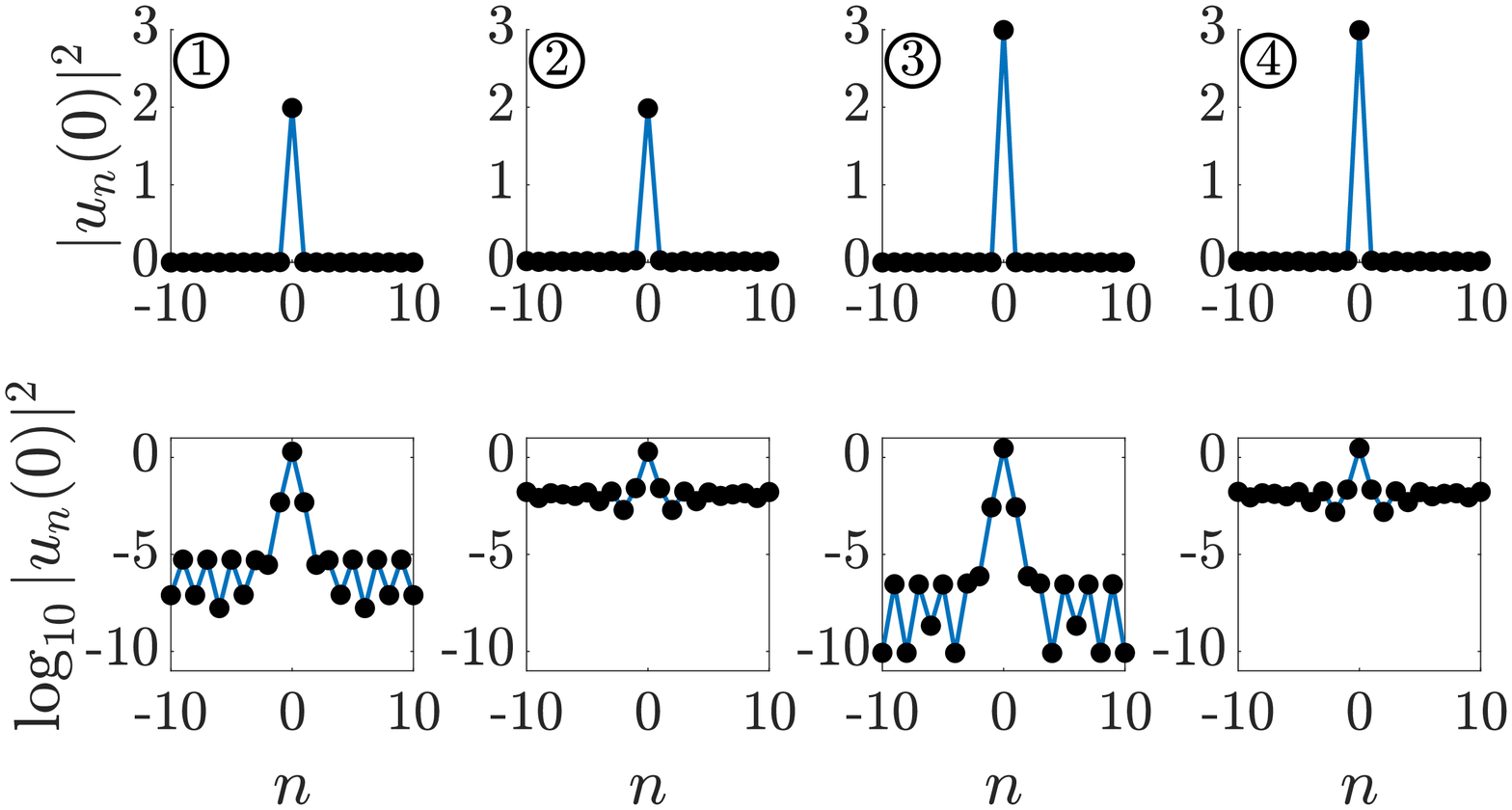}
    \caption{Branches of stationary solutions with spatial period (in $Z$) of 1, obtained from numerical parameter continuation starting with DNLS soliton at $C=0$. Bottom plots show intensity (top) and log of the intensity (bottom) of initial condition, and correspond to $C=0.05$ at labeled points on bifurcation diagram. Solid lines correspond to solutions with Floquet spectrum contained in the unit circle, dotted lines correspond to solutions with some Floquet spectrum outside of unit circle. Solutions on other branches at these values of $C$ are qualitatively similar.} 
    \label{fig:statAC}
\end{figure}

Using numerical parameter continuation, we start with the DNLS soliton at $C=0$ and slowly vary $C$, tracing the curves in \cref{fig:statAC} ($J_0 = 0.05$ throughout). Since we are looking for solutions with period 1, the starting single site intensity must take integer values so that the frequency of the DNLS standing wave at $C=0$ is commensurate with this period. In all cases, a turning point is reached, as $C$ is increased, at which point the parameter continuation in the
coupling parameter $C$ reverses direction. This turning point occurs at a larger value of $C$ for solutions which start at a higher initial intensities at $C=0$. All stationary solutions initially have their Floquet spectrum confined to the unit circle, thus are spectrally stable. Spectral stability is lost at some point before the turning point observed in the graph, when Floquet multipliers collide and leave the unit circle, creating an oscillatory instability. Solutions on the upper branches of the bifurcation diagram are periodic solutions to the DNLS which are not pure standing waves. To leading order, these upper solutions are the sum of two Fourier modes,
as opposed to standing waves, which comprise a single Fourier mode. Substituting the finite Fourier ansatz 
\[
u_n(Z) = \sum_{k=-N}^N a_{n,k} e^{2 \pi i k z}
\]
into \cref{eq:modelZ} and projecting onto each of the Fourier basis functions, we can obtain expressions for the coefficients $a_{n,k}$ for each wavenumber $k$. An FFT of the numerical solution on the upper branches suggests that the solutions at each site are composed predominantly of the modes with wavenumbers 0 and 1. Thus, to leading order, these solutions are of the form $u_n(Z) = a_{n,0} + a_{n,1} e^{2 \pi i \omega Z}$, where the coefficients $a_{n,0}$ and $a_{n,1}$ satisfy 
\begin{align*}
&J_0(a_{n+1,0}+a_{n-1,0}) + a_{n,0}^3 + 2 a_{n,1}^2 a_{n,0} = 0 \\
&J_0(a_{n+1,1}+a_{n-1,1}) + a_{n,1}^3 - w a_{n,1} + 2 a_{n,1} a_{n,0}^2 = 0.
\end{align*}
We note that if $a_{n,0} = 0$ for all $n$, the second equation reduces to the DNLS equation, in which case the solution is a standing wave.

\begin{figure}
    \centering
    \includegraphics[width=8cm]{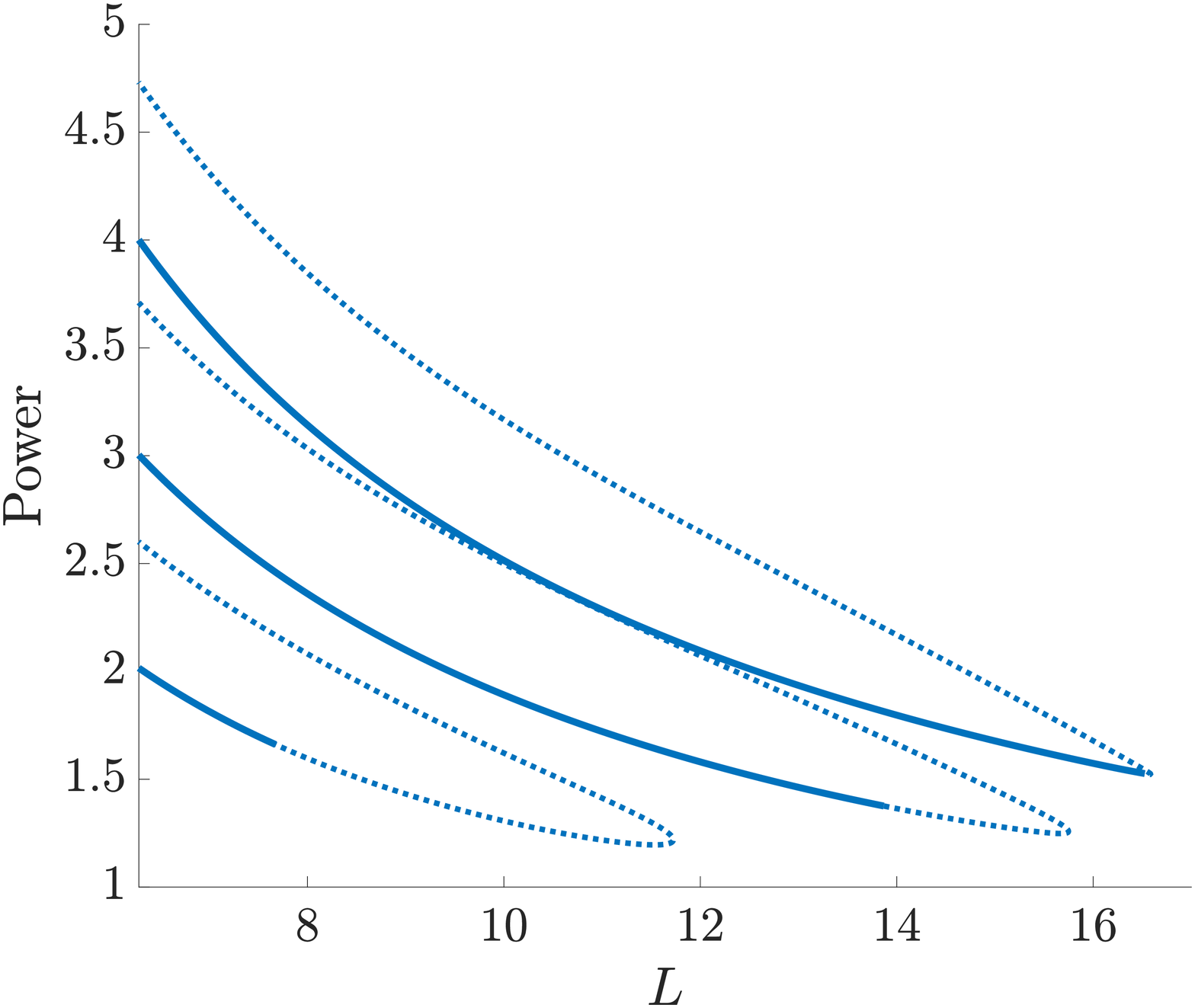}
    \caption{Parameter continuation in coupling period $L$. Plot shows power of solution vs. $L$, starting with solutions of approximate power 2, 3, and 4 at $L = 2\pi$. Solid lines correspond to solutions with Floquet spectrum contained in the unit circle, dotted lines correspond to solutions with some Floquet spectrum outside of unit circle. Diagram is only shown for $L\geq 2\pi$. Parameters $J_0 = 0.05$ and $C=0.25$.}
    \label{fig:statcontL}
\end{figure}

We can also continue solutions in the coupling period $L$ (\cref{fig:statcontL}). The intensity of the central peak, hence the overall power of the solution,
decreases with increasing $L$, thus solutions with greater starting power at $L=2\pi$ persist for higher $L$. 
For example, solutions with (approximate) starting power of 2, 3, and 4 at $L=2\pi$ persist up to $L=11.7$, 15.7, and 16.6, respectively (see \cref{fig:statcontL}).
Again, there is a turning point where the continuation reverses directions, which occurs at larger $L$ for higher power branches. The central site for the upper and lower branches of each loop has approximately the same intensity; the higher power of the upper branches is due to larger intensity in the tails of the solutions. For contrast, the spatial period of the solutions in \cite[Figure 2]{Jurgensen2021} is $L=8000$, which simulates the adiabatic regime; 
since the power of the solution decreases as $L$ is increased by parameter continuation,
we would have to start with a solution with extremely high power at $L=2\pi$ to be able to reach such a large $L$ using this method. 
Obtaining such solutions with a very large spatial period $L$ directly by a shooting method is similarly computationally 
impractical.

Finally, we note that, while we have only considered non-moving solutions with period of 1, stationary solutions do exist for other positive integer periods. For example, if we start with a single-site initial condition with intensity $k/2$ for positive, odd integer $k$, we expect that we will obtain a stationary solution with period 2. (We have verified that this is the case for single-site initial conditions with intensities 3/2 and 5/2). While, in principle, this can be done for any integer period, it becomes computationally intractable for larger periods.

\subsubsection{Moving solutions}\label{sec:movingsol}

\begin{figure}
    \centering
    \includegraphics[width=9cm]{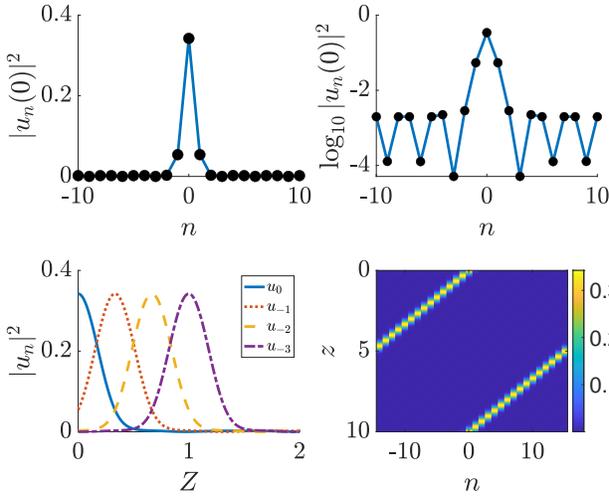}
    \caption{Initial intensity $|u_n(0)|^2$ (top left) and log of initial intensity (top right) for left-moving solution. The intensity of the solution evolved in $Z$ over a period is presented for a few select sites (bottom left), and the space-time contour 
    plot evolution of the intensity for the traveling wave is also shown (bottom right).}
    \label{fig:leftsol}
\end{figure}

\begin{figure}
    \centering
    \includegraphics[width=9cm]{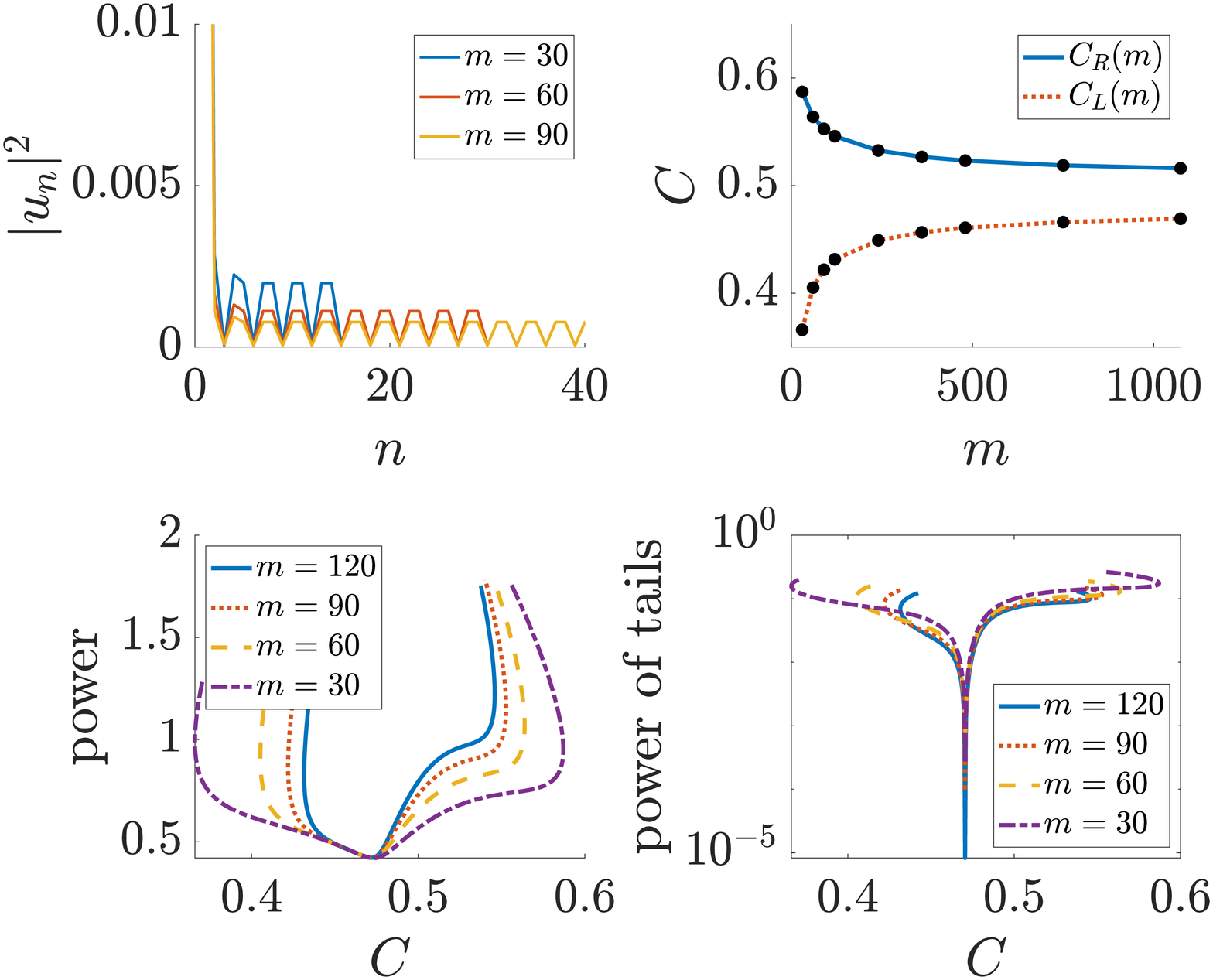}
    \caption{Top left: plot of intensity of the tails for the left-moving solution and for 3 values of the lattice size $m$. Top right: interval of existence $[C_L(m),C_R(m)]$ of left-moving solution. Bottom left: power of left-moving solution vs. $C$ for parameter
    continuation in $C$. Bottom right: maximum intensity of the tails for the left-moving solution vs. $C$. Minimum is at $C^* = 0.4709$ for all lattice sizes $m$.}
    \label{fig:leftdiag}
\end{figure}

\begin{figure}
    \centering
    \includegraphics[width=9cm]{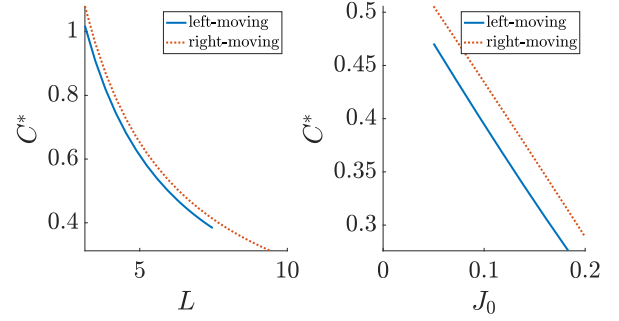}
    \caption{Plot of critical value $C^*$ of $C$ at which the intensity of the 
    tails of left- and right-moving solutions is a minimum vs. $L$ (left), $J_0 = 0.05$. Plot of $C^*$ vs $J_0$ (right), $L=2\pi$.
    }
    \label{fig:Cstar}
\end{figure}

Next, we look for moving solutions. For a given lattice size $m$, we find that leftward moving solutions exist (\cref{fig:leftsol}) for all values of $C$ within an interval $[C_L(m), C_R(m)]$ (see top right and bottom left of \cref{fig:leftdiag}).
These are true coherent structures, in that the entire solution reproduces itself exactly after one period, shifted three sites to the left. (In the numerical simulation, where we are using periodic boundary conditions on the lattice, we can think of this as a ``circular shift''). 
We note that is possible to find solutions which reproduce themselves modulo a phase multiplier $e^{i\theta}$ after one period, and these will have different power from the true coherent structures; we will not consider these solutions herein.
Generically, these solutions have oscillatory tails (\cref{fig:leftsol}, top right), and the amplitude of these oscillations depends on the lattice size (\cref{fig:leftdiag}, top left). 
Notice, however, that the corresponding wavenumber in the far field does not.
At a critical value $C^*$ of $C$ ($C^* = 0.4709$ for $J_0 = 0.05$ and $L=2\pi$), the tail oscillations vanish, leaving a localized traveling solution (\cref{fig:leftdiag}, bottom right). Most notably, the value of $C^*$ is independent of the lattice size $m$, although it does depend on both $L$ and $J_0$ (\cref{fig:Cstar}). The left moving solution appears to be stable when $C=C^*$ (see \cref{fig:movelong}, left); at minimum, it persists unchanged for at least 1000 periods. In addition, the solution appears to be stable for an interval in $C$ containing $C^*$ (not shown). 
The presence of $C^*$ seems to suggest an analogy with the so-called Stokes
constant calculation in similar traveling (DNLS-type) problems, as in the
work of~\cite{igorb}. Further expanding on this connection could be an 
interesting problem for the future (but is outside the scope of the present work).
Since the traveling solution is not a periodic orbit, we cannot use standard Floquet analysis to determine its spectral stability. That being said, since the traveling solution is periodic modulo a shift by an integer number of lattice points, it might be possible to adapt some aspects of Floquet theory to this case.
We note that while the parameter continuation in the bottom left of \cref{fig:leftdiag} continues past the turning points at $C_L(m)$ and $C_R(m)$
(so that there are solutions with different powers for the same value of $C$),
this merely represents growth of the tail oscillations, while the intensity of the central site remains essentially unchanged; since none of these solutions are stable, the continuation diagram is not shown past these turning points. 

\begin{figure}
    \centering
    \includegraphics[width=9cm]{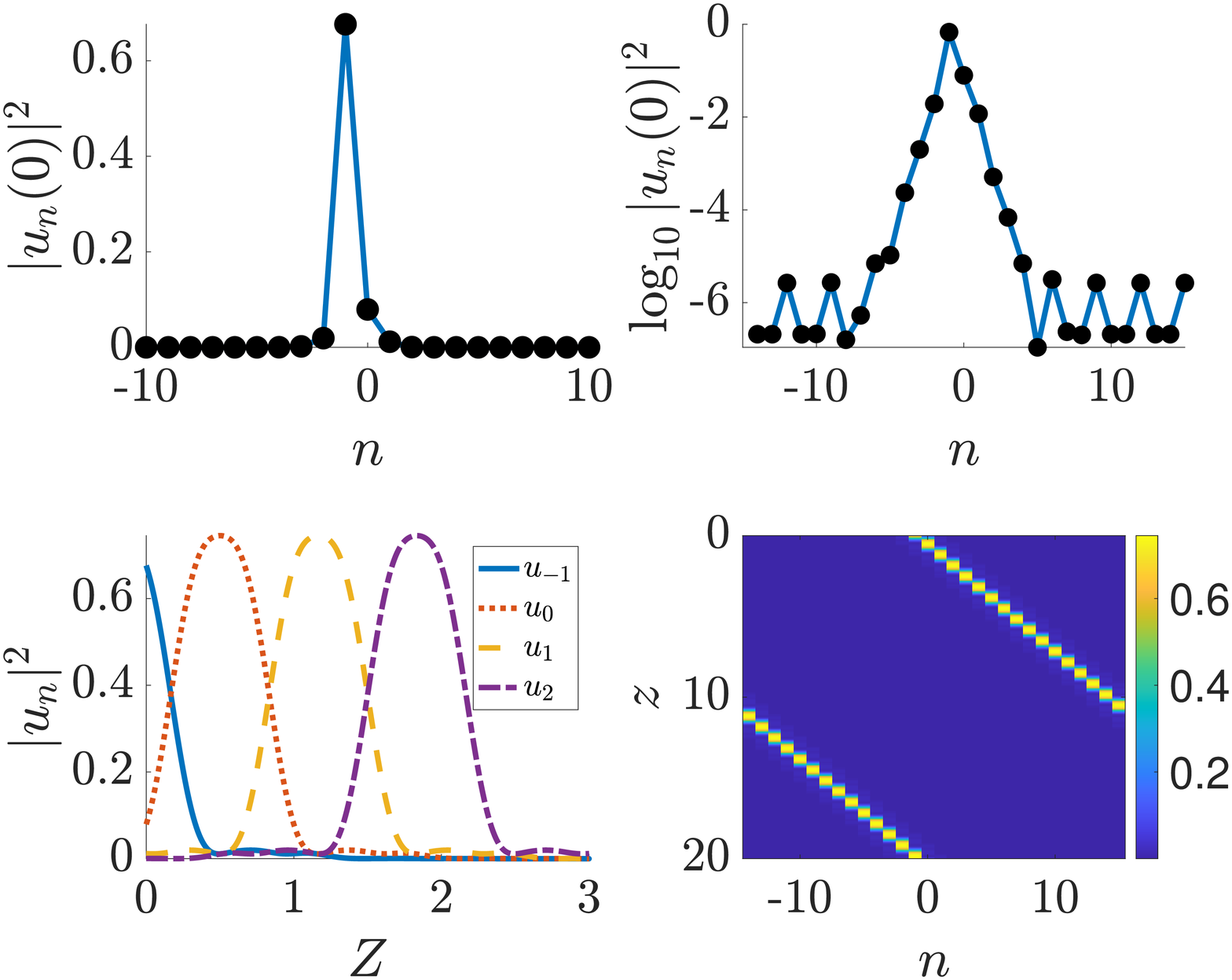}
    \caption{Initial intensity $|u_n(0)|^2$ (top left) and log of initial intensity (top right) for right moving solution. The intensity of the solution evolved in $Z$ over a period is presented for a few select sites (bottom left), and the space-time contour 
    plot evolution of the intensity for the traveling wave is also shown (bottom right).}
    \label{fig:rightsol}
\end{figure}

\begin{figure}
    \centering
    \includegraphics[width=9cm]{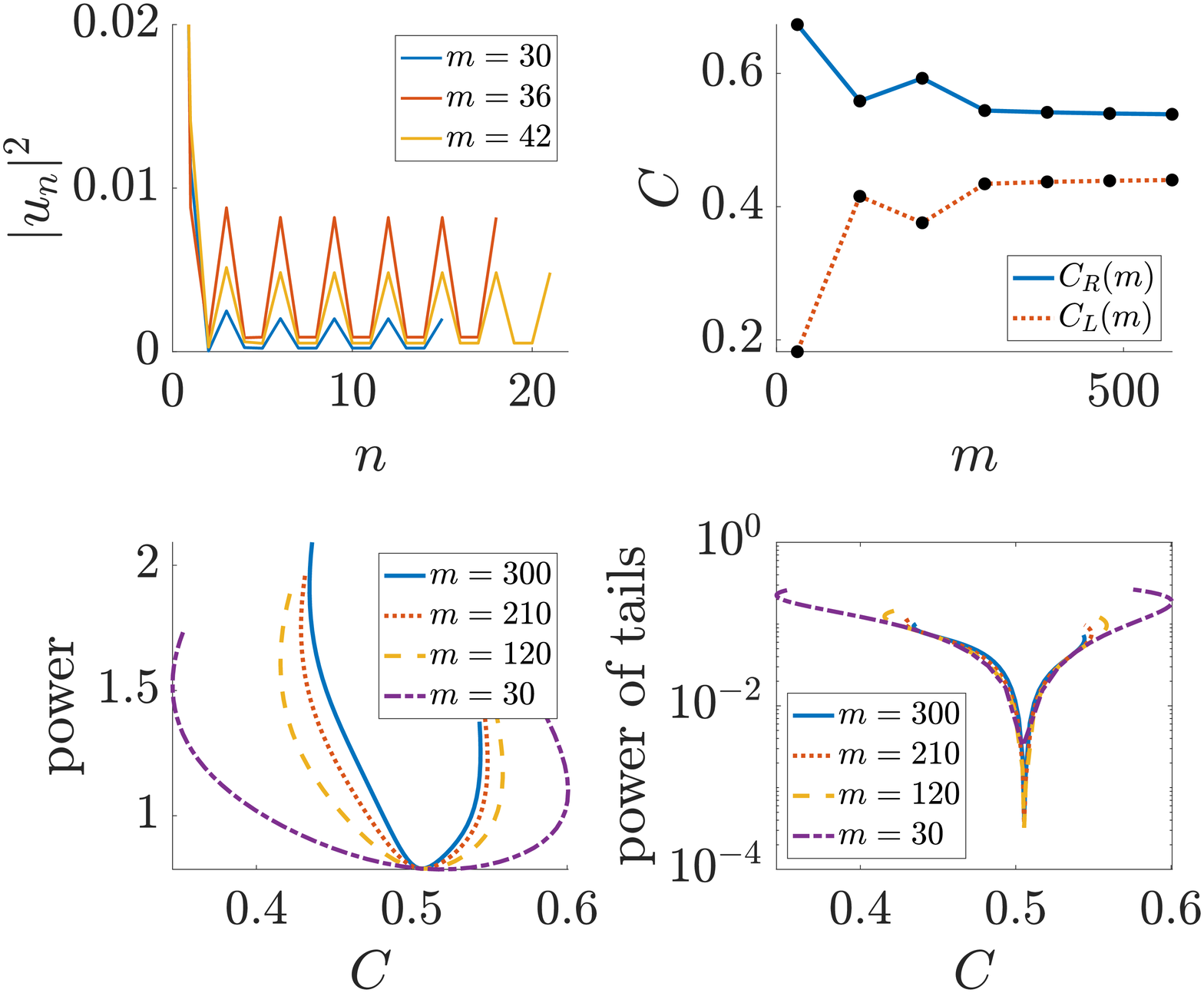}
    \caption{Top left: plot of intensity of of tails of of right-moving solution for 3 values of the lattice size $m$. Top right: interval of existence $[C_L(m),C_R(m)]$ of right-moving solution. Bottom left: power of right-moving solution vs. $C$ for parameter continuation in $C$. Bottom right: maximum intensity of tails of right-moving solution vs. $C$. Minimum is at $C^* = 0.5054$ for all lattice sizes $m$.}
    \label{fig:rightdiag}
\end{figure}

\begin{figure}
    \centering
    \hspace{-0.15cm}
    \includegraphics[width=4cm]{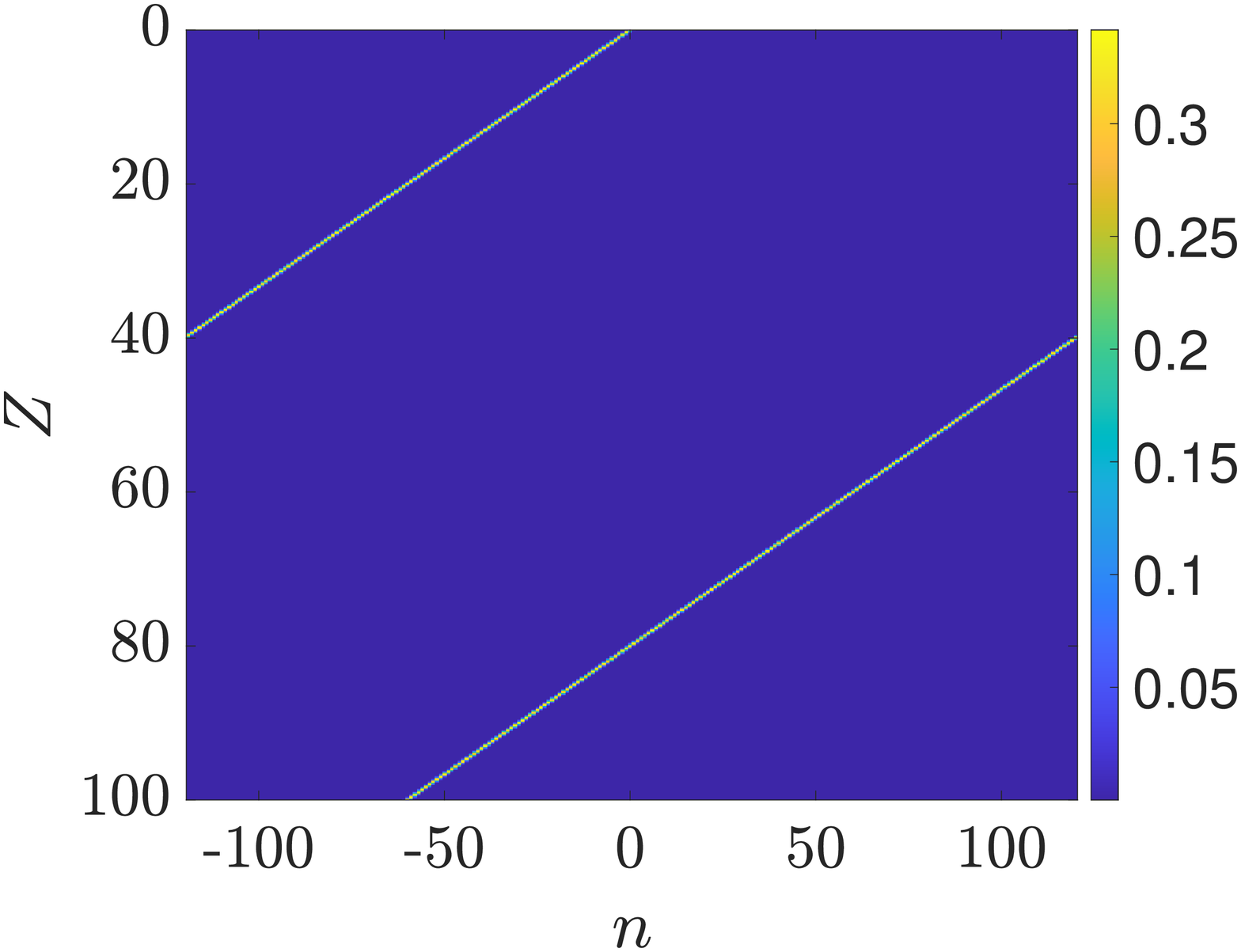}\hspace{-0.15cm}
    \includegraphics[width=4cm]{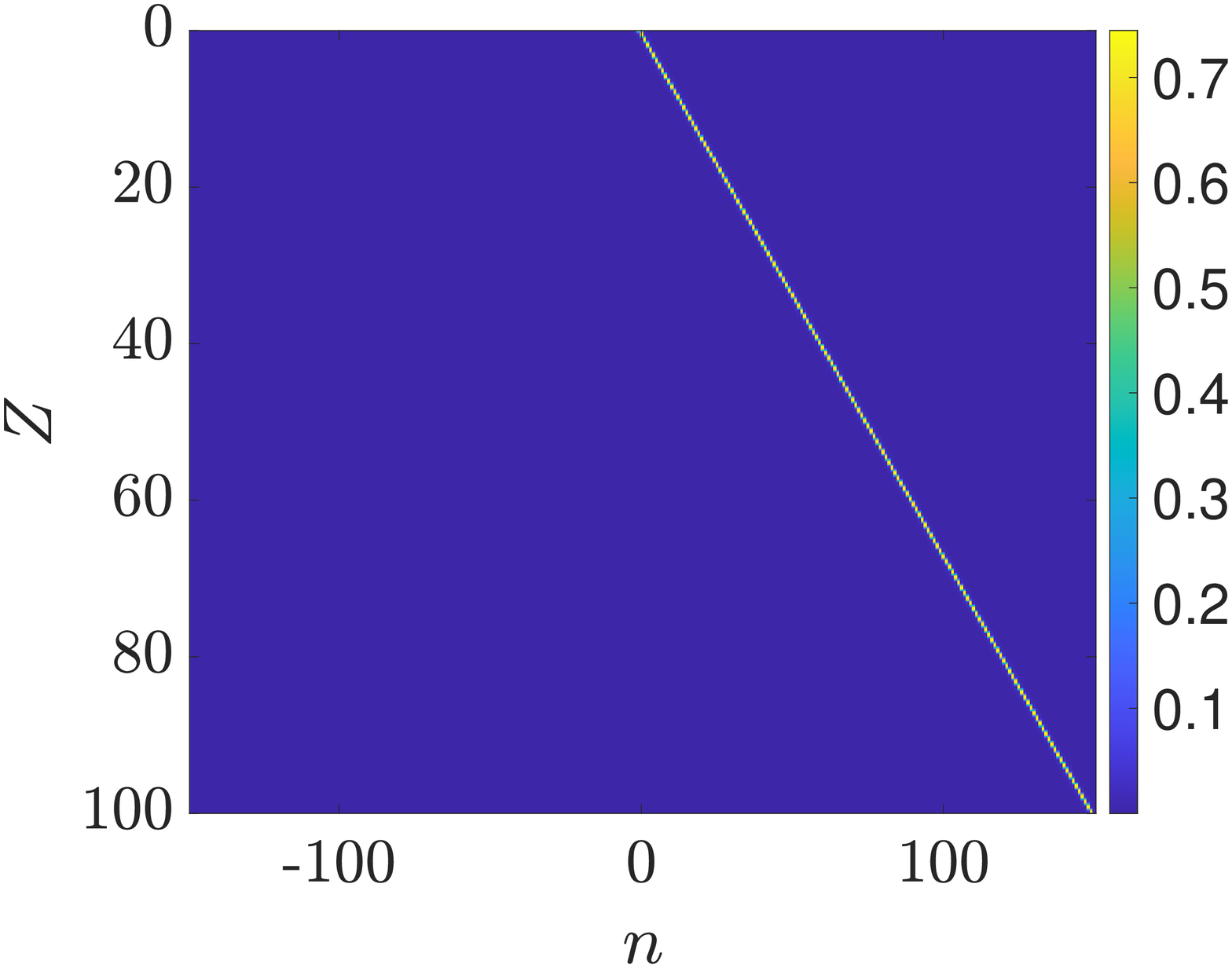}
    \includegraphics[width=3.95cm]{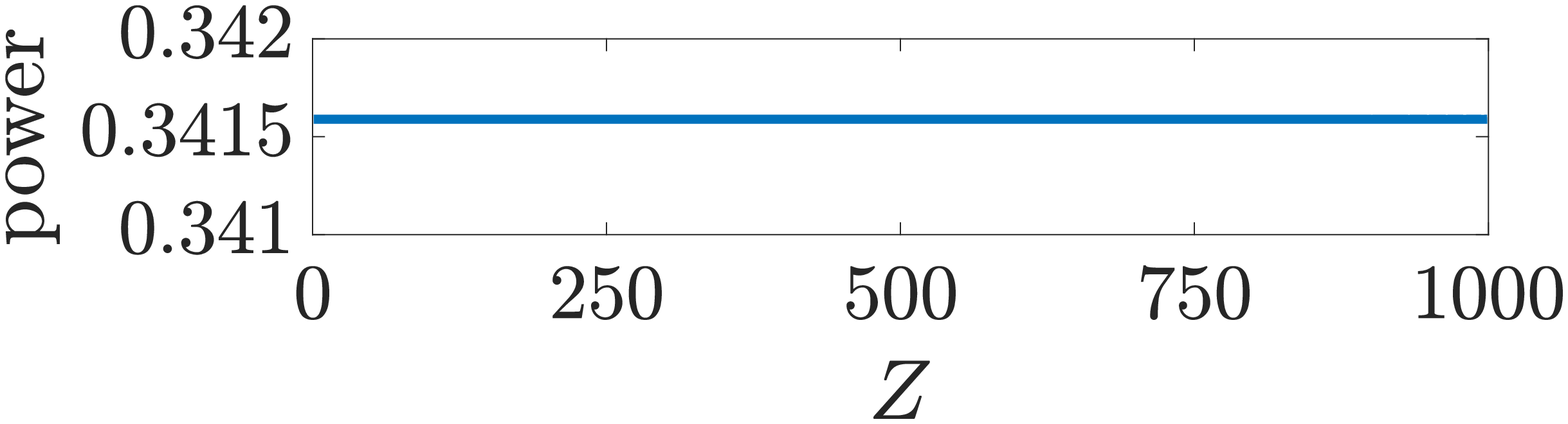}\hspace{-0.15cm}
    \includegraphics[width=3.95cm]{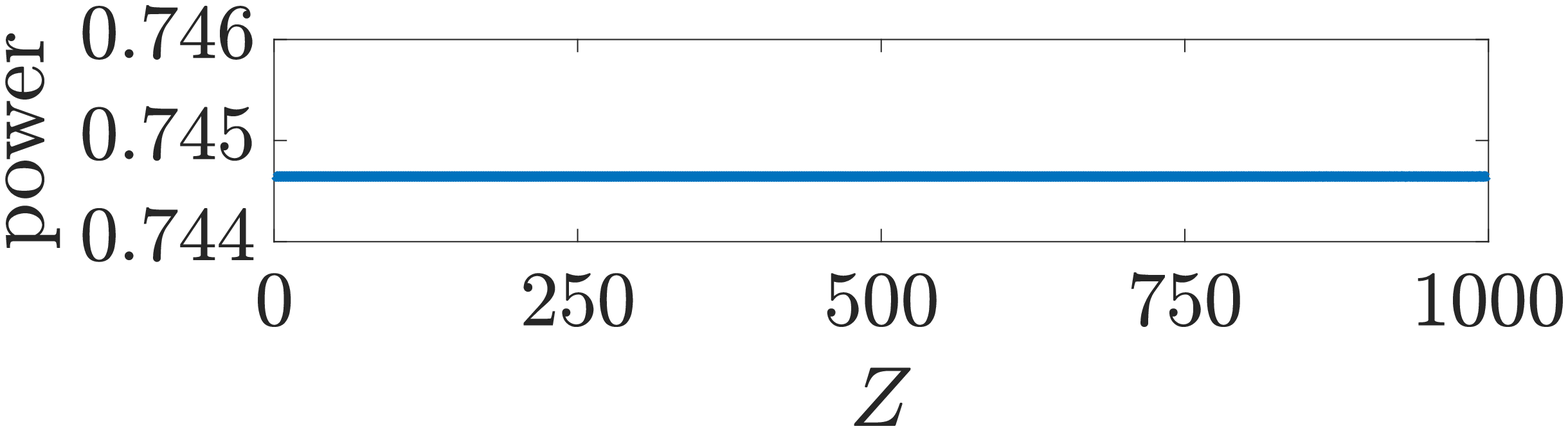}
    \caption{Top: colormap of long term evolution in $Z$ of left-moving solution (left, $C=0.4703$, $m=240$) and right-moving solution (right, $C=0.5054$, $m=300$) for $C=C^*$. Bottom: intensity of site with peak intensity of moving solution over 1000 periods.}
    \label{fig:movelong}
\end{figure}

Similar results are obtained for the right-moving solutions (\cref{fig:rightsol} and \cref{fig:rightdiag}). Once again, the tail oscillations vanish at a critical value $C^*$ of $C$ ($C^* = 0.5054$ for $J_0 = 0.05$ and $L=2\pi$), which is close, but not equal, to the value for the left-moving solution. The right-moving solution also appears to be stable at (and near) $C=C^*$ (see \cref{fig:movelong}, right). Unlike the left-moving solution, which is symmetric (\cref{fig:leftsol}, top left), the right-moving solution is asymmetric (\cref{fig:rightsol}, top left). For the initial condition of the right-moving solution, the intensity profile is skewed to the right. 
In addition, the
intensity of the central site for the right-moving solution (approximately 0.6842) is significantly higher than that of the left-moving solution (approximately 0.3416).

\subsection{Collisions}\label{sec:collisions}

\begin{figure}
    \centering
    \includegraphics[width=9cm]{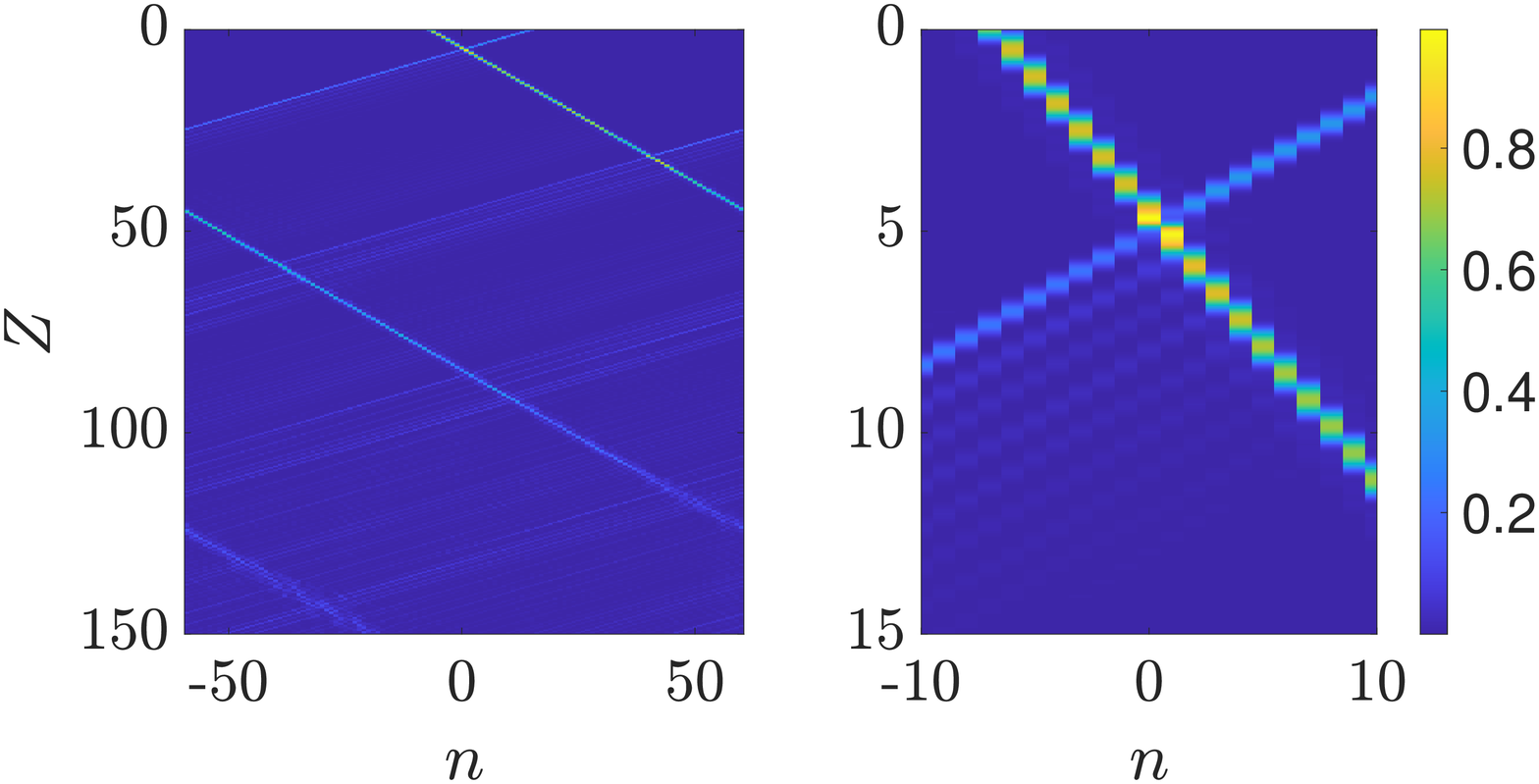}
    \includegraphics[width=9cm]{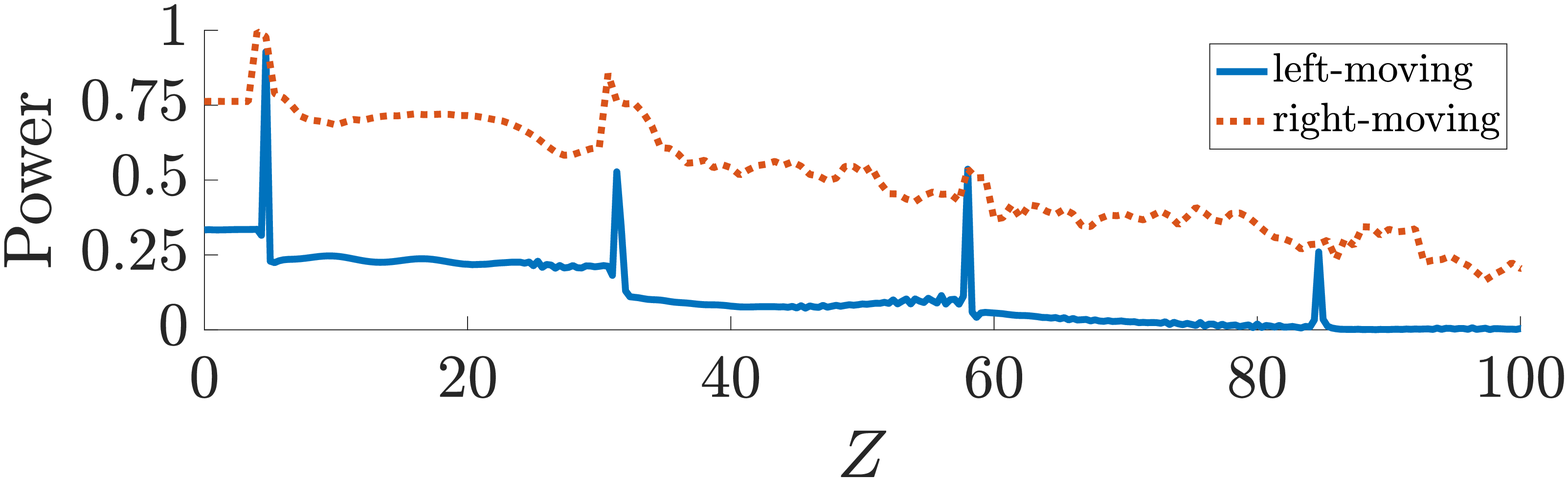}
    \caption{Top: colormap of evolution in $Z$ and space denoted by $n$ of the collision between left-moving and right-moving solution. The right panel is a zoomed-in view of the first collision, representing more clearly the intensity loss that the waves incur as a result. Bottom: evolution of the site with peak intensity for the profile bearing the left- and right-moving solutions.
    $m=120$, $C=0.4703$.}
    \label{fig:collision1}
\end{figure}

Finally, we briefly explore the resulting phenomenology when a left-moving and a right-moving solution collide, an event shown in \cref{fig:collision1}. For the
relevant initial condition, we splice together well-separated copies of the left-moving and right-moving solutions. To avoid combining the tail oscillations of the two solutions, we choose to simulate such a scenario when $C=C^*$ for the left-moving solution, so that its tail oscillations are suppressed. 
The reasons for this are three-fold. First, since we are interested in collisions between the localized structures, we seek to minimize effects stemming from the small, but nonzero background oscillations. Second, we wish to minimize the effect of lattice size, since these tail oscillations depend on the size of the underlying lattice (and hence would impact
the reproducibility of the results for different size lattices).
Finally, in a different case, the tail oscillations would superpose, producing
more drastic events of dispersive radiation wavepackets throughout the course
of our simulations.
Numerical evolution experiments show that although both structures emerge from the first collision, they both lose intensity in the form of radiation of intensity to the left (recall that the overall power of the solution is conserved). Intensity is lost with each subsequent collision (\cref{fig:collision1}, bottom) within the periodic ring of our domain. Accordingly,
the waveforms keep disintegrating (a feature possibly due to the non-integrability
of the solitary waves) as a progressive outcome of the relevant collisions.

\section{Conclusions and Future Directions}\label{sec:conclusions}

In the present work, we have studied coherent structures in a one-dimensional 
optical waveguide array with periodically modulated coupling, which was directly
motivated by a sequence of impactful physical experiments in the work of~\cite{Mukherjee2020,recht21,PhysRevLett.128.113901,Jurgensen2021}. We have found that the system exhibits two fundamental coherent structures in which the bulk of the intensity is concentrated on a single site. At low intensity, we find moving solutions, in which the intensity propagates leftward or rightward along the lattice. The direction and speed of propagation depend on which site is initially excited; this can be explained in terms of the coupling function which is most active at a given propagation distance $z$. At high intensity, we find stationary solutions, which are periodic orbits of the system. By analyzing a simplification of the model where the couplings between waveguides are given by step functions, we are able to explain this behavior by looking at an effective dimer setting, which features
the celebrated self-trapping transition.  Indeed, in the dimer, when the couplings do not change,  there is a sharp transition between solutions in which intensity is completely transferred back and forth between the two adjacent nodes and ones in which the intensity is mainly confined to one of the waveguides. For larger lattices, the couplings change three times every spatial period. This ``interrupts'' the intensity transfer in the dimer, which explains the fact that the sharp transition is now ``smoothened'' (i.e., is more gradual) in larger lattices. 
Nevertheless, the principal phenomenology is still present, as is also revealed by the
direct comparison of the two models (the original one and the variant with the step functions).
Using Floquet analysis, we find that the stationary coherent structures are stable for a wide range of parameters. The moving solutions are characterized by small-amplitude, oscillatory tails, whose amplitude and configuration depend on the lattice size. There is, however, a critical set of parameters for which these tail oscillations disappear. Interestingly, these critical parameters do not depend on the lattice size, and the moving solution appears to be stable for these parameters.

One potential avenue for future investigation would be examine what happens for very large spatial period $L$, which is the regime studied in \cite{Jurgensen2021}. Using our numerical parameter continuation methods, this would require starting with very large power single-site solutions at $C=0$ (see \cref{fig:statcontL}). So far, this has not been found to be computationally feasible, and would likely require a different numerical approach
(indeed, an adiabaticity-based one was used earlier in~\cite{Jurgensen2021}). Another direction would be to explore the solutions on the upper branches of the bifurcation diagram in \cref{fig:statAC}. These solutions, to leading order, involve Fourier modes of two different wavenumbers. Although we expect that the qualitative behavior will be the same for similar coupling functions, we could explore similar systems with unit cells comprising different numbers of sites. For a two-site unit cell, there would be left-right symmetry, and we expect that the rightward- and leftward-moving solutions would be mirror images of each other. It would be interesting to investigate what occurs if the unit cell comprises more than three sites, a setting that has also been explored in the above experiments.
Furthermore, in the vein of the earlier work of~\cite{PhysRevE.105.044211}, understanding
the impact of higher-dimensions (and possibly topological lattices therein) in the relevant
phenomenology would also be of substantial interest.
Lastly, it is relevant to point out that the study of non-Hermitian
systems is gaining considerable traction in recent years; see, e.g., the review
of~\cite{yangkonotop} and the book of~\cite{yangchristo}. It would be interesting
to explore the impact of different types of boundary conditions (including of
ones violating Hermiticity) to the non-autonomous lattice settings considered herein.

\begin{acknowledgments}
This material is based upon work supported by the U.S. National Science Foundation under the RTG grant DMS-1840260 (R.P. and A.A.), PHY-2110030, DMS-2204702 (P.G.K.), and DMS-1909559 (A.A.). J.C.-M. acknowledges support from EU (FEDER program 2014-2020) through both Consejería de Economía, Conocimiento, Empresas y Universidad de la Junta de Andaluc\'{\i}a (under the project US-1380977), and MICINN/AEI/10.13039/501100011033 (under the projects PID2019-110430GB-C21 and PID2020-112620GB-I00). We want to thank Mikael Rechtsman for his insights and suggestions which motivated many of the questions considered in the present manuscript.
\end{acknowledgments}

\bibliographystyle{apsrev4-2}
\bibliography{main.bib}

\begin{thebibliography}{49}%
\makeatletter
\providecommand \@ifxundefined [1]{%
 \@ifx{#1\undefined}
}%
\providecommand \@ifnum [1]{%
 \ifnum #1\expandafter \@firstoftwo
 \else \expandafter \@secondoftwo
 \fi
}%
\providecommand \@ifx [1]{%
 \ifx #1\expandafter \@firstoftwo
 \else \expandafter \@secondoftwo
 \fi
}%
\providecommand \natexlab [1]{#1}%
\providecommand \enquote  [1]{``#1''}%
\providecommand \bibnamefont  [1]{#1}%
\providecommand \bibfnamefont [1]{#1}%
\providecommand \citenamefont [1]{#1}%
\providecommand \href@noop [0]{\@secondoftwo}%
\providecommand \href [0]{\begingroup \@sanitize@url \@href}%
\providecommand \@href[1]{\@@startlink{#1}\@@href}%
\providecommand \@@href[1]{\endgroup#1\@@endlink}%
\providecommand \@sanitize@url [0]{\catcode `\\12\catcode `\$12\catcode
  `\&12\catcode `\#12\catcode `\^12\catcode `\_12\catcode `\%12\relax}%
\providecommand \@@startlink[1]{}%
\providecommand \@@endlink[0]{}%
\providecommand \url  [0]{\begingroup\@sanitize@url \@url }%
\providecommand \@url [1]{\endgroup\@href {#1}{\urlprefix }}%
\providecommand \urlprefix  [0]{URL }%
\providecommand \Eprint [0]{\href }%
\providecommand \doibase [0]{https://doi.org/}%
\providecommand \selectlanguage [0]{\@gobble}%
\providecommand \bibinfo  [0]{\@secondoftwo}%
\providecommand \bibfield  [0]{\@secondoftwo}%
\providecommand \translation [1]{[#1]}%
\providecommand \BibitemOpen [0]{}%
\providecommand \bibitemStop [0]{}%
\providecommand \bibitemNoStop [0]{.\EOS\space}%
\providecommand \EOS [0]{\spacefactor3000\relax}%
\providecommand \BibitemShut  [1]{\csname bibitem#1\endcsname}%
\let\auto@bib@innerbib\@empty
\bibitem [{\citenamefont {Lederer}\ \emph {et~al.}(2008)\citenamefont
  {Lederer}, \citenamefont {Stegeman}, \citenamefont {Christodoulides},
  \citenamefont {Assanto}, \citenamefont {Segev},\ and\ \citenamefont
  {Silberberg}}]{LEDERER20081}%
  \BibitemOpen
  \bibfield  {author} {\bibinfo {author} {\bibfnamefont {F.}~\bibnamefont
  {Lederer}}, \bibinfo {author} {\bibfnamefont {G.~I.}\ \bibnamefont
  {Stegeman}}, \bibinfo {author} {\bibfnamefont {D.~N.}\ \bibnamefont
  {Christodoulides}}, \bibinfo {author} {\bibfnamefont {G.}~\bibnamefont
  {Assanto}}, \bibinfo {author} {\bibfnamefont {M.}~\bibnamefont {Segev}},\
  and\ \bibinfo {author} {\bibfnamefont {Y.}~\bibnamefont {Silberberg}},\
  }\href {https://doi.org/https://doi.org/10.1016/j.physrep.2008.04.004}
  {\bibfield  {journal} {\bibinfo  {journal} {Physics Reports}\ }\textbf
  {\bibinfo {volume} {463}},\ \bibinfo {pages} {1} (\bibinfo {year}
  {2008})}\BibitemShut {NoStop}%
\bibitem [{\citenamefont {Morsch}\ and\ \citenamefont
  {Oberthaler}(2006)}]{RevModPhys.78.179}%
  \BibitemOpen
  \bibfield  {author} {\bibinfo {author} {\bibfnamefont {O.}~\bibnamefont
  {Morsch}}\ and\ \bibinfo {author} {\bibfnamefont {M.}~\bibnamefont
  {Oberthaler}},\ }\href {https://doi.org/10.1103/RevModPhys.78.179} {\bibfield
   {journal} {\bibinfo  {journal} {Rev. Mod. Phys.}\ }\textbf {\bibinfo
  {volume} {78}},\ \bibinfo {pages} {179} (\bibinfo {year} {2006})}\BibitemShut
  {NoStop}%
\bibitem [{\citenamefont {Remoissenet}(1999)}]{remoissenet}%
  \BibitemOpen
  \bibfield  {author} {\bibinfo {author} {\bibfnamefont {M.}~\bibnamefont
  {Remoissenet}},\ }\href@noop {} {\emph {\bibinfo {title} {Waves Called
  Solitons}}}\ (\bibinfo  {publisher} {Springer-Verlag, Berlin},\ \bibinfo
  {year} {1999})\BibitemShut {NoStop}%
\bibitem [{\citenamefont {Dauxois}\ and\ \citenamefont {Peyrard}(2006)}]{DP06}%
  \BibitemOpen
  \bibfield  {author} {\bibinfo {author} {\bibfnamefont {T.}~\bibnamefont
  {Dauxois}}\ and\ \bibinfo {author} {\bibfnamefont {M.}~\bibnamefont
  {Peyrard}},\ }\href@noop {} {\emph {\bibinfo {title} {Physics of solitons}}}\
  (\bibinfo  {publisher} {Cambridge University Press},\ \bibinfo {address}
  {Cambridge},\ \bibinfo {year} {2006})\BibitemShut {NoStop}%
\bibitem [{\citenamefont {Kevrekidis}(2009)}]{kev09}%
  \BibitemOpen
  \bibfield  {author} {\bibinfo {author} {\bibfnamefont {P.}~\bibnamefont
  {Kevrekidis}},\ }\href
  {https://link.springer.com/book/10.1007/978-3-540-89199-4} {\emph {\bibinfo
  {title} {{The discrete nonlinear Schr{\"o}dinger Equation}}}},\ \bibinfo
  {edition} {1st}\ ed.\ (\bibinfo  {publisher} {Springer-Verlag},\ \bibinfo
  {address} {Heidelberg},\ \bibinfo {year} {2009})\BibitemShut {NoStop}%
\bibitem [{\citenamefont {Ozawa}\ \emph {et~al.}(2019)\citenamefont {Ozawa},
  \citenamefont {Price}, \citenamefont {Amo}, \citenamefont {Goldman},
  \citenamefont {Hafezi}, \citenamefont {Lu}, \citenamefont {Rechtsman},
  \citenamefont {Schuster}, \citenamefont {Simon}, \citenamefont {Zilberberg},\
  and\ \citenamefont {Carusotto}}]{Ozawa2019}%
  \BibitemOpen
  \bibfield  {author} {\bibinfo {author} {\bibfnamefont {T.}~\bibnamefont
  {Ozawa}}, \bibinfo {author} {\bibfnamefont {H.~M.}\ \bibnamefont {Price}},
  \bibinfo {author} {\bibfnamefont {A.}~\bibnamefont {Amo}}, \bibinfo {author}
  {\bibfnamefont {N.}~\bibnamefont {Goldman}}, \bibinfo {author} {\bibfnamefont
  {M.}~\bibnamefont {Hafezi}}, \bibinfo {author} {\bibfnamefont
  {L.}~\bibnamefont {Lu}}, \bibinfo {author} {\bibfnamefont {M.~C.}\
  \bibnamefont {Rechtsman}}, \bibinfo {author} {\bibfnamefont {D.}~\bibnamefont
  {Schuster}}, \bibinfo {author} {\bibfnamefont {J.}~\bibnamefont {Simon}},
  \bibinfo {author} {\bibfnamefont {O.}~\bibnamefont {Zilberberg}},\ and\
  \bibinfo {author} {\bibfnamefont {I.}~\bibnamefont {Carusotto}},\ }\href
  {https://doi.org/10.1103/RevModPhys.91.015006} {\bibfield  {journal}
  {\bibinfo  {journal} {Rev. Mod. Phys.}\ }\textbf {\bibinfo {volume} {91}},\
  \bibinfo {pages} {015006} (\bibinfo {year} {2019})}\BibitemShut {NoStop}%
\bibitem [{\citenamefont {Ma}\ \emph {et~al.}(2019)\citenamefont {Ma},
  \citenamefont {Xiao},\ and\ \citenamefont {Chan}}]{Ma2019}%
  \BibitemOpen
  \bibfield  {author} {\bibinfo {author} {\bibfnamefont {G.}~\bibnamefont
  {Ma}}, \bibinfo {author} {\bibfnamefont {M.}~\bibnamefont {Xiao}},\ and\
  \bibinfo {author} {\bibfnamefont {C.~T.}\ \bibnamefont {Chan}},\ }\href
  {http://www.nature.com/articles/s42254-019-0030-x} {\bibfield  {journal}
  {\bibinfo  {journal} {Nat. Rev. Phys.}\ }\textbf {\bibinfo {volume} {1}},\
  \bibinfo {pages} {281} (\bibinfo {year} {2019})}\BibitemShut {NoStop}%
\bibitem [{\citenamefont {S{\"{u}}sstrunk}\ and\ \citenamefont
  {Huber}(2016)}]{Susstrunk2016}%
  \BibitemOpen
  \bibfield  {author} {\bibinfo {author} {\bibfnamefont {R.}~\bibnamefont
  {S{\"{u}}sstrunk}}\ and\ \bibinfo {author} {\bibfnamefont {S.~D.}\
  \bibnamefont {Huber}},\ }\href {https://doi.org/10.1073/pnas.1605462113}
  {\bibfield  {journal} {\bibinfo  {journal} {Proc. Natl. Acad. Sci. USA}\
  }\textbf {\bibinfo {volume} {113}},\ \bibinfo {pages} {E4767} (\bibinfo
  {year} {2016})}\BibitemShut {NoStop}%
\bibitem [{\citenamefont {Deng}\ \emph {et~al.}(2021)\citenamefont {Deng},
  \citenamefont {Li}, \citenamefont {Tournat}, \citenamefont {Purohit},\ and\
  \citenamefont {Bertoldi}}]{Bertoldi}%
  \BibitemOpen
  \bibfield  {author} {\bibinfo {author} {\bibfnamefont {B.}~\bibnamefont
  {Deng}}, \bibinfo {author} {\bibfnamefont {J.}~\bibnamefont {Li}}, \bibinfo
  {author} {\bibfnamefont {V.}~\bibnamefont {Tournat}}, \bibinfo {author}
  {\bibfnamefont {P.~K.}\ \bibnamefont {Purohit}},\ and\ \bibinfo {author}
  {\bibfnamefont {K.}~\bibnamefont {Bertoldi}},\ }\href@noop {} {\bibfield
  {journal} {\bibinfo  {journal} {Journal of the Mechanics and Physics of
  Solids}\ }\textbf {\bibinfo {volume} {147}},\ \bibinfo {pages} {104233}
  (\bibinfo {year} {2021})}\BibitemShut {NoStop}%
\bibitem [{\citenamefont {Cooper}\ \emph {et~al.}(2019)\citenamefont {Cooper},
  \citenamefont {Dalibard},\ and\ \citenamefont {Spielman}}]{Cooper2019}%
  \BibitemOpen
  \bibfield  {author} {\bibinfo {author} {\bibfnamefont {N.~R.}\ \bibnamefont
  {Cooper}}, \bibinfo {author} {\bibfnamefont {J.}~\bibnamefont {Dalibard}},\
  and\ \bibinfo {author} {\bibfnamefont {I.~B.}\ \bibnamefont {Spielman}},\
  }\href {https://doi.org/10.1103/RevModPhys.91.015005} {\bibfield  {journal}
  {\bibinfo  {journal} {Rev. Mod. Phys.}\ }\textbf {\bibinfo {volume} {91}},\
  \bibinfo {pages} {015005} (\bibinfo {year} {2019})}\BibitemShut {NoStop}%
\bibitem [{\citenamefont {Zhou}\ \emph {et~al.}(2022)\citenamefont {Zhou},
  \citenamefont {Rocklin}, \citenamefont {Leamy},\ and\ \citenamefont
  {Yao}}]{ssh}%
  \BibitemOpen
  \bibfield  {author} {\bibinfo {author} {\bibfnamefont {D.}~\bibnamefont
  {Zhou}}, \bibinfo {author} {\bibfnamefont {D.~Z.}\ \bibnamefont {Rocklin}},
  \bibinfo {author} {\bibfnamefont {M.}~\bibnamefont {Leamy}},\ and\ \bibinfo
  {author} {\bibfnamefont {Y.}~\bibnamefont {Yao}},\ }\href
  {https://doi.org/10.1038/s41467-022-31084-y} {\bibfield  {journal} {\bibinfo
  {journal} {Nature Communications}\ }\textbf {\bibinfo {volume} {13}},\
  \bibinfo {pages} {3379} (\bibinfo {year} {2022})}\BibitemShut {NoStop}%
\bibitem [{\citenamefont {Lumer}\ \emph {et~al.}(2013)\citenamefont {Lumer},
  \citenamefont {Plotnik}, \citenamefont {Rechtsman},\ and\ \citenamefont
  {Segev}}]{Lumer2013}%
  \BibitemOpen
  \bibfield  {author} {\bibinfo {author} {\bibfnamefont {Y.}~\bibnamefont
  {Lumer}}, \bibinfo {author} {\bibfnamefont {Y.}~\bibnamefont {Plotnik}},
  \bibinfo {author} {\bibfnamefont {M.~C.}\ \bibnamefont {Rechtsman}},\ and\
  \bibinfo {author} {\bibfnamefont {M.}~\bibnamefont {Segev}},\ }\href
  {https://doi.org/10.1103/PhysRevLett.111.243905} {\bibfield  {journal}
  {\bibinfo  {journal} {Phys. Rev. Lett.}\ }\textbf {\bibinfo {volume} {111}},\
  \bibinfo {pages} {243905} (\bibinfo {year} {2013})}\BibitemShut {NoStop}%
\bibitem [{\citenamefont {Solnyshkov}\ \emph {et~al.}(2017)\citenamefont
  {Solnyshkov}, \citenamefont {Bleu}, \citenamefont {Teklu},\ and\
  \citenamefont {Malpuech}}]{Solnyshkov2017}%
  \BibitemOpen
  \bibfield  {author} {\bibinfo {author} {\bibfnamefont {D.~D.}\ \bibnamefont
  {Solnyshkov}}, \bibinfo {author} {\bibfnamefont {O.}~\bibnamefont {Bleu}},
  \bibinfo {author} {\bibfnamefont {B.}~\bibnamefont {Teklu}},\ and\ \bibinfo
  {author} {\bibfnamefont {G.}~\bibnamefont {Malpuech}},\ }\href
  {https://doi.org/10.1103/PhysRevLett.118.023901} {\bibfield  {journal}
  {\bibinfo  {journal} {Phys. Rev. Lett.}\ }\textbf {\bibinfo {volume} {118}},\
  \bibinfo {pages} {023901} (\bibinfo {year} {2017})}\BibitemShut {NoStop}%
\bibitem [{\citenamefont {Smirnova}\ \emph {et~al.}(2019)\citenamefont
  {Smirnova}, \citenamefont {Smirnov}, \citenamefont {Leykam},\ and\
  \citenamefont {Kivshar}}]{Smirnova2019}%
  \BibitemOpen
  \bibfield  {author} {\bibinfo {author} {\bibfnamefont {D.~A.}\ \bibnamefont
  {Smirnova}}, \bibinfo {author} {\bibfnamefont {L.~A.}\ \bibnamefont
  {Smirnov}}, \bibinfo {author} {\bibfnamefont {D.}~\bibnamefont {Leykam}},\
  and\ \bibinfo {author} {\bibfnamefont {Y.~S.}\ \bibnamefont {Kivshar}},\
  }\href {https://doi.org/10.1002/lpor.201900223} {\bibfield  {journal}
  {\bibinfo  {journal} {Laser Photonics Rev.}\ }\textbf {\bibinfo {volume}
  {13}},\ \bibinfo {pages} {1900223} (\bibinfo {year} {2019})}\BibitemShut
  {NoStop}%
\bibitem [{\citenamefont {Marzuola}\ \emph {et~al.}(2019)\citenamefont
  {Marzuola}, \citenamefont {Rechtsman}, \citenamefont {Osting},\ and\
  \citenamefont {Bandres}}]{Marzuola2019}%
  \BibitemOpen
  \bibfield  {author} {\bibinfo {author} {\bibfnamefont {J.~L.}\ \bibnamefont
  {Marzuola}}, \bibinfo {author} {\bibfnamefont {M.}~\bibnamefont {Rechtsman}},
  \bibinfo {author} {\bibfnamefont {B.}~\bibnamefont {Osting}},\ and\ \bibinfo
  {author} {\bibfnamefont {M.}~\bibnamefont {Bandres}},\ }\href
  {http://arxiv.org/abs/1904.10312} {\bibfield  {journal} {\bibinfo  {journal}
  {arXiv:1904.10312}\ } (\bibinfo {year} {2019})}\BibitemShut {NoStop}%
\bibitem [{\citenamefont {Mukherjee}\ and\ \citenamefont
  {Rechtsman}(2020)}]{Mukherjee2020}%
  \BibitemOpen
  \bibfield  {author} {\bibinfo {author} {\bibfnamefont {S.}~\bibnamefont
  {Mukherjee}}\ and\ \bibinfo {author} {\bibfnamefont {M.~C.}\ \bibnamefont
  {Rechtsman}},\ }\href {https://doi.org/10.1126/science.aba8725} {\bibfield
  {journal} {\bibinfo  {journal} {Science}\ }\textbf {\bibinfo {volume}
  {368}},\ \bibinfo {pages} {856} (\bibinfo {year} {2020})}\BibitemShut
  {NoStop}%
\bibitem [{\citenamefont {Chen}\ \emph {et~al.}(2014)\citenamefont {Chen},
  \citenamefont {Upadhyaya},\ and\ \citenamefont {Vitelli}}]{Chen2014}%
  \BibitemOpen
  \bibfield  {author} {\bibinfo {author} {\bibfnamefont {B.~G.-g.}\
  \bibnamefont {Chen}}, \bibinfo {author} {\bibfnamefont {N.}~\bibnamefont
  {Upadhyaya}},\ and\ \bibinfo {author} {\bibfnamefont {V.}~\bibnamefont
  {Vitelli}},\ }\href {https://doi.org/10.1073/pnas.1405969111} {\bibfield
  {journal} {\bibinfo  {journal} {Proc. Natl. Acad. Sci. USA}\ }\textbf
  {\bibinfo {volume} {111}},\ \bibinfo {pages} {13004} (\bibinfo {year}
  {2014})}\BibitemShut {NoStop}%
\bibitem [{\citenamefont {Hadad}\ \emph {et~al.}(2017)\citenamefont {Hadad},
  \citenamefont {Vitelli},\ and\ \citenamefont {Alu}}]{Hadad2017}%
  \BibitemOpen
  \bibfield  {author} {\bibinfo {author} {\bibfnamefont {Y.}~\bibnamefont
  {Hadad}}, \bibinfo {author} {\bibfnamefont {V.}~\bibnamefont {Vitelli}},\
  and\ \bibinfo {author} {\bibfnamefont {A.}~\bibnamefont {Alu}},\ }\href
  {https://doi.org/10.1021/acsphotonics.7b00303} {\bibfield  {journal}
  {\bibinfo  {journal} {ACS Photon.}\ }\textbf {\bibinfo {volume} {4}},\
  \bibinfo {pages} {1974} (\bibinfo {year} {2017})}\BibitemShut {NoStop}%
\bibitem [{\citenamefont {Poddubny}\ and\ \citenamefont
  {Smirnova}(2018)}]{Poddubny2018}%
  \BibitemOpen
  \bibfield  {author} {\bibinfo {author} {\bibfnamefont {A.~N.}\ \bibnamefont
  {Poddubny}}\ and\ \bibinfo {author} {\bibfnamefont {D.~A.}\ \bibnamefont
  {Smirnova}},\ }\href {https://doi.org/10.1103/PhysRevA.98.013827} {\bibfield
  {journal} {\bibinfo  {journal} {Phys. Rev. A}\ }\textbf {\bibinfo {volume}
  {98}},\ \bibinfo {pages} {013827} (\bibinfo {year} {2018})}\BibitemShut
  {NoStop}%
\bibitem [{\citenamefont {Ablowitz}\ \emph {et~al.}(2014)\citenamefont
  {Ablowitz}, \citenamefont {Curtis},\ and\ \citenamefont {Ma}}]{Ablowitz2014}%
  \BibitemOpen
  \bibfield  {author} {\bibinfo {author} {\bibfnamefont {M.~J.}\ \bibnamefont
  {Ablowitz}}, \bibinfo {author} {\bibfnamefont {C.~W.}\ \bibnamefont
  {Curtis}},\ and\ \bibinfo {author} {\bibfnamefont {Y.-P.}\ \bibnamefont
  {Ma}},\ }\href {https://doi.org/10.1103/PhysRevA.90.023813} {\bibfield
  {journal} {\bibinfo  {journal} {Phys. Rev. A}\ }\textbf {\bibinfo {volume}
  {90}},\ \bibinfo {pages} {023813} (\bibinfo {year} {2014})}\BibitemShut
  {NoStop}%
\bibitem [{\citenamefont {Leykam}\ and\ \citenamefont
  {Chong}(2016)}]{Leykam2016}%
  \BibitemOpen
  \bibfield  {author} {\bibinfo {author} {\bibfnamefont {D.}~\bibnamefont
  {Leykam}}\ and\ \bibinfo {author} {\bibfnamefont {Y.~D.}\ \bibnamefont
  {Chong}},\ }\href {https://doi.org/10.1103/PhysRevLett.117.143901} {\bibfield
   {journal} {\bibinfo  {journal} {Phys. Rev. Lett.}\ }\textbf {\bibinfo
  {volume} {117}},\ \bibinfo {pages} {143901} (\bibinfo {year}
  {2016})}\BibitemShut {NoStop}%
\bibitem [{\citenamefont {Kartashov}\ and\ \citenamefont
  {Skryabin}(2016)}]{Kartashov2016}%
  \BibitemOpen
  \bibfield  {author} {\bibinfo {author} {\bibfnamefont {Y.~V.}\ \bibnamefont
  {Kartashov}}\ and\ \bibinfo {author} {\bibfnamefont {D.~V.}\ \bibnamefont
  {Skryabin}},\ }\href {https://doi.org/10.1364/optica.3.001228} {\bibfield
  {journal} {\bibinfo  {journal} {Optica}\ }\textbf {\bibinfo {volume} {3}},\
  \bibinfo {pages} {1228} (\bibinfo {year} {2016})}\BibitemShut {NoStop}%
\bibitem [{\citenamefont {Snee}\ and\ \citenamefont {Ma}(2019)}]{Snee2019}%
  \BibitemOpen
  \bibfield  {author} {\bibinfo {author} {\bibfnamefont {D.~D.}\ \bibnamefont
  {Snee}}\ and\ \bibinfo {author} {\bibfnamefont {Y.-P.}\ \bibnamefont {Ma}},\
  }\href {https://doi.org/https://doi.org/10.1016/j.eml.2019.100487} {\bibfield
   {journal} {\bibinfo  {journal} {Extreme Mech. Lett.}\ }\textbf {\bibinfo
  {volume} {30}},\ \bibinfo {pages} {100487} (\bibinfo {year}
  {2019})}\BibitemShut {NoStop}%
\bibitem [{\citenamefont {Tao}\ \emph {et~al.}(2020)\citenamefont {Tao},
  \citenamefont {Dai}, \citenamefont {Yang}, \citenamefont {Zeng},\ and\
  \citenamefont {Xu}}]{Tao2020}%
  \BibitemOpen
  \bibfield  {author} {\bibinfo {author} {\bibfnamefont {Y.-L.}\ \bibnamefont
  {Tao}}, \bibinfo {author} {\bibfnamefont {N.}~\bibnamefont {Dai}}, \bibinfo
  {author} {\bibfnamefont {Y.-B.}\ \bibnamefont {Yang}}, \bibinfo {author}
  {\bibfnamefont {Q.-B.}\ \bibnamefont {Zeng}},\ and\ \bibinfo {author}
  {\bibfnamefont {Y.}~\bibnamefont {Xu}},\ }\href
  {http://arxiv.org/abs/2005.04433} {\bibfield  {journal} {\bibinfo  {journal}
  {arXiv:2005.04433}\ } (\bibinfo {year} {2020})}\BibitemShut {NoStop}%
\bibitem [{\citenamefont {Ablowitz}\ \emph {et~al.}(2021)\citenamefont
  {Ablowitz}, \citenamefont {Cole}, \citenamefont {Hu},\ and\ \citenamefont
  {Rosenthal}}]{Abl21a}%
  \BibitemOpen
  \bibfield  {author} {\bibinfo {author} {\bibfnamefont {M.~J.}\ \bibnamefont
  {Ablowitz}}, \bibinfo {author} {\bibfnamefont {J.~T.}\ \bibnamefont {Cole}},
  \bibinfo {author} {\bibfnamefont {P.}~\bibnamefont {Hu}},\ and\ \bibinfo
  {author} {\bibfnamefont {P.}~\bibnamefont {Rosenthal}},\ }\href
  {https://doi.org/10.1103/PhysRevE.103.042214} {\bibfield  {journal} {\bibinfo
   {journal} {Phys. Rev. E}\ }\textbf {\bibinfo {volume} {103}},\ \bibinfo
  {pages} {042214} (\bibinfo {year} {2021})}\BibitemShut {NoStop}%
\bibitem [{\citenamefont {Ablowitz}\ and\ \citenamefont {Cole}(2019)}]{Abl19a}%
  \BibitemOpen
  \bibfield  {author} {\bibinfo {author} {\bibfnamefont {M.~J.}\ \bibnamefont
  {Ablowitz}}\ and\ \bibinfo {author} {\bibfnamefont {J.~T.}\ \bibnamefont
  {Cole}},\ }\href {https://doi.org/10.1103/PhysRevA.99.033821} {\bibfield
  {journal} {\bibinfo  {journal} {Phys. Rev. A}\ }\textbf {\bibinfo {volume}
  {99}},\ \bibinfo {pages} {033821} (\bibinfo {year} {2019})}\BibitemShut
  {NoStop}%
\bibitem [{\citenamefont {Smirnova}\ \emph {et~al.}(2020)\citenamefont
  {Smirnova}, \citenamefont {Leykam}, \citenamefont {Chong},\ and\
  \citenamefont {Kivshar}}]{Smirnova2020}%
  \BibitemOpen
  \bibfield  {author} {\bibinfo {author} {\bibfnamefont {D.}~\bibnamefont
  {Smirnova}}, \bibinfo {author} {\bibfnamefont {D.}~\bibnamefont {Leykam}},
  \bibinfo {author} {\bibfnamefont {Y.}~\bibnamefont {Chong}},\ and\ \bibinfo
  {author} {\bibfnamefont {Y.}~\bibnamefont {Kivshar}},\ }\href
  {https://aip.scitation.org/doi/10.1063/1.5142397} {\bibfield  {journal}
  {\bibinfo  {journal} {Appl. Phys. Rev.}\ }\textbf {\bibinfo {volume} {7}},\
  \bibinfo {pages} {021306} (\bibinfo {year} {2020})}\BibitemShut {NoStop}%
\bibitem [{\citenamefont {Ma}\ \emph {et~al.}(2021)\citenamefont {Ma},
  \citenamefont {Grushin},\ and\ \citenamefont {Burch}}]{Ma2021}%
  \BibitemOpen
  \bibfield  {author} {\bibinfo {author} {\bibfnamefont {Q.}~\bibnamefont
  {Ma}}, \bibinfo {author} {\bibfnamefont {A.}~\bibnamefont {Grushin}},\ and\
  \bibinfo {author} {\bibfnamefont {K.}~\bibnamefont {Burch}},\ }\bibfield
  {journal} {\bibinfo  {journal} {Nat. Mater.}\ }\href
  {https://doi.org/10.1038/s41563-021-00992-7} {10.1038/s41563-021-00992-7}
  (\bibinfo {year} {2021})\BibitemShut {NoStop}%
\bibitem [{\citenamefont {Ablowitz}\ and\ \citenamefont {Cole}(2022)}]{cole}%
  \BibitemOpen
  \bibfield  {author} {\bibinfo {author} {\bibfnamefont {M.~J.}\ \bibnamefont
  {Ablowitz}}\ and\ \bibinfo {author} {\bibfnamefont {J.~T.}\ \bibnamefont
  {Cole}},\ }\href
  {https://doi.org/https://doi.org/10.1016/j.physd.2022.133440} {\bibfield
  {journal} {\bibinfo  {journal} {Physica D: Nonlinear Phenomena}\ }\textbf
  {\bibinfo {volume} {440}},\ \bibinfo {pages} {133440} (\bibinfo {year}
  {2022})}\BibitemShut {NoStop}%
\bibitem [{\citenamefont {Mukherjee}\ and\ \citenamefont
  {Rechtsman}(2021)}]{recht21}%
  \BibitemOpen
  \bibfield  {author} {\bibinfo {author} {\bibfnamefont {S.}~\bibnamefont
  {Mukherjee}}\ and\ \bibinfo {author} {\bibfnamefont {M.~C.}\ \bibnamefont
  {Rechtsman}},\ }\href@noop {} {\bibfield  {journal} {\bibinfo  {journal}
  {Phys. Rev. X (accepted)}\ } (\bibinfo {year} {2021})}\BibitemShut {NoStop}%
\bibitem [{\citenamefont {J\"urgensen}\ and\ \citenamefont
  {Rechtsman}(2022)}]{PhysRevLett.128.113901}%
  \BibitemOpen
  \bibfield  {author} {\bibinfo {author} {\bibfnamefont {M.}~\bibnamefont
  {J\"urgensen}}\ and\ \bibinfo {author} {\bibfnamefont {M.~C.}\ \bibnamefont
  {Rechtsman}},\ }\href {https://doi.org/10.1103/PhysRevLett.128.113901}
  {\bibfield  {journal} {\bibinfo  {journal} {Phys. Rev. Lett.}\ }\textbf
  {\bibinfo {volume} {128}},\ \bibinfo {pages} {113901} (\bibinfo {year}
  {2022})}\BibitemShut {NoStop}%
\bibitem [{\citenamefont {J\"{u}rgensen}\ \emph {et~al.}(2021)\citenamefont
  {J\"{u}rgensen}, \citenamefont {Mukherjee},\ and\ \citenamefont
  {Rechtsman}}]{Jurgensen2021}%
  \BibitemOpen
  \bibfield  {author} {\bibinfo {author} {\bibfnamefont {M.}~\bibnamefont
  {J\"{u}rgensen}}, \bibinfo {author} {\bibfnamefont {S.}~\bibnamefont
  {Mukherjee}},\ and\ \bibinfo {author} {\bibfnamefont {M.~C.}\ \bibnamefont
  {Rechtsman}},\ }\href {https://doi.org/10.1038/s41586-021-03688-9} {\bibfield
   {journal} {\bibinfo  {journal} {Nature}\ }\textbf {\bibinfo {volume}
  {596}},\ \bibinfo {pages} {63} (\bibinfo {year} {2021})}\BibitemShut
  {NoStop}%
\bibitem [{\citenamefont {Haldane}\ and\ \citenamefont {Raghu}(2008)}]{raghu1}%
  \BibitemOpen
  \bibfield  {author} {\bibinfo {author} {\bibfnamefont {F.~D.~M.}\
  \bibnamefont {Haldane}}\ and\ \bibinfo {author} {\bibfnamefont
  {S.}~\bibnamefont {Raghu}},\ }\href
  {https://doi.org/10.1103/PhysRevLett.100.013904} {\bibfield  {journal}
  {\bibinfo  {journal} {Phys. Rev. Lett.}\ }\textbf {\bibinfo {volume} {100}},\
  \bibinfo {pages} {013904} (\bibinfo {year} {2008})}\BibitemShut {NoStop}%
\bibitem [{\citenamefont {Raghu}\ and\ \citenamefont {Haldane}(2008)}]{raghu2}%
  \BibitemOpen
  \bibfield  {author} {\bibinfo {author} {\bibfnamefont {S.}~\bibnamefont
  {Raghu}}\ and\ \bibinfo {author} {\bibfnamefont {F.~D.~M.}\ \bibnamefont
  {Haldane}},\ }\href@noop {} {\bibfield  {journal} {\bibinfo  {journal}
  {Physical Review A}\ }\textbf {\bibinfo {volume} {78}},\ \bibinfo {pages}
  {033834} (\bibinfo {year} {2008})}\BibitemShut {NoStop}%
\bibitem [{\citenamefont {Parker}\ \emph {et~al.}(2022)\citenamefont {Parker},
  \citenamefont {Aceves}, \citenamefont {Cuevas-Maraver},\ and\ \citenamefont
  {Kevrekidis}}]{PhysRevE.105.044211}%
  \BibitemOpen
  \bibfield  {author} {\bibinfo {author} {\bibfnamefont {R.}~\bibnamefont
  {Parker}}, \bibinfo {author} {\bibfnamefont {A.}~\bibnamefont {Aceves}},
  \bibinfo {author} {\bibfnamefont {J.}~\bibnamefont {Cuevas-Maraver}},\ and\
  \bibinfo {author} {\bibfnamefont {P.~G.}\ \bibnamefont {Kevrekidis}},\ }\href
  {https://doi.org/10.1103/PhysRevE.105.044211} {\bibfield  {journal} {\bibinfo
   {journal} {Phys. Rev. E}\ }\textbf {\bibinfo {volume} {105}},\ \bibinfo
  {pages} {044211} (\bibinfo {year} {2022})}\BibitemShut {NoStop}%
\bibitem [{\citenamefont {Thouless}(1983)}]{PhysRevB.27.6083}%
  \BibitemOpen
  \bibfield  {author} {\bibinfo {author} {\bibfnamefont {D.~J.}\ \bibnamefont
  {Thouless}},\ }\href {https://doi.org/10.1103/PhysRevB.27.6083} {\bibfield
  {journal} {\bibinfo  {journal} {Phys. Rev. B}\ }\textbf {\bibinfo {volume}
  {27}},\ \bibinfo {pages} {6083} (\bibinfo {year} {1983})}\BibitemShut
  {NoStop}%
\bibitem [{\citenamefont {Kronig}\ and\ \citenamefont {Penney}(1931)}]{kronig}%
  \BibitemOpen
  \bibfield  {author} {\bibinfo {author} {\bibfnamefont {R.~D.~L.}\
  \bibnamefont {Kronig}}\ and\ \bibinfo {author} {\bibfnamefont {P.~W.~G.}\
  \bibnamefont {Penney}},\ }\href@noop {} {\bibfield  {journal} {\bibinfo
  {journal} {Proc. R. Soc. Lond. A}\ }\textbf {\bibinfo {volume} {130}},\
  \bibinfo {pages} {499} (\bibinfo {year} {1931})}\BibitemShut {NoStop}%
\bibitem [{\citenamefont {Kenkre}\ and\ \citenamefont
  {Campbell}(1986)}]{Kenkre1986}%
  \BibitemOpen
  \bibfield  {author} {\bibinfo {author} {\bibfnamefont {V.~M.}\ \bibnamefont
  {Kenkre}}\ and\ \bibinfo {author} {\bibfnamefont {D.~K.}\ \bibnamefont
  {Campbell}},\ }\href {https://doi.org/10.1103/PhysRevB.34.4959} {\bibfield
  {journal} {\bibinfo  {journal} {Phys. Rev. B}\ }\textbf {\bibinfo {volume}
  {34}},\ \bibinfo {pages} {4959} (\bibinfo {year} {1986})}\BibitemShut
  {NoStop}%
\bibitem [{\citenamefont {Aubry}\ and\ \citenamefont
  {André}(1980)}]{Aubry1980}%
  \BibitemOpen
  \bibfield  {author} {\bibinfo {author} {\bibfnamefont {S.}~\bibnamefont
  {Aubry}}\ and\ \bibinfo {author} {\bibfnamefont {G.}~\bibnamefont {André}},\
  }\href@noop {} {\bibfield  {journal} {\bibinfo  {journal} {Proceedings, VIII
  International Colloquium on Group-Theoretical Methods in Physics}\ }\textbf
  {\bibinfo {volume} {3}} (\bibinfo {year} {1980})}\BibitemShut {NoStop}%
\bibitem [{\citenamefont {Harper}(1955)}]{Harper1955}%
  \BibitemOpen
  \bibfield  {author} {\bibinfo {author} {\bibfnamefont {P.~G.}\ \bibnamefont
  {Harper}},\ }\href {https://doi.org/10.1088/0370-1298/68/10/304} {\bibfield
  {journal} {\bibinfo  {journal} {Proc. Phys. Soc. A}\ }\textbf {\bibinfo
  {volume} {68}},\ \bibinfo {pages} {874} (\bibinfo {year} {1955})}\BibitemShut
  {NoStop}%
\bibitem [{\citenamefont {Centurion}\ \emph
  {et~al.}(2006{\natexlab{a}})\citenamefont {Centurion}, \citenamefont
  {Porter}, \citenamefont {Kevrekidis},\ and\ \citenamefont
  {Psaltis}}]{PhysRevLett.97.033903}%
  \BibitemOpen
  \bibfield  {author} {\bibinfo {author} {\bibfnamefont {M.}~\bibnamefont
  {Centurion}}, \bibinfo {author} {\bibfnamefont {M.~A.}\ \bibnamefont
  {Porter}}, \bibinfo {author} {\bibfnamefont {P.~G.}\ \bibnamefont
  {Kevrekidis}},\ and\ \bibinfo {author} {\bibfnamefont {D.}~\bibnamefont
  {Psaltis}},\ }\href {https://doi.org/10.1103/PhysRevLett.97.033903}
  {\bibfield  {journal} {\bibinfo  {journal} {Phys. Rev. Lett.}\ }\textbf
  {\bibinfo {volume} {97}},\ \bibinfo {pages} {033903} (\bibinfo {year}
  {2006}{\natexlab{a}})}\BibitemShut {NoStop}%
\bibitem [{\citenamefont {Centurion}\ \emph
  {et~al.}(2006{\natexlab{b}})\citenamefont {Centurion}, \citenamefont
  {Porter}, \citenamefont {Pu}, \citenamefont {Kevrekidis}, \citenamefont
  {Frantzeskakis},\ and\ \citenamefont {Psaltis}}]{PhysRevLett.97.234101}%
  \BibitemOpen
  \bibfield  {author} {\bibinfo {author} {\bibfnamefont {M.}~\bibnamefont
  {Centurion}}, \bibinfo {author} {\bibfnamefont {M.~A.}\ \bibnamefont
  {Porter}}, \bibinfo {author} {\bibfnamefont {Y.}~\bibnamefont {Pu}}, \bibinfo
  {author} {\bibfnamefont {P.~G.}\ \bibnamefont {Kevrekidis}}, \bibinfo
  {author} {\bibfnamefont {D.~J.}\ \bibnamefont {Frantzeskakis}},\ and\
  \bibinfo {author} {\bibfnamefont {D.}~\bibnamefont {Psaltis}},\ }\href
  {https://doi.org/10.1103/PhysRevLett.97.234101} {\bibfield  {journal}
  {\bibinfo  {journal} {Phys. Rev. Lett.}\ }\textbf {\bibinfo {volume} {97}},\
  \bibinfo {pages} {234101} (\bibinfo {year} {2006}{\natexlab{b}})}\BibitemShut
  {NoStop}%
\bibitem [{\citenamefont {Kenkre}(1989)}]{Kenkre1989}%
  \BibitemOpen
  \bibfield  {author} {\bibinfo {author} {\bibfnamefont {V.~M.}\ \bibnamefont
  {Kenkre}},\ }in\ \href@noop {} {\emph {\bibinfo {booktitle} {Singular
  Behavior and Nonlinear Dynamics}}},\ \bibinfo {editor} {edited by\ \bibinfo
  {editor} {\bibfnamefont {S.}~\bibnamefont {Pnevmatikos}}, \bibinfo {editor}
  {\bibfnamefont {T.}~\bibnamefont {Bountis}},\ and\ \bibinfo {editor}
  {\bibfnamefont {S.}~\bibnamefont {Pnevmatikos}}}\ (\bibinfo  {publisher}
  {World Scientific},\ \bibinfo {address} {Singapore},\ \bibinfo {year}
  {1989})\BibitemShut {NoStop}%
\bibitem [{{\relax DLMF}()}]{NIST:DLMF}%
  \BibitemOpen
  {\relax DLMF},\ \href {https://dlmf.nist.gov/} {\bibinfo {title} {{\it NIST
  Digital Library of Mathematical Functions}}},\ \bibinfo {howpublished}
  {\url{https://dlmf.nist.gov/}, Release 1.1.9 of 2023-03-15},\ \bibinfo {note}
  {f.~W.~J. Olver, A.~B. {Olde Daalhuis}, D.~W. Lozier, B.~I. Schneider, R.~F.
  Boisvert, C.~W. Clark, B.~R. Miller, B.~V. Saunders, H.~S. Cohl, and M.~A.
  McClain, eds.}\BibitemShut {Stop}%
\bibitem [{\citenamefont {Carlson}(2006)}]{Carlson2006}%
  \BibitemOpen
  \bibfield  {author} {\bibinfo {author} {\bibfnamefont {B.~C.}\ \bibnamefont
  {Carlson}},\ }\href {https://doi.org/10.1016/j.jmaa.2005.10.063} {\bibfield
  {journal} {\bibinfo  {journal} {Journal of Mathematical Analysis and
  Applications}\ }\textbf {\bibinfo {volume} {323}},\ \bibinfo {pages} {522}
  (\bibinfo {year} {2006})}\BibitemShut {NoStop}%
\bibitem [{\citenamefont {Doedel}\ \emph {et~al.}(2007)\citenamefont {Doedel},
  \citenamefont {Fairgrieve}, \citenamefont {Sandstede}, \citenamefont
  {Champneys}, \citenamefont {Kuznetsov},\ and\ \citenamefont
  {Wang}}]{auto07p}%
  \BibitemOpen
  \bibfield  {author} {\bibinfo {author} {\bibfnamefont {E.~J.}\ \bibnamefont
  {Doedel}}, \bibinfo {author} {\bibfnamefont {T.~F.}\ \bibnamefont
  {Fairgrieve}}, \bibinfo {author} {\bibfnamefont {B.}~\bibnamefont
  {Sandstede}}, \bibinfo {author} {\bibfnamefont {A.~R.}\ \bibnamefont
  {Champneys}}, \bibinfo {author} {\bibfnamefont {Y.~A.}\ \bibnamefont
  {Kuznetsov}},\ and\ \bibinfo {author} {\bibfnamefont {X.}~\bibnamefont
  {Wang}},\ }\href@noop {} {\  (\bibinfo {year} {2007})}\BibitemShut {NoStop}%
\bibitem [{\citenamefont {Oxtoby}\ and\ \citenamefont
  {Barashenkov}(2007)}]{igorb}%
  \BibitemOpen
  \bibfield  {author} {\bibinfo {author} {\bibfnamefont {O.~F.}\ \bibnamefont
  {Oxtoby}}\ and\ \bibinfo {author} {\bibfnamefont {I.~V.}\ \bibnamefont
  {Barashenkov}},\ }\href {https://doi.org/10.1103/PhysRevE.76.036603}
  {\bibfield  {journal} {\bibinfo  {journal} {Phys. Rev. E}\ }\textbf {\bibinfo
  {volume} {76}},\ \bibinfo {pages} {036603} (\bibinfo {year}
  {2007})}\BibitemShut {NoStop}%
\bibitem [{\citenamefont {Konotop}\ \emph {et~al.}(2016)\citenamefont
  {Konotop}, \citenamefont {Yang},\ and\ \citenamefont
  {Zezyulin}}]{yangkonotop}%
  \BibitemOpen
  \bibfield  {author} {\bibinfo {author} {\bibfnamefont {V.~V.}\ \bibnamefont
  {Konotop}}, \bibinfo {author} {\bibfnamefont {J.}~\bibnamefont {Yang}},\ and\
  \bibinfo {author} {\bibfnamefont {D.~A.}\ \bibnamefont {Zezyulin}},\ }\href
  {https://doi.org/10.1103/RevModPhys.88.035002} {\bibfield  {journal}
  {\bibinfo  {journal} {Rev. Mod. Phys.}\ }\textbf {\bibinfo {volume} {88}},\
  \bibinfo {pages} {035002} (\bibinfo {year} {2016})}\BibitemShut {NoStop}%
\bibitem [{\citenamefont {Christodoulides}\ and\ \citenamefont
  {Yang}(2018)}]{yangchristo}%
  \BibitemOpen
  \bibfield  {author} {\bibinfo {author} {\bibfnamefont {D.}~\bibnamefont
  {Christodoulides}}\ and\ \bibinfo {author} {\bibfnamefont {J.}~\bibnamefont
  {Yang}},\ }\href@noop {} {\emph {\bibinfo {title} {Parity-time Symmetry and
  Its Applications}}}\ (\bibinfo  {publisher} {Springer Singapore},\ \bibinfo
  {year} {2018})\BibitemShut {NoStop}%
\end{thebibliography}%

\end{document}